\documentclass[nofootinbib,aps,prd,groupedaddress,preprintnumbers,%
  showpacs,showkeys,floatfix,amssymb,amsfonts,twocolumn]{revtex4-1}  
\usepackage{hyperref}
\usepackage[usenames]{color}
\usepackage{todonotes}
\usepackage{bbold}
\usepackage{amsmath}
\usepackage{multirow}
\usepackage{graphicx}
\usepackage{xfrac}
\usepackage[utf8]{inputenc}
\newcommand{\tr}{\operatorname{Tr}}
\usepackage{dsfont}  
\usepackage{amsfonts}
\usepackage{amssymb}

\newcommand{\be}{\begin{equation}}  
\newcommand{\ee}{\end{equation}}  
\newcommand{\beq}{\begin{eqnarray}}  
\newcommand{\eeq}{\end{eqnarray}}

\begin{document}
\title{Proton and neutron electromagnetic form factors from lattice QCD}
\preprint{DESY 18-033}
\author{
  C.~Alexandrou$^{1,2}$,
  S.~Bacchio$^1$,
  M.~Constantinou$^{3}$,
  J.~Finkenrath$^{2}$
  K.~Hadjiyiannakou$^{2}$,
  K.~Jansen$^{4}$,
  G.~Koutsou$^{2}$, and
  A.~Vaquero Aviles-Casco$^5$
}
\affiliation{
  $^1$Department of Physics, University of Cyprus,  P.O. Box 20537,  1678 Nicosia, Cyprus\\
  $^2$Computation-based Science and Technology Research Center,
  The Cyprus Institute, 20 Kavafi Str., Nicosia 2121, Cyprus \\
  $^3$Department of Physics, Temple University, 1925 N. 12th Street, Philadelphia, PA 19122-1801, USA\\
  $^4$NIC, DESY, Platanenallee 6,  D-15738 Zeuthen,  Germany\\
  $^5$Department of Physics and Astronomy, University of Utah, Salt Lake City, UT 84112, USA}

\begin{abstract}
  The electromagnetic form factors of the proton and the neutron are computed within lattice QCD using simulations with  quarks masses fixed to their physical values. Both connected and disconnected contributions are computed.
  We analyze two new ensembles of $N_f=2$ and $N_f=2+1+1$ twisted mass clover-improved fermions  and determine   the proton and neutron form factors, the electric and magnetic radii, and the magnetic moments. We   use  several values of the sink-source time 
  separation in the range of 1.0~fm to 1.6~fm to ensure ground state identification.  Disconnected contributions are calculated to an unprecedented accuracy  at the physical point. Although they constitute a small correction, they are  non-negligible and contribute up to 15\% for the case of the neutron electric charge radius.

\end{abstract}
\pacs{11.15.Ha, 12.38.Gc, 24.85.+p, 12.38.Aw, 12.38.-t}
\keywords{Nucleon structure, Nucleon electromagnetic form factors, Disconnected, Lattice QCD}
\maketitle

\newcommand{\Op}{\mathcal{O}} 
\newcommand{\C}{\mathcal{C}} 
\newcommand{\eins}{\mathds{1}} 

\section{Introduction}
Nucleons, being composite particles, have a non-trivial internal structure that can be probed by measuring  their electromagnetic form factors. These fundamental quantities have been extensively
studied both theoretically and experimentally. However, open issues still persist and there are on-going experimental efforts to determine them at higher precision and over a wider range of  momentum transfers and to describe them theoretically.  
 The proton electric form factor is extracted to high precision from electron proton scattering~\cite{Bernauer:2013tpr}. Its slope at vanishing momentum transfer squared yields the proton charge 
root mean square (r.m.s) radius.
Prior to 2010, the charge r.m.s. radius of the proton  was considered a well-determined quantity (see Ref.~\cite{Punjabi:2015bba} for a recent review). A pioneering experiment using Lamb shifts in muonic hydrogen surprisingly found a value smaller by five standard deviations~\cite{Pohl:2010zza},
 triggering the so-called proton radius puzzle. The origin of this discrepancy is not yet  understood,
 and potential systematic uncertainties related to the analysis methodologies in the two types of experiments have not been excluded.
 Another quantity of interest is the neutron electric form factor~\cite{Golak:2000nt}, which is accessed indirectly experimentally through electron-deuteron or electron-helium scattering and therefore remains poorly-known.
It is of substantial importance to compute these fundamental quantities from first principles using lattice QCD, which provides an ideal formulation for such an investigation and with simulations at physical values of the QCD parameters.

Within this work, we compute the proton and neutron electromagnetic form factors including light quark disconnected 
contributions. We  use an ensemble of  twisted mass fermions  with two degenerate light quarks, a 
strange and a charm quark ($N_f{=}2{+}1{+}1$) with masses fixed to their physical value (referred to hereafter as physical point). A clover term is added to the action to suppress isospin breaking effects  that come quadratically with the lattice spacing.  Details on the simulation can be found in Ref.~\cite{Alexandrou:2018egz}. We will refer to this ensemble as cB211.072.64.
In addition, we present an analysis of a twisted mass ensemble of two degenerate light quarks with masses fixed to their physical values ($N_f=2$) to assess finite volume artifacts by comparing to previous results  obtained using an $N_f=2$ ensemble with a smaller volume and same pion mass and lattice spacing~\cite{Alexandrou:2018zdf,Alexandrou:2017ypw}. Comparison between $N_f=2$ and $N_f=2+1+1$ also sheds light on any possible unquenching effect of the strange and charm quarks. 
The momentum dependence of the form factors is fitted using two Ans\"atze,  namely either a dipole
or the  Galster-like parameterization~\cite{Galster:1971kv} and the model independent z-expansion~\cite{Hill:2010yb}. The fits  allow 
for the extraction of the magnetic moment and the electric and magnetic r.m.s radii of the
proton and neutron and provide a measure of the systematics due to the choice of the fit method.

A crucial component of our analysis is the use of hierarchical probing~\cite{Stathopoulos:2013aci} combined with deflation of the lower lying eigenvalues~\cite{Gambhir:2016uwp} that enables us to
calculate the light quark disconnected contributions to the form factors at an unprecedented accuracy at the physical point. This allows us to obtain the proton and neutron form factors at the physical point without neglecting disconnected contributions. 

The remainder of this paper is organized as follows:
In Section~\ref{sec:EMff}, we describe the  nucleon matrix elements 
required to extract the electromagnetic form factors and
in Section~\ref{sec:LMeth} we provide details on the lattice QCD techniques employed for the computation of the  connected and 
disconnected diagrams. In Section~\ref{sec:Analysis}, we discuss the analysis of the data paying particular attention to the identification of the ground state matrix element. In Section~\ref{sec:latArtifacts} we include an assessment of finite volume and unquenching effects using results from the analysis of the two $N_f{=}2$ ensembles.
In Section~\ref{sec:Isov Isos}, we fit the isovector and isoscalar form factors to extract the  magnetic moments and radii. We compare to other lattice QCD studies using simulations close to physical pion masses in Section~\ref{sec:comp} \cite{Capitani:2015sba,Green:2014xba,Hasan:2017wwt,Ishikawa:2018rew}.
Our final results for the proton and neutron electromagnetic form factors  are given in Section~\ref{sec:PNFF}.
Finally, in Section~\ref{sec:conclusions}, we summarize our findings and conclude. For completeness, we summarize  in Appendix~\ref{sec:appendix equations} 
the decomposition of the nucleon matrix elements in terms of the form factors and in Appendix~\ref{sec:appendix results} we provide 
a table with the numerical results for the electric and magnetic form factors as a function of the momentum 
transfer squared.

\section{Electromagnetic form factors} \label{sec:EMff}

The nucleon matrix element of the electromagnetic current is parameterized in terms of the Dirac ($F_1$) and Pauli ($F_2$) form
factors given in Minkowski space by,
\begin{eqnarray}
  && \langle N(p',s') \vert j_\mu \vert N(p,s) \rangle = \sqrt{\frac{m_N^2}{E_N(\vec{p}\,') E_N(\vec{p})}} \times \nonumber \\
  && \bar{u}_N(p',s') \left[ \gamma_\mu F_1(q^2) + \frac{i \sigma_{\mu\nu} q^\nu}{2 m_N} F_2(q^2) \right] u_N(p,s)\,.
  \label{Eq:Decomposition}
\end{eqnarray}
$N(p,s)$ is the nucleon state with initial (final) momentum $p$ ($p'$)
and spin $s$ ($s'$), with energy $E_N(\vec{p})$ ($E_N(\vec{p}\,') $)
and mass $m_N$. $q^2{\equiv}q_\mu q^\mu$ is the momentum transfer squared $q_\mu{=}(p_\mu'-p_\mu)$ and
$u_N$ is the nucleon spinor.
The  local vector current $j_\mu$ is given by,
\begin{equation}
  j_\mu = \sum_f e_f \; j_\mu^f = \sum_f e_f \; \bar{q}_f \gamma_\mu q_f\,,
\end{equation}
where $q_f$ is the quark field of flavor $f$ and $e_f$  its electric charge, and the summation runs over all the quark flavors.
Instead of the local vector current, we instead  use  the symmetrized lattice conserved vector current given by
\begin{eqnarray*}
  j_\mu^f(x) = \frac{1}{4} &[& \bar{q}_f(x+\hat{\mu}) U^\dagger_\mu (x) (1+\gamma_\mu) q_f(x) \nonumber \\
    &-& \bar{q}_f(x) U_\mu(x) (1-\gamma_\mu) q_f(x+\hat{\mu}) \nonumber \\
    &+& \bar{q}_f(x) U^\dagger_\mu(x-\hat{\mu}) (1+\gamma_\mu) q_f(x-\hat{\mu}) \nonumber \\
    &-& \bar{q}_f(x-\hat{\mu}) U_\mu(x-\hat{\mu}) (1-\gamma_\mu) q_f(x) \;]\,,
  \label{Eq:conserved}
\end{eqnarray*}
which, unlike the local vector current, does not need  renormalization.
The electric and magnetic Sachs form factors $G_E(q^2)$ and $G_M(q^2)$ are
alternative Lorentz invariant quantities and are expressed in terms of $F_1(q^2)$ and $F_2(q^2)$ via the 
relations,
\begin{equation}
  G_E(q^2) = F_1(q^2) + \frac{q^2}{4m_N^2} F_2(q^2)\,,
\end{equation}
\begin{equation}
  G_M(q^2) = F_1(q^2) + F_2(q^2)\,.
\end{equation}
In the isospin limit, where the up and down quarks are degenerate, we consider the 
isovector combination $\langle p \vert j_\mu^u - j_\mu^d \vert p \rangle$ that gives  the difference between the proton and neutron form factors and the
isoscalar combination  $\langle p \vert j_\mu^u + j_\mu^d \vert p \rangle/3$ for the sum
of the proton and neutron form factors.
The electric form factor at zero momentum yields the nucleon charge, i.e. $G_E^p(0){=}1$ and $G_E^n(0){=}0$ which, when using the lattice conserved current, holds by symmetry, even prior to gauge averaging.   
The magnetic form factor at $q^2{=}0$ gives the magnetic moment, while the  radii 
can be extracted from the slope of the electric and magnetic form factors as $q^2\rightarrow 0$, namely:
\begin{equation}
  \langle r^2 \rangle = \frac{6}{G(0)} \frac{\partial G(q^2)}{\partial q^2} \Big \vert_{q^2=0}\,.
  \label{Eq:radius}
\end{equation}

\section{Calculation on the Lattice} \label{sec:LMeth}

\subsection{Nucleon matrix element}

Extraction of nucleon matrix elements within the lattice QCD formulation requires the evaluation
of two- and three-point correlation functions in Euclidean space. We thus give all quantities in Euclidean space from here on.
We use the standard nucleon interpolating field
\begin{equation}
  J_N(\vec{x},t)=\epsilon^{abc}u^a(x)[u^{b\intercal}(x)\mathcal{C}\gamma_5d^c(x)]\,,
\end{equation}
where $u$ and $d$ are up- and down-quark spinors and $\mathcal{C}{=}\gamma_0 \gamma_2$ is the charge conjugation matrix.
The two-point function in momentum space is given by
\begin{equation}
C(\Gamma_0,\vec{p};t_s,t_0) {=} \hspace{-0.1cm} \sum_{\vec{x}_s} \hspace{-0.1cm}  
\tr \left[ \Gamma_0 {\langle}J_N(t_s,\vec{x}_s) \bar{J}_N(t_0,\vec{x}_0) {\rangle} \right]e^{{-}i (\vec{x}_s{-}\vec{x}_0) \cdot \vec{p}} \,,
\label{Eq:2pf}
\end{equation}
and the three-point function is given by
\begin{eqnarray}
 && C_\mu(\Gamma_\nu,\vec{q},\vec{p}\,';t_s,t_{\rm ins},t_0) {=} 
 \hspace{-0.1cm} {\sum_{\vec{x}_{\rm ins},\vec{x}_s}} \hspace{-0.1cm} e^{i (\vec{x}_{\rm ins} {-} \vec{x}_0)  \cdot \vec{q}}  e^{-i(\vec{x}_s {-} \vec{x}_0)\cdot \vec{p}\,'} {\times} \nonumber \\
  && \hspace{1.75cm} \tr \left[ \Gamma_\nu \langle J_N(t_s,\vec{x}_s) j_\mu(t_{\rm ins},\vec{x}_{\rm ins}) \bar{J}_N(t_0,\vec{x}_0) \rangle \right].
  \label{Eq:3pf}
\end{eqnarray}
The initial position and time, $x_0$, is referred to as the \textit{source}, the position and time of 
the current $j_\mu$ couples to a quark is denoted by  $x_{\rm ins}$ and referred to  as the \textit{insertion} and the final position, $x_s$, as the \textit{sink}. $\Gamma_\nu$ is a projector acting on 
spin indices, with $\Gamma_0 {=} \frac{1}{2}(1{+}\gamma_0)$ yielding the unpolarized and 
$\Gamma_k{=}\Gamma_0 i \gamma_5 \gamma_k$ the polarized matrix elements. Inserting complete 
sets of states in Eq.~(\ref{Eq:3pf}), one obtains the nucleon matrix element as well as additional matrix elements of higher energy states with the quantum numbers of the nucleon
multiplied by overlap terms and time dependent exponentials. For large enough
time separations, the excited state contributions are suppressed compared to the nucleon ground state and one can then extract the desired 
matrix element.
In order to increase the overlap with the nucleon state and  
decrease overlap with excited states we use Gaussian smeared quark  
fields~\cite{Alexandrou:1992ti,Gusken:1989qx} for the construction of  
the interpolating fields:  
\beq  
q_{\rm smear}^a(t,{\bf x}) &=& \sum_{\bf y} F^{ab}({\bf x},{\bf y};U(t))\ q^b(t,{\bf y})\,,\\  
F &=& (\eins + {\alpha} H)^{n} \,, \nonumber\\  
H({\bf x},{\bf y}; U(t)) &=& \sum_{i=1}^3[U_i(x) \delta_{x,y-\hat\imath} + U_i^\dagger(x-\hat\imath) \delta_{x,y+\hat\imath}]\,. \nonumber  
\eeq  
In addition, we apply APE-smearing to the gauge fields $U_\mu$ entering   
the hopping matrix $H$.

The Gaussian smearing parameters   are optimized using the nucleon two-point function. We set $\alpha=0.2$ and $n=125$~\cite{Alexandrou:2008tn}.
The values are: $\alpha=4.0$ and $n=50$, $70$ and $90$   
for $\beta=3.9$, $4.05$ and $4.2$ respectively.  
For the APE smearing~\cite{Albanese:1987ds} we use 50 iteration steps and $\alpha_{APE}{=}0.5$.

An optimized ratio~\cite{Alexandrou:2013joa,Alexandrou:2011db,Alexandrou:2006ru} of the three-point function
over a combination of two-point functions is used to cancel time dependent exponentials and overlaps, given by
\begin{eqnarray}
&&  R_\mu(\Gamma_\nu,\vec{p}\,',\vec{p};t_s,t_{\rm ins}) = \frac{C_\mu(\Gamma_\nu,\vec{p}\,',\vec{p};t_s,t_{\rm ins})}{C(\Gamma_0,\vec{p}\,';t_s)} \times \nonumber \\
&&  \sqrt{\frac{C(\Gamma_0,\vec{p};t_s-t_{\rm ins}) C(\Gamma_0,\vec{p}\,';t_{\rm ins}) C(\Gamma_0,\vec{p}\,';t_s)}{C(\Gamma_0,\vec{p}\,';t_s-t_{\rm ins}) C(\Gamma_0,\vec{p};t_{\rm ins}) C(\Gamma_0,\vec{p};t_s)}} \,,
\label{Eq:ratio}
\end{eqnarray}
where $t_s$ and $t_{\rm ins}$ are taken to be relative to the source $t_0$ for simplicity.
In the limit of large time separations, $(t_s{-}t_{\rm ins}) \gg 1$ and $t_{\rm ins} \gg 1$, the lowest state dominates and
the ratio becomes time independent 
\begin{equation}
R_\mu(\Gamma_\nu;\vec{p}\,',\vec{p};t_s;t_{\rm ins})\xrightarrow[t_{\rm ins}\gg 1]{t_s-t_{\rm ins}\gg 1}\Pi_\mu(\Gamma_\nu;\vec{p}\,',\vec{p})\,.
\end{equation}

$G_E(Q^2)$ and $G_M(Q^2)$ are extracted from linear combinations of $\Pi_\mu(\Gamma_\nu;\vec{p}\,',\vec{p})$ as expressed in Appendix~\ref{sec:appendix equations}, with $Q^2 {\equiv} -q^2$ the Euclidean momentum transfer squared.

\begin{figure}[ht!]
  \includegraphics[scale=0.75]{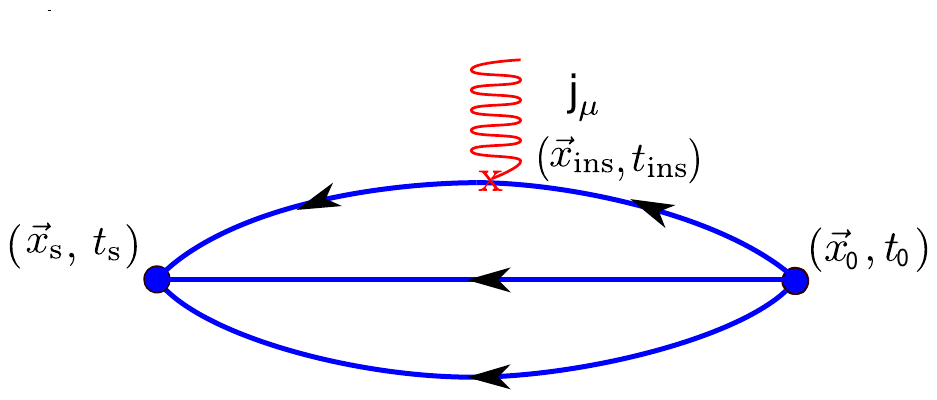}  \\
  \vspace{0.5cm}
  \includegraphics[scale=0.75]{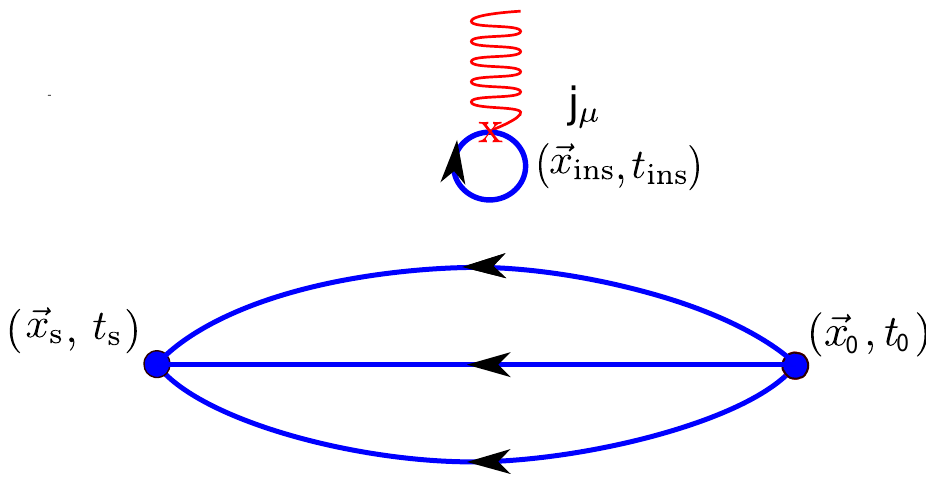}
  \vspace*{-0.2cm}
  \caption{Connected (upper panel) and disconnected (lower panel) contributions to
    the nucleon three-point function, with the source at $x_0$, the sink at $x_s$ and the current
    insertion ($j_\mu$) at $x_{\rm ins}$.}
  \label{Fig:ConnDisc}
\end{figure}

Contracting the quark fields in Eq.~(\ref{Eq:3pf}) gives rise to two types of diagrams depicted in Fig.~\ref{Fig:ConnDisc}, 
namely the so-called connected and disconnected contributions.
In the case of the connected diagram, the insertion operator couples to a valence
quark and an all-to-all propagator arises between sink and insertion. We use sequential inversions
through the sink that require keeping the sink-source time separation $t_s$, the projector, and the sink momentum $\vec{p}^\prime$ fixed. We perform additional sets of inversions to compute the three-point function for several values of $t_s$, for both the unpolarized
and polarized projectors. We set $\vec{p}^\prime{=}\vec{0}$.
We  use an appropriately tuned multigrid algorithm~\cite{Bacchio:2017pcp,Bacchio:2016bwn,Alexandrou:2016izb} for the efficient inversion of the Dirac operator entering in the computation of the connected diagram.
The disconnected diagram involves  the disconnected quark loop correlated with 
the nucleon two-point correlator. The disconnected quark loop is given by
\begin{equation}
  L(t_{\rm ins},\vec{q}) = \sum_{\vec{x}_{\rm ins}} \tr \left[ D^{-1}(x_{\rm ins};x_{\rm ins}) \mathcal{G} \right] e^{i \vec{q} \cdot \vec{x}_{\rm ins}},\label{Eq:looptrace}
\end{equation}
where $D^{-1}(x_{\rm ins};x_{\rm ins})$ is the quark propagator that starts and ends at the same point $x_{\rm ins}$
and $\mathcal{G}$  is an appropriately chosen $\gamma$-structure. For the local vector current, which we use for the disconnected diagram,  $\mathcal{G}=\gamma_\mu$.
A direct computation of quark loops would need inversions from all spatial points on the lattice,
making the evaluation unfeasible for our lattice size. We therefore employ stochastic
techniques to estimate it combined with dilution schemes \cite{Wilcox:1999ab}  that take into account the sparsity of the Dirac operator and
its decay properties. Namely, 
in this work, we employ the \emph{hierarchical probing} technique~\cite{Stathopoulos:2013aci}, which
provides a partitioning scheme that eliminates contributions from neighboring points in the trace of Eq.~\ref{Eq:looptrace} up to a certain coloring distance $2^k$. Using Hadamard vectors as the basis vectors for the partitioning, one needs $2^{d*(k-1)+1}$ vectors, where $d{=}4$
for a 4-dimensional partitioning. Note that the computational resources required are proportional to the number of Hadamard vectors, and therefore in $d$=4 dimensions increase 16-fold each time the probing distance $2^k$ doubles. Contributions entering from points beyond the probing distance are expected to be suppressed due to the exponential decay of the quark propagator and are treated with standard noise vectors which suppress all off-diagonal contributions by $1/\sqrt{N_r}$, i.e.
\begin{equation}
  \frac{1}{N_r} \sum_r \vert \xi_r \rangle \langle \xi_r \vert = 1 + \mathcal{O} \left( \frac{1}{\sqrt{N_r}} \right),
\end{equation}
where $N_r$ is the size of the stochastic ensemble.
Hierarchical probing has been employed
with great success in previous studies~\cite{Green:2015wqa,Green:2017keo} for an ensemble with  a pion mass of 317~MeV.
For simulations at the physical point, it is expected that a  larger probing distance is required since the light quark propagator decays more slowly at smaller quark masses. We avoid the need of increasing the distance by combining hierarchical probing with deflation of the
low modes~\cite{Gambhir:2016uwp}. Namely, we construct
the low mode contribution to the light quark loops by computing exactly the 200 smallest eigenvalues and corresponding eigenvectors of the squared Dirac operator and combine them with the contribution
from the remaining modes, which are estimated using hierarchical probing. Additionally, we employ the \emph{one-end trick}~\cite{McNeile:2006bz},
used in our previous studies~\cite{Alexandrou:2013wca,Alexandrou:2017hac,Alexandrou:2017qyt,Alexandrou:2017oeh}
and fully dilute in spin and color.

\subsection{Gauge Ensembles and Statistics} \label{sec:Stats}

For the extraction of the electromagnetic form factors we analyze one $N_f{=}2{+}1{+}1$~\cite{Alexandrou:2018egz}  and one $N_f=2$ ensemble. For both ensembles the quark masses are tuned to their physical values.
The fermion action is the twisted mass fermion action 
with a clover term. Automatic ${\cal O}(a)$ improvement is achieved by tuning to maximal twist
\cite{Frezzotti:2003ni,Frezzotti:2003xj}. The $N_f=2+1+1$ cB211.072.64 ensemble is simulated using a lattice of size $64^3{\times}128$ with $L m_\pi{=}3.62$~\cite{Alexandrou:2018egz}, where $L$ is the spatial extent of the lattice.  We determine the nucleon mass by fitting the effective mass in the large-time limit where the ground state dominates. The final value is chosen within a fit range where the value extracted is within half a standard deviation from the one determined  by including in the fit the first excited state (two-state fit). 
The ratio of the nucleon to pion mass is $m_N/m_\pi=6.74(3)$ compared to the physical ratio of 6.8. Therefore, we use directly the average proton and neutron mass of 0.9389~GeV to set the scale. We find $a=0.0801(4)$~fm. For the pion mass we find  $m_\pi=0.1393(7)$~MeV consistent with the average physical pion mass. Our current value of the lattice spacing is an update compared to the one given in Ref.~\cite{Alexandrou:2018egz} using higher statistics where the values are consistent.

To assess finite volume effects, we use two $N_f{=}2$ ensembles, which only differ in their volume, namely one has $L m_\pi=2.98$ and the other $L m_\pi{=}3.97$. We will refer to them as the cA2.09.48 and cA2.09.64 ensembles, respectively. We note that since the pion mass is not exactly at the physical value we interpolate to the physical pion mass using one-loop chiral perturbation theory. We include a systematic error on the extracted lattice spacing, determined as the difference in the mean value obtained using one-loop chiral perturbation theory and heavy baryon chiral perturbation theory. This systematic error on the lattice spacing appears for the two $N_f{=}2$ ensembles, while it is absent in the case of the cB211.072.64 ensemble. Results on the form factors for the $N_f=2$  ensemble with $L m_\pi{=}2.98$ are from Ref.~\cite{Alexandrou:2017ypw} while results for the other two ensembles are reported here for the first time. The simulation parameters of all three ensembles considered in this work are tabulated in Table~\ref{table:sim}. 

 \begin{widetext}
   \begin{center}   
 \begin{table}[ht!]
  \caption{Simulation parameters for the $N_f{=}2{+}1{+}1$~\cite{Alexandrou:2018egz} and $N_f{=}2$~\cite{Abdel-Rehim:2015pwa} ensembles used in this work. When two errors are given, the first error is statistical and the second is systematic. The systematic in the lattice spacing of the $N_f=2$ ensembles is obtained as described in the text, while the systematic in the pion mass provided in physical units is propagated from the lattice spacing.}
  \label{table:sim}
  \begin{tabular}{l r@{.}l r@{.}l l c r@{$\times$}l cccccr}
    \hline\hline
    ensemble &\multicolumn{2}{c}{$c_{\rm SW}$} & \multicolumn{2}{c}{$\beta$} & \multicolumn{1}{c}{$N_f$} & $a$ [fm] & \multicolumn{2}{c}{Vol.} & $a m_\pi$ & $m_\pi L$ & $a m_N$ & $m_N/m_\pi$ & $m_\pi$ [GeV] &$L$ [fm] \\
    \hline
    cB211.072.64 & 1&69    & 1&778 & 2+1+1 & 0.0801(4) & $64^3$&$128$ & 0.05658(6)  & 3.62   & 0.3813(19)     & 6.74(3)     & 0.1393(7)          &5.12(3) \\
    cA2.09.64 & 1&57551 & 2&1   & 2     & 0.0938(3)(1) & $48^3$&$96$  & 0.06208(2)  & 2.98   & 0.4436(11)     & 7.15(2)     & 0.1306(4)(2)       &4.50(1) \\
    cA2.09.48 & 1&57551 & 2&1   & 2     & 0.0938(3)(1) & $64^3$&$128$ & 0.06193(7)  & 3.97   & 0.4421(25)     & 7.14(4)     & 0.1303(4)(2)       &6.00(2) \\\hline\hline
  \end{tabular}
 \end{table}
  \end{center}
 
\end{widetext}

\begin{table}[ht!]
  \caption{Statistics for the evaluation of the connected three-point functions for the $N_f{=}2{+}1{+}1$ cB211.072.64 ensemble.
    Columns from left to right are the sink-source time separations, the number of configurations
    analyzed, the number of source positions per configuration chosen randomly
    and the total number of measurements for each time separation.}
  \label{table:StatsCon211}
  \vspace{0.2cm}
  \begin{tabular}{c|c|c|c}
    \hline\hline
    $t_s/a$ & $N_{\rm cnfs}$ & $N_{\rm srcs}$ & $N_{\rm meas}$ \\
    \hline
    12 & 750 & 4 & 3000 \\
    14 & 750 & 6 & 4500 \\
    16 & 750 & 16 & 12000 \\
    18 & 750 & 48 & 36000 \\
    20 & 750 & 64 & 48000 \\
    \hline\hline
  \end{tabular}
\end{table}
For the analysis of  the cB211.072.64 ensemble we use
750 configurations separated by 4 trajectories. For the connected contributions
we evaluate the three-point function for five sink-source time separations in the range 0.96~fm to 1.60~fm increasing the number of source positions per configuration as we increase the time separation so as to keep the statistical
error approximately constant. In Table~\ref{table:StatsCon211} we give the statistics used in the calculation of the connected
three-point functions.

For the evaluation of the disconnected contributions we use  $N_{\rm srcs}=200$ source positions to
generate  the nucleon two-point functions that are correlated with
 the quark loop to produce the disconnected contribution to the three-point function. We find that the volume is sufficiently large so that the data extracted from  this large number of randomly distributed source positions on the same
configuration is statistically independent. Nevertheless, we average over all source positions for each configuration and take the averaged correlation function as one statistic in our jackknife error analysis. As mentioned in the previous section, for the evaluation of the light quark 
loops we use the first 200 low modes of the squared Dirac operator to reconstruct exactly  part of the loop. The contribution from the high modes is estimated stochastically using one noise 
vector per configuration combining  hierarchical probing, one-end trick and spin-color dilution. 
For the hierarchical probing we use distance eight coloring resulting in 512 Hadamard vectors, 
which when combined with spin-color dilution leads to 6144 inversions per configuration. We note that 
the next coloring distance would demand 8192 Hadamard vectors, resulting in 98304 inversions per 
configurations after combining with spin-color dilution, making such a computation more than an order of magnitude more 
expensive.

For the computation of the disconnected contributions for  $N_f=2$ cA2.09.48  ensemble computed previously we used only the one-end trick and 2250 noise vectors for the calculation. Two-point functions were computed for 100 source positions. More detail can  be found   in Ref.~\cite{Alexandrou:2018zdf}.
in Table \ref{table:StatsDisc} we summarize the parameters for the computation 
of the disconnected three-point functions.

\begin{table}[ht!]

  \caption{Details on the set-up for the evaluation of the light disconnected diagrams.
    $N_{\rm cnfs}$ is the number of configurations analyzed,
    $N_{\rm def}$ is the number of low modes we deflate, $N_r$ the number of noise vectors, and
    $N_{\rm Had}$ the number of Hadamard vectors. $N_{\rm sc}$ corresponds to spin-color dilution and
    $N_{\rm inv}$/conf is the total number of inversions per configuration. $N_{\rm srcs}$ is the number of 
    randomly distributed smeared point sources per configuration used to obtain  the nucleon two-point functions
    and $N_{\rm meas}$ the total number of measurements.}
  \label{table:StatsDisc}
  \begin{center}
  \scalebox{0.9}{
  \begin{tabular}{c||c||c|c|c|c|c||c|c}
    \hline
    & \multicolumn{5}{c||}{Loops} & \multicolumn{2}{c}{Two-point}  \\
    \hline 
   ensemble & $N_{\rm cnfs}$ & $N_{\rm def}$ & $N_r$ & $N_{\rm Had}$ & $N_{\rm sc}$ &  $N_{\rm  inv}$/conf & $N_{\rm srcs}$ & $N_{\rm meas}$ \\
    \hline
    cB211.072.64 & 750 & 200 & 1 & 512 & 12 & 6144 & 200 & 150000 \\
    \hline
   cA2.09.48 & 2120 & - & 2250 & - & - & 2250 & 100 & 212000 \\
    \hline\hline
  \end{tabular}
  }
  \end{center}
\end{table}

The cA2.09.64 ensemble is used to check for finite volume effects, comparing the connected contributions to those of the cA2.09.48  ensemble. For the latter, the setup is reported in Ref.~\cite{Alexandrou:2017ypw} and summarized in Table~\ref{table:sim}. For the larger lattice size ensemble,
 we analyze three sink-source
time separations in the range of 1.1~fm to 1.5~fm. We fix the number of source positions
per configuration to 16 and we use more configurations for the larger time separations to control
statistical error. In Table~\ref{table:StatsCon2} we summarize the statistics for both $N_f=2$ ensembles.

\begin{table}[ht!]
  \caption{Statistics for the evaluation of the connected three-point functions for the cA2.09.64 and cA2.09.48 ensembles. For the latter, for $t_s/a=16,18$ the connected
    three-point functions have been computed only for the unpolarized projector.
  The notation is as in Table~\ref{table:StatsCon211}.}
  \label{table:StatsCon2}
  \vspace{0.2cm}
  \begin{tabular}{c|c|c|c}
    \hline
    $t_s/a$ & $N_{\rm cnfs}$ & $N_{\rm srcs}$ & $N_{\rm meas}$ \\
    \hline
    \multicolumn{4}{c}{cA2.09.64: $N_f=2\;\; 64^3 \times 128$ ensemble} \\
    \hline
    12 & 333 & 16 & 5328 \\
    14 & 515 & 16 & 8240 \\
    16 & 1040 & 16 & 16640 \\
    \hline
    \multicolumn{4}{c}{cA2.09.48: $N_f=2\;\; 48^3 \times 96$ ensemble} \\
    \hline
    10,12,14 & 578 & 16 & 9248 \\
    16       & 530 & 88 & 46640 \\
    18       & 725 & 88 & 63800 \\
    \hline\hline
  \end{tabular}
\end{table}

\subsection{Excited states contamination}\label{sec:excited states}
Assessment  of excited state effects is imperative for the proper extraction of the desired nucleon matrix element. 
However, ensuring ground state dominance is a delicate process due to the exponentially increasing statistical noise with increasing  sink-source separation.
We use four methods to study the effect of excited  states and identify the final results based on a critical comparison among these methods. Only by employing these different methods can one reach a reliable assessment of excited state contributions and extract the nucleon matrix element of interest. The methods employed are as follows:

\vspace{0.15cm}
\noindent \emph{Plateau method:} In this method we use the ratio in Eq.~(\ref{Eq:ratio}) and identify a time-independent 
window (plateau) as
we increase $t_s$. The converged plateau value  then yields  the desired matrix element.

\vspace{0.15cm}
\noindent \emph{Two-state method:}
Within this method we fit the two- and three-point functions considering contributions up to the first excited state  using the expressions
\begin{equation}
C(\vec{p},t_s) = c_0(\vec{p}) e^{-E_0(\vec{p}) t_s} + c_1(\vec{p}) e^{-E_1(\vec{p}) t_s}\,,
\label{Eq:Twp_tsf}
\end{equation}
\begin{align}
  &C_\mu(\Gamma_\nu,\vec{p}\,',\vec{p},t_s,t_{\rm ins}) = \nonumber \\
  &  A_{00}^\mu(\Gamma_\nu,\vec{p}\,',\vec{p}) e^{-E_0(\vec{p}\,')(t_s-t_{\rm ins})-E_0(\vec{p})t_{\rm ins}} \nonumber \\
  &+ A_{01}^\mu(\Gamma_\nu,\vec{p}\,',\vec{p}) e^{-E_0(\vec{p}\,')(t_s-t_{\rm ins})-E_1(\vec{p})t_{\rm ins}} \nonumber \\
  &+ A_{10}^\mu(\Gamma_\nu,\vec{p}\,',\vec{p}) e^{-E_1(\vec{p}\,')(t_s-t_{\rm ins})-E_0(\vec{p})t_{\rm ins}} \nonumber \\
  &+ A_{11}^\mu(\Gamma_\nu,\vec{p}\,',\vec{p}) e^{-E_1(\vec{p}\,')(t_s-t_{\rm ins})-E_1(\vec{p})t_{\rm ins}}.
\label{Eq:Thrp_tsf}
\end{align}
In Eqs.~(\ref{Eq:Twp_tsf}) -  (\ref{Eq:Thrp_tsf}) $E_0(\vec{p})$ and $E_1(\vec{p})$ are the energies of the ground and first excited states with total momentum $\vec{p}$, respectively. The ground state corresponds to a single particle state and therefore one can use the continuum dispersion relation, $E_0(\vec{p})=\sqrt{\vec{p}^2+m_N^2}$, with $\vec{p}=\frac{2\pi}{L}\vec{n}$ with $\vec{n}$ a lattice vector with components $n_i\in (-\frac{L}{2a}, \frac{L}{2a}]$. The continuum dispersion relation is satisfied for all $Q^2$ values considered in this work. The first excited state, on the other hand, can be a two-particle state. 
We, thus, fit simultaneously the two-point functions with momenta $\vec{p}$ and $\vec{p}^\prime$  and the three-point function involving in total eleven  
parameters.     Note that for non-zero momentum transfer, $A_{01}^\mu(\Gamma_\nu,\vec{p}\,',\vec{p}) \neq A_{10}^\mu(\Gamma_\nu,\vec{p}\,',\vec{p})$. This allows us to extract the matrix element  given by 
\begin{equation}
\Pi_\mu(\Gamma_\nu;\vec{p}\,',\vec{p})=\frac{A_{00}^\mu(\Gamma_\nu,\vec{p}\,',\vec{p})}{\sqrt{c_0(\vec{p}\,') c_0(\vec{p})}}.
\label{Eq:Tsf}
\end{equation}

\vspace{0.15cm}
\noindent \emph{Summation method:} Summing over $t_{\rm ins}$ in the ratio of Eq.~(\ref{Eq:ratio}) yields a geometric sum~ \cite{Maiani:1987by,Capitani:2012gj} from which we obtain,
\begin{align}
  R_\mu^{sum}(\Gamma_\nu;\vec{p}\,',\vec{p};t_s) &= \sum_{t_{\rm ins}=a}^{t_s-a} R_\mu(\Gamma_\nu;\vec{p}\,',\vec{p};t_s;t_{\rm ins}) = \nonumber \\
  &\hspace*{0.35cm}c + \Pi_\mu(\Gamma_\nu;\vec{p}\,',\vec{p}) {\times}t_s + \cdots
  \label{Eq:Summ}
\end{align}
where the ground state contribution, $\Pi_\mu(\Gamma_\nu;\vec{p}\,',\vec{p})$, is extracted from the slope of a linear fit with respect to $t_s$. The sink-source time separation $t_s$ considered in the fit should be large enough to suppress higher order contributions.

\vspace{0.15cm}
\noindent \emph{Derivative Summation method:} Instead of performing a linear fit in Eq.~(\ref{Eq:Summ}) to extract the matrix element, one can
take finite differences to the summed ratio~\cite{Savage:2016kon} as follows
\begin{align}
  & d R_\mu^{sum}(\Gamma_\nu;\vec{p}\,',\vec{p};t_s) = \nonumber \\
  &\frac{R_\mu^{sum}(\Gamma_\nu;\vec{p}\,',\vec{p};t_s+dt_s) - R_\mu^{sum}(\Gamma_\nu;\vec{p}\,',\vec{p};t_s)}{dt_s}
  \label{Eq:Dsumm}
\end{align}
and fit to a constant  to extract the desired matrix element.

\section{Analysis of lattice results} \label{sec:Analysis}
\subsection{Isovector and connected isoscalar form factors}
\label{subsec:analysis}
\begin{figure}[ht!]
  \includegraphics[width=\linewidth]{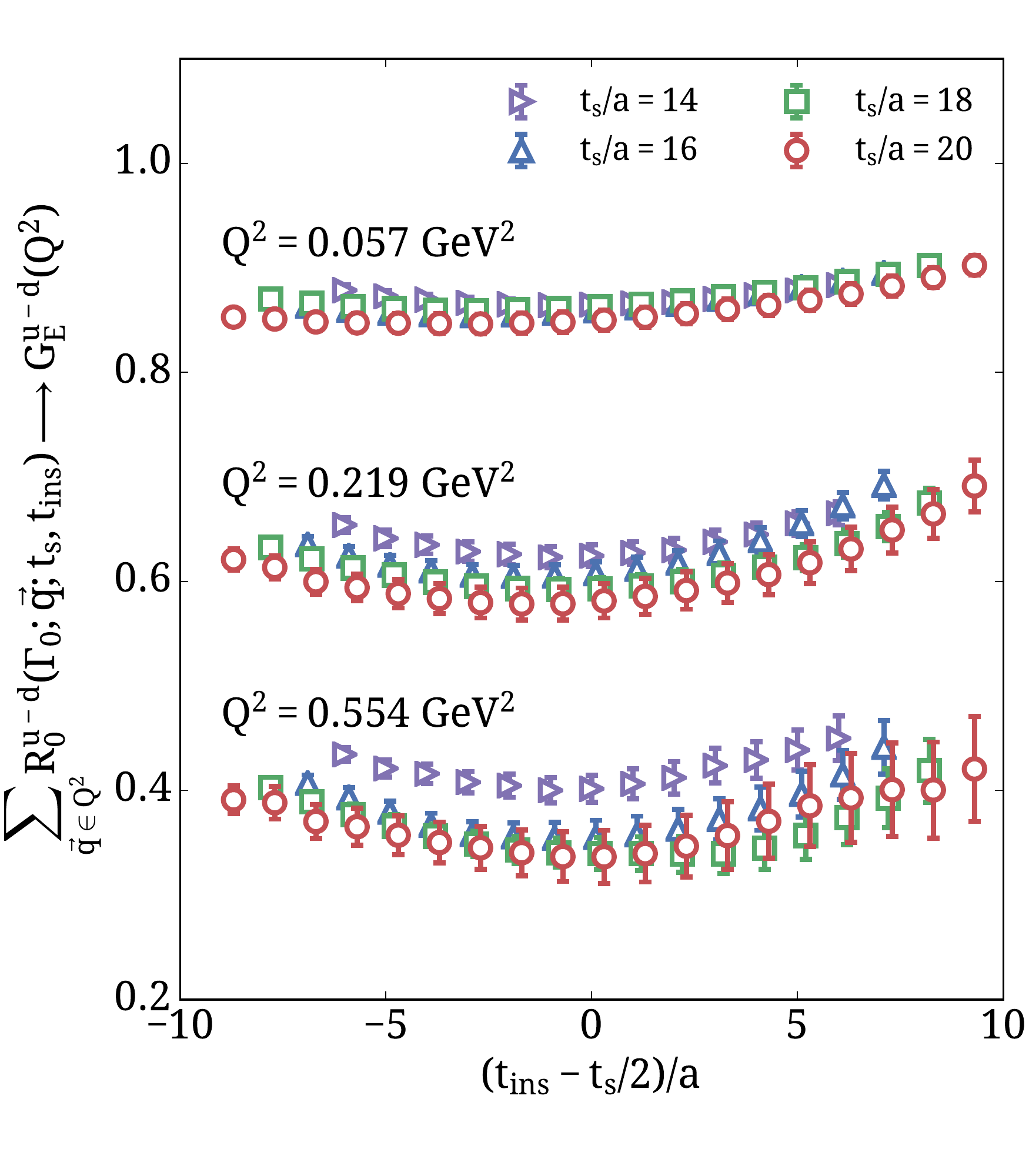}
  \vspace{-1.2cm}
  \caption{Ratio yielding the isovector electric form factor. The
    sink-source time separations are $t_s/a{=}14$ (right triangles), $t_s/a{=}16$ (triangles),
    $t_s/a{=}18$ (squares) and $t_s/a{=}20$ (circles). We
    present the ratio for three $Q^2$ values, namely 0.057, 0.219 and 0.554 GeV$^2$,
    from top to bottom.} 
  \label{fig:RGE}
\end{figure}

\begin{figure}[ht!]
  \includegraphics[width=\linewidth]{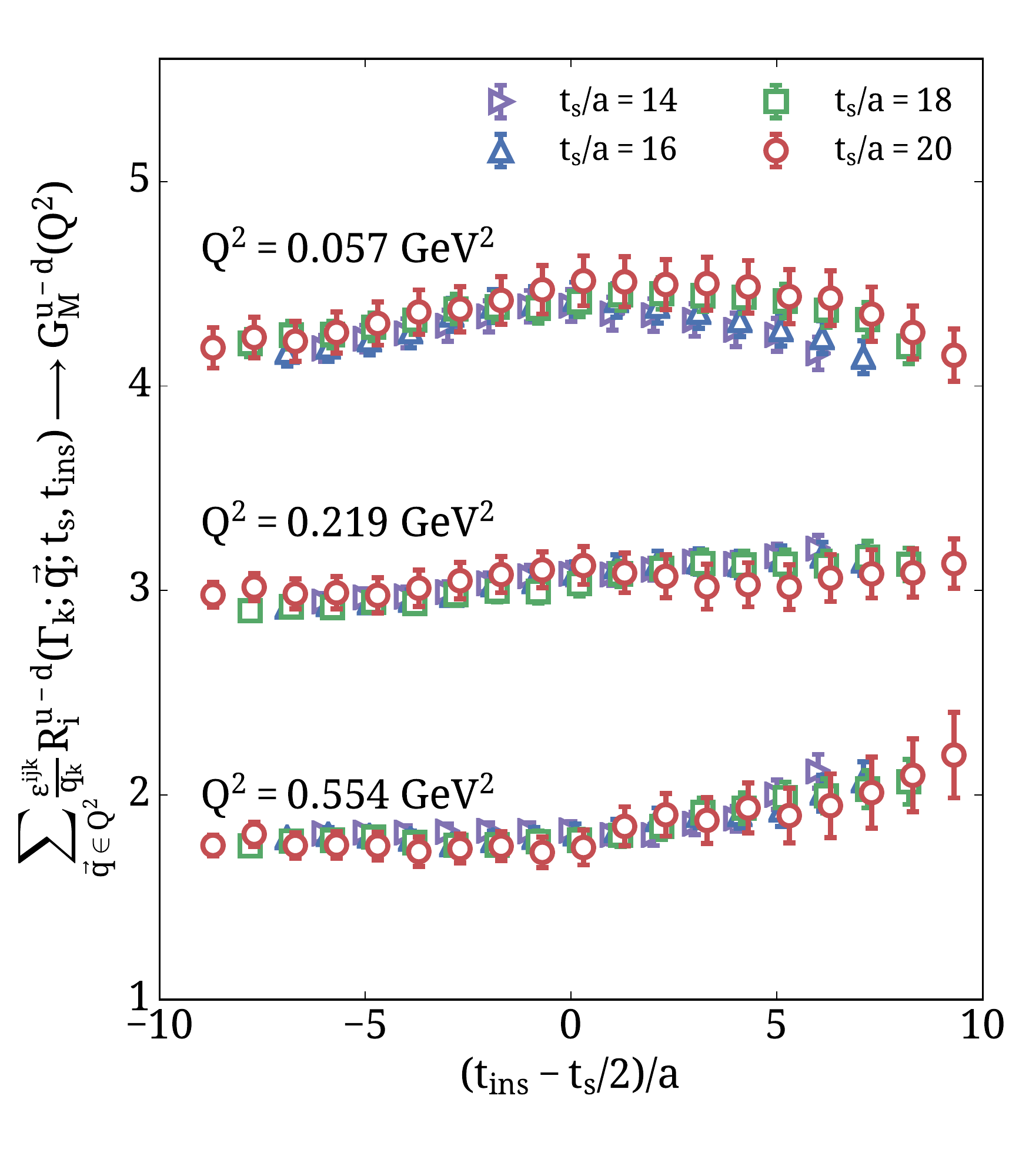}
  \vspace{-1.2cm}
  \caption{Ratio yielding the isovector magnetic form factor. The notation is
  the same as in Fig.~\ref{fig:RGE}.}
  \label{fig:RGM}
\end{figure}

The isovector combination  gives the difference between the proton 
and neutron form factors, and in this case, only the connected diagram contributes since disconnected 
contributions cancel, up to cut-off effects of ${\cal O}(a^2)$. For the connected diagram we use a frame where the nucleon 
final momentum is zero, thus $\vec{q} {=} {-} \vec{p}$. 
In Figs.~\ref{fig:RGE} and \ref{fig:RGM} we show the ratios defined in Eq.~(\ref{Eq:ratio}) as a function
of the sink-source time separation, and for three values of the momentum transfer squared,
that is for $Q^2{=}0.057$ GeV$^2$, $Q^2{=} 0.219$~GeV$^2$  and $Q^2{=}0.554$~GeV$^2$. 
In a frame where the final momentum of the nucleon  $\vec{p}^\prime$ is zero, the expressions in Appendix~\ref{sec:appendix equations} given by Eqs.~(\ref{Eq:GEGMEQ1}) and (\ref{Eq:GEGMEQ2}) reduce to  Eqs.~(\ref{Eq:GEQ2_1},\ref{Eq:GEQ2_2}), and (\ref{Eq:GMQ2}), giving separately the electric and magnetic form factors. We note that for the electric form factor Eq.~(\ref{Eq:GEQ2_1}) leads to much more precise results compared to Eq.~(\ref{Eq:GEQ2_2}) and therefore we use only Eq.~(\ref{Eq:GEQ2_1}). In the case of the ratio determining  $G_E^{u-d}(Q^2)$, as $t_s$ increases,
the plateau value  decreases with larger deviations as  $Q^2$ increases. This shows  that at larger $Q^2$ values contamination due to excited states is more severe.
In the case of $G_M(Q^2)$, excited states are suppressed and only a small variation with $t_s$ is observed.

We further investigate effects due to
excited states by employing the summation and two-state fits.
In  Fig.~\ref{fig:Summ_dSumm}
we show  linear fits to the summed ratio for three different values of $Q^2$. The slope gives the nucleon matrix element.
All three momenta follow well the linear behavior, within the statistical error, indicating
that contributions from higher order terms are suppressed. In the right panel of Fig.~\ref{fig:Summ_dSumm},
we demonstrate the plateaus for the derivative summation method, fitting to a constant to extract
the matrix element of the ground state. Within statistical
accuracy all three momenta are indeed flat, and are thus described well by a constant.

\begin{figure}[ht!]
  \includegraphics[width=\linewidth]{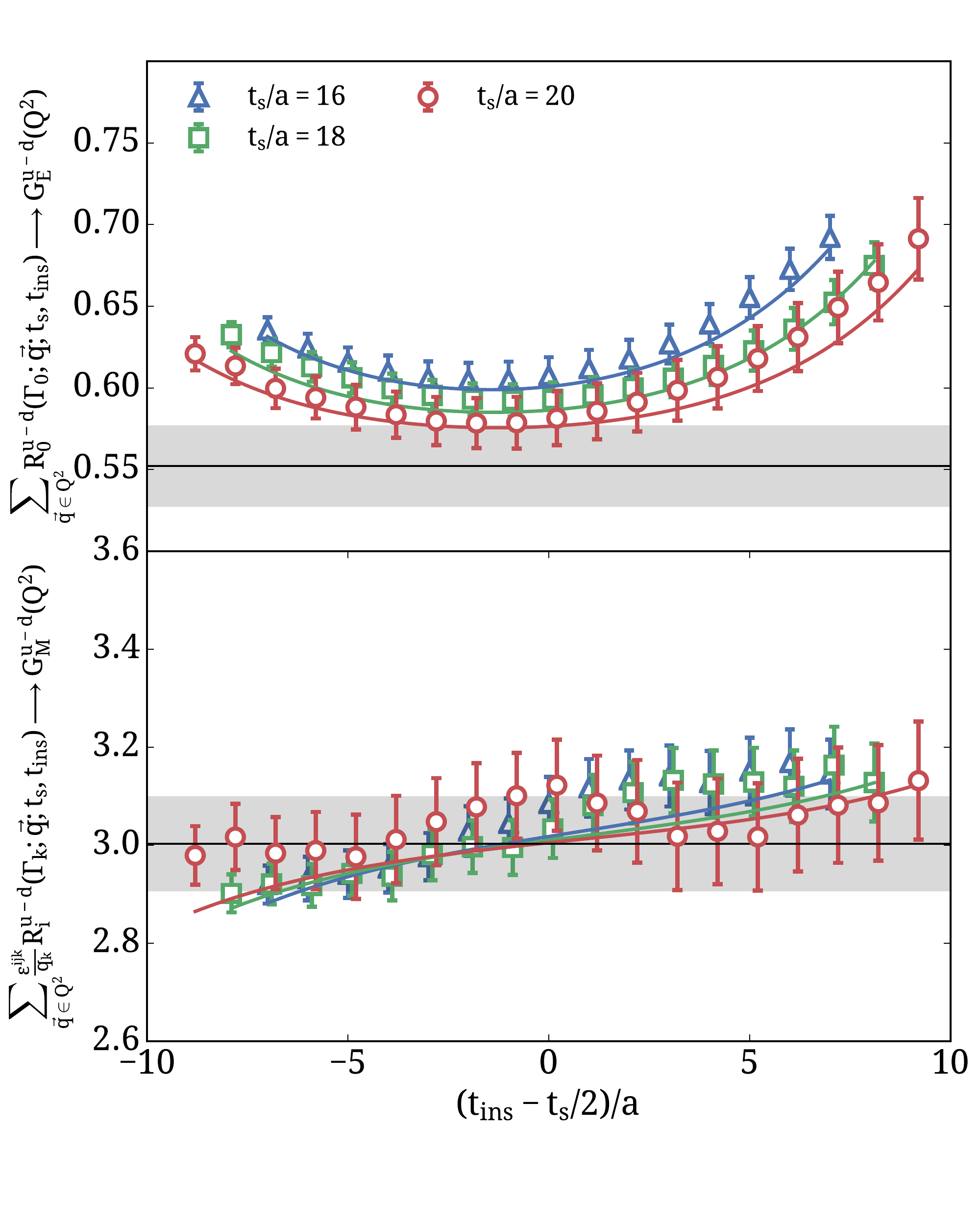}
        \vspace*{-.8cm}
  \caption{Ratio yielding the   isovector electric (top) and the isovector magnetic (bottom) form factors,  for the three
    largest sink-source separations, following the notation of Fig.~\ref{fig:RGE}. The curves show the result of the two-state fit method, while the gray horizontal
    band is the extracted value of the nucleon matrix elements and its error. We show the case for $Q^2{=}0.219$ GeV$^2$.}
    \label{fig:TwoSt}
\end{figure}

In Fig.~\ref{fig:TwoSt} we show the results extracted using two-state fits for both electric and magnetic
form factors. The data correspond to the ratio of Eq.~(\ref{Eq:ratio}) and the curves are obtained by fitting simultaneously the three- and two-point functions to Eqs.~(\ref{Eq:Thrp_tsf}) and~(\ref{Eq:Twp_tsf}). The gray horizontal band shows the nucleon matrix element value and error extracted from
the two-state fit as in Eq.~(\ref{Eq:Tsf}). For the electric form factor, the ratio shows a trend towards lower values as we increase the sink-source separation, with $t_s/a{=}20$
becoming compatible with the value extracted from the two-state fit. In the case of the magnetic form factor,
the value extracted from the two-state fit is compatible with the ratio for all time separations considered confirming the weak dependence of the matrix element on the sink-source time separation observed in the plateau method.

\begin{widetext}
  \begin{center}
\begin{figure}[!h]
  \includegraphics[width=0.8\textwidth]{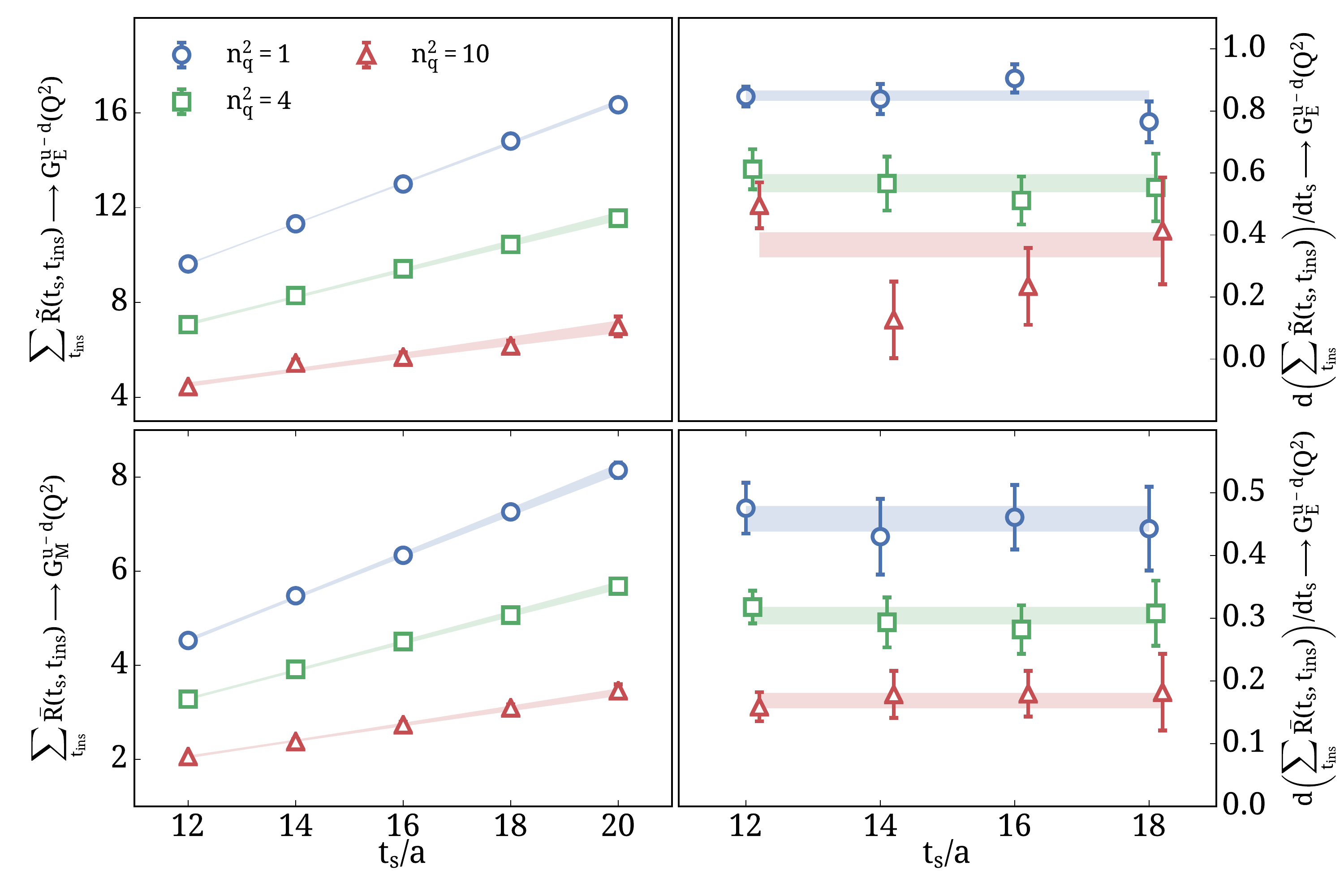}
  \caption{We define $\displaystyle \tilde{R}(t_s,t_{\rm {ins}}) \equiv \sum_{\vec{q} \in Q^2}R_0^{u-d}(\Gamma_0;\vec{q};t_s,t_{\rm ins})$ and $\displaystyle \bar{R}(t_s,t_{\rm {ins}}) \equiv \sum_{\vec{q} \in Q^2} \frac{\epsilon^{ijk}}{q_k} R_i^{u-d}(\Gamma_k;\vec{q};t_s,t_{\rm ins})$.
    Left panel: the summed ratio of Eq.~(\ref{Eq:Summ}) as
    a function of the sink-source time  separation for three momenta, namely $n_q^2{=}1$ (blue circles),
    $n_q^2{=}4$ (green squares) and $n_q^2{=}10$ (red triangles) corresponding to $Q^2{=}0.057$, 0.219, and 
    0.554~GeV$^2$ for the isovector electric (top) and isovector magnetic (bottom) 
    form factors. The bands are fits to a linear form. Right panel: the derivative of the summed ratio as in Eq.~(\ref{Eq:Dsumm}), using the same notation as that of the left panel. The bands are fits to a constant.}
  \label{fig:Summ_dSumm}
\end{figure}
\end{center}
\end{widetext}

\begin{widetext}
  \begin{center}
\begin{figure}[h]
  \includegraphics[width=0.9\textwidth]{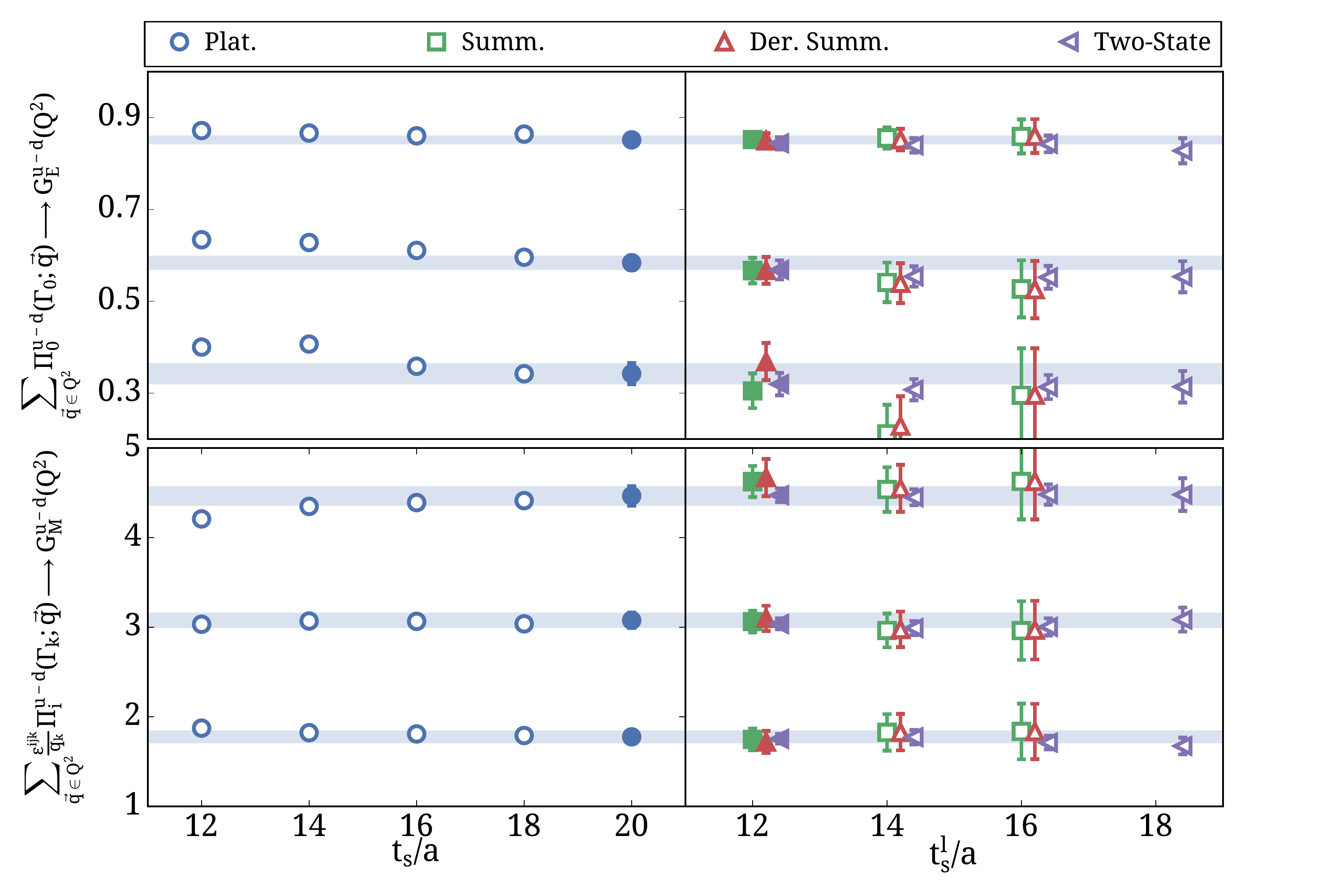}
        \vspace*{-.8cm}
  \caption{Results for the matrix element yielding $G_E^{u-d}$ (upper panel) and $G_M^{u-d}$ (lower panel), for $Q^2{=}0.057$~GeV, $Q^2{=}0.219$~GeV and $Q^2{=}0.554$~GeV$^2$  from top to bottom. In the left column 
    we show the extracted values using the plateau method (blue circles) for all five separations analyzed, while in the right panel
    we show the values extracted using the summation (green squares), derivative summation (red upper triangles) and two-state fit (purple left triangles) approaches
    as we change the lower fit range (${\rm t^l_s/a}$) keeping the upper fit range fixed to $t_s/a{=}20$. The filled circle and band show the value
    and statistical error used to quote our final result, while the other filled symbols show the fit ranges for the two-state fit and summation methods that will be used in the figures that follow. 
  }
  \label{fig:PlatSummDSumm}
\end{figure}
  \end{center}
\end{widetext}

In Fig.~\ref{fig:PlatSummDSumm} we show the extracted values for the  matrix element yielding the isovector electromagnetic form factors.
We compare the plateau, summation, derivative summation and two-state fit methods. For the plateau method we show the
value extracted from the constant fit for all sink-source separations available. For the other cases we vary the
lower fit range, keeping the upper range fixed to $t_s/a{=}20$. We seek for the earliest agreement between the  plateau
method and the other three cases. As already pointed out, 
the isovector electric form factor shows more severe excited state effects for large $Q^2$-values and we therefore take the largest
time separation for the plateau method to fulfill our criterion for agreement with the other methods. For the isovector magnetic form factor, although excited
state effects are mild, we still observe a shift to larger values for the smallest $Q^2$ and, therefore,
we conservatively use the largest time separation available also in this case. An additional observation is that summation and derivative summation methods produce compatible results with similar
accuracy, as can be seen in Fig.~\ref{fig:PlatSummDSumm}, and thus from now on we will restrict to showing results
only from the summation method.

\begin{figure}[ht!]
  \includegraphics[width=\linewidth]{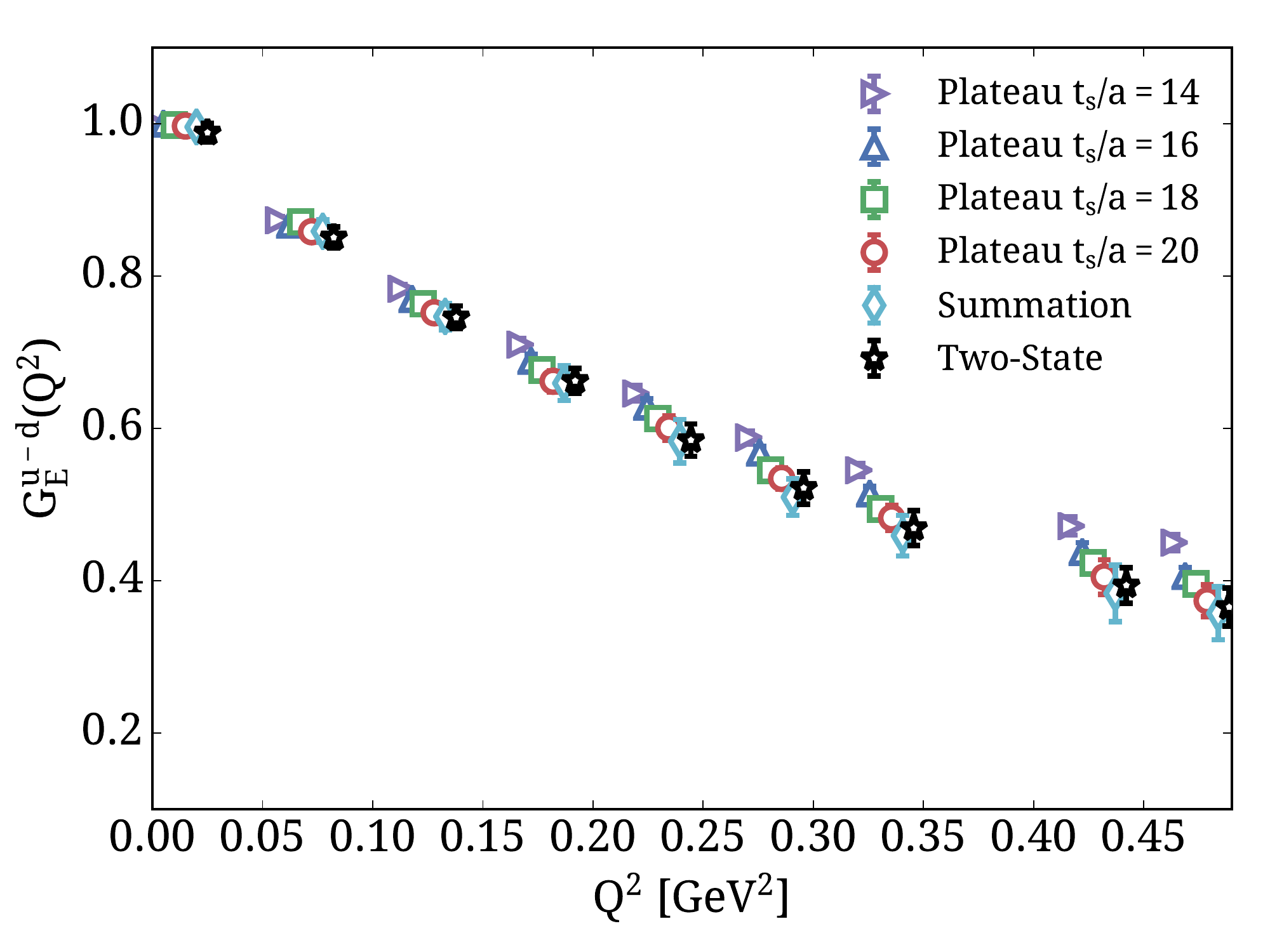}\\[-1.5ex]
    \includegraphics[width=\linewidth]{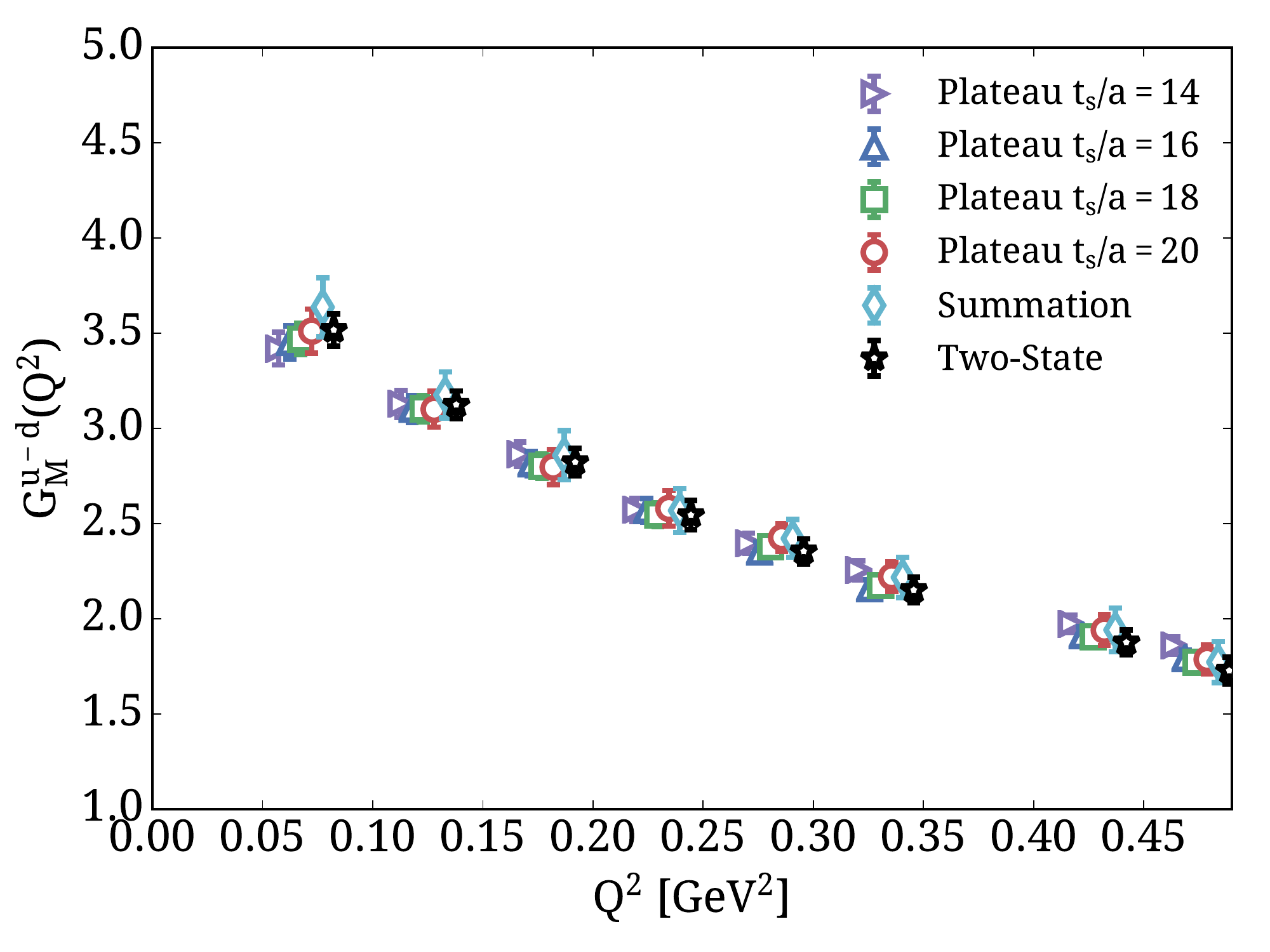}
        \vspace*{-.8cm}
  \caption{The isovector electric (upper panel)  and magnetic (lower panel) form factor as a function of $Q^2$. We show the values extracted from fitting the plateau  for the four largest $t_s$ values, namely $t_s/a{=}14$ (right triangles), $t_s/a{=}16$ (triangles),
    $t_s/a{=}18$ (squares) and $t_s/a{=}20$ (circles), compared to the 
    summation method (diamonds) and using two-state fits (stars). Results from different methods are slightly shifted to the right for clarity.}
  \label{fig:GEGMQ2_methods}
\end{figure}

In Fig.~\ref{fig:GEGMQ2_methods}, we present our results for $G_E^{u-d}(Q^2)$  and $G_M^{u-d}(Q^2)$ as a function of the momentum transfer squared $Q^2$. We limit the plot to  up  $Q^2{=}0.5$ GeV$^2$ to make visible  the values extracted using the plateau at the four largest separations,
the summation, and the two-state fit approaches. For the summation and two-state fit we show the values indicated by the filled symbols in Fig.~\ref{fig:PlatSummDSumm}.
As can be seen,  for the electric form factor  effects of  excited states are small
for small values of  $Q^2$ but become more severe for higher $Q^2$ values, with the extracted value decreasing with increasing time separation in line with  the observation made in Fig.~\ref{fig:RGE}.
For $G_M^{u-d}(Q^2)$,  excited state effects are small for larger $Q^2$ values whereas for smaller $Q^2$ values there is a systematic increase in the values of the form factor  with the time separation.

\begin{figure}[ht!]
  \includegraphics[width=\linewidth]{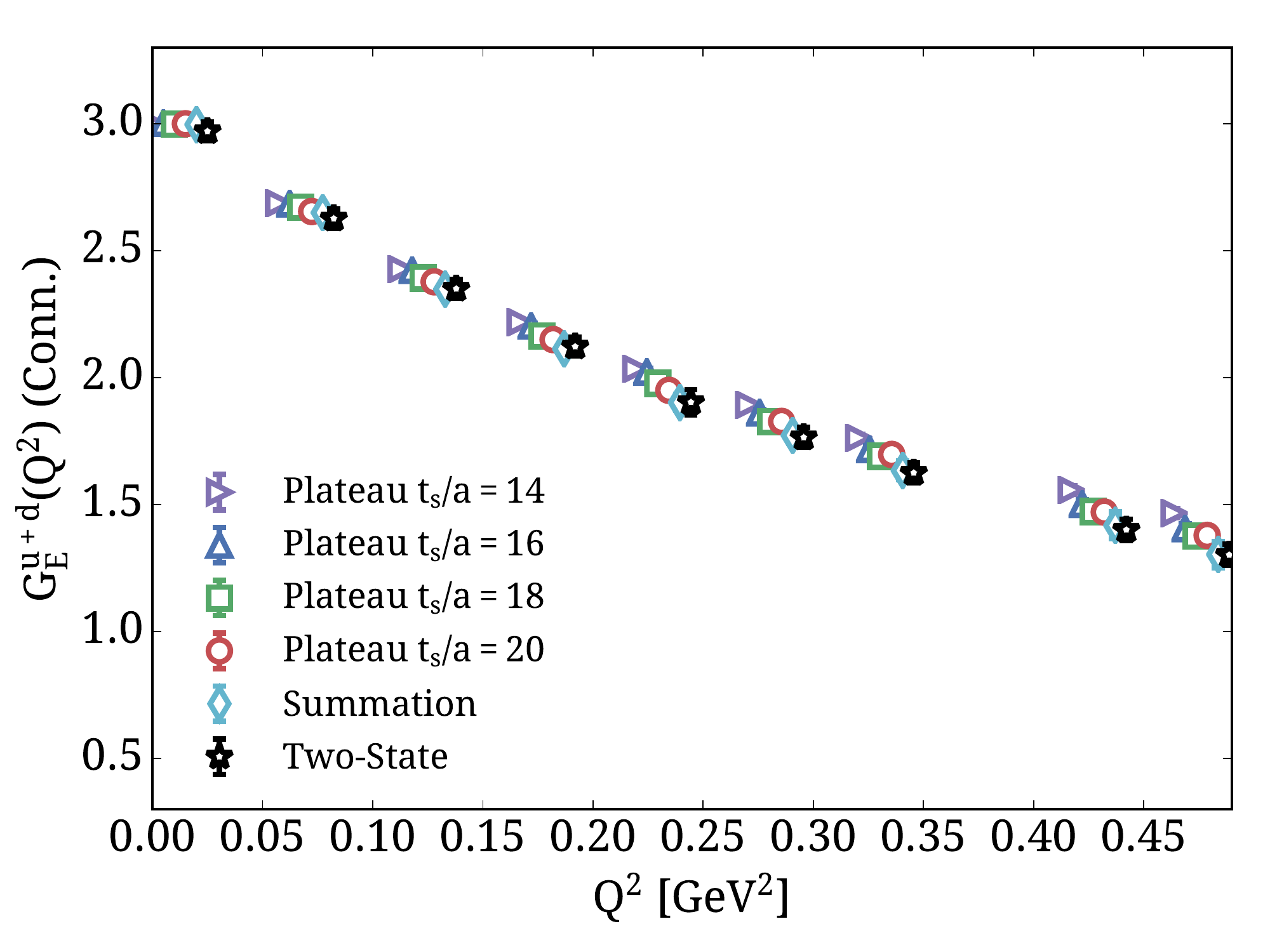}\\[-1.5ex]
    \includegraphics[width=\linewidth]{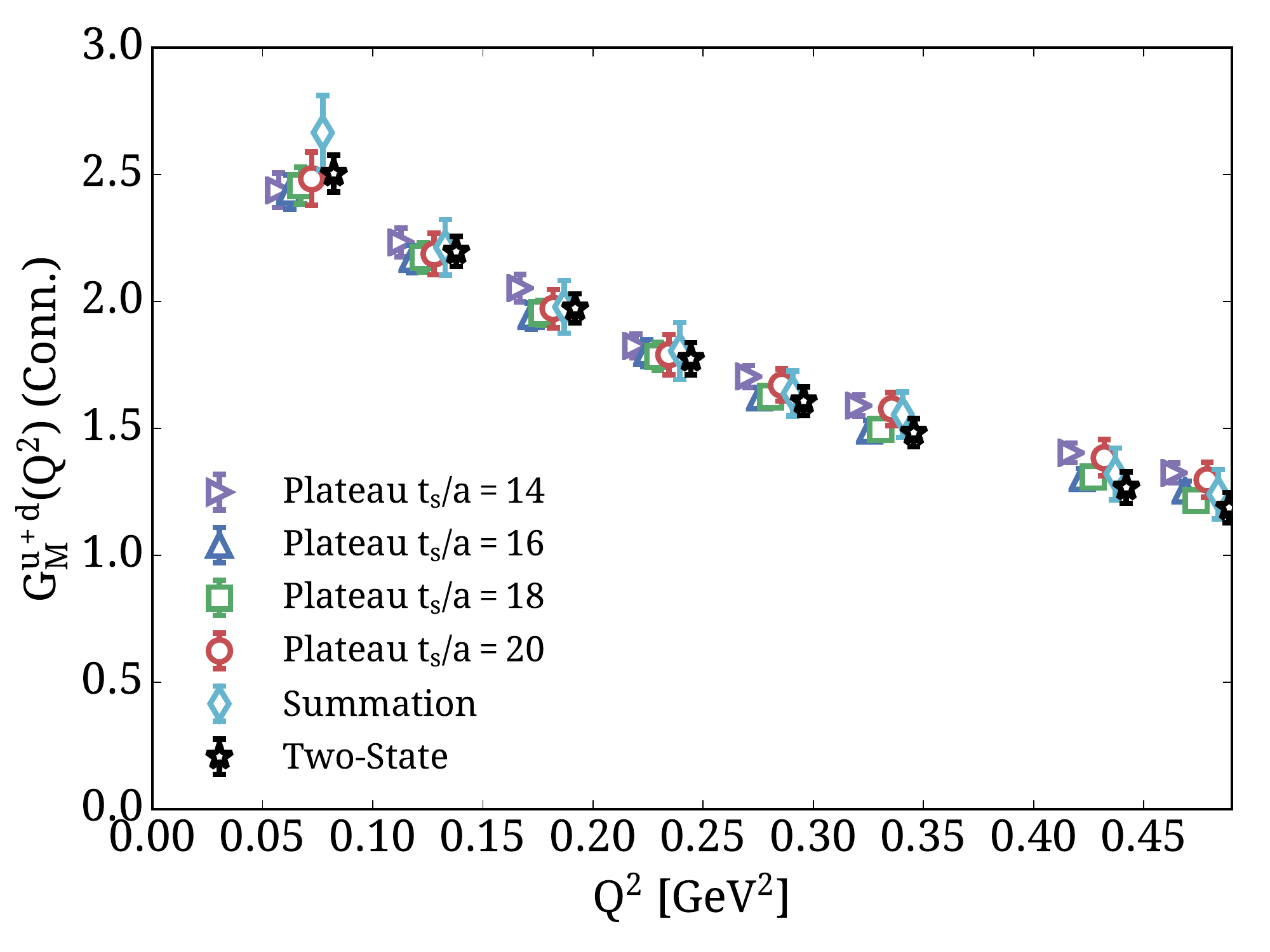}
            \vspace*{-.8cm}
  \caption{Connected contribution to the isoscalar electric (upper panel) and magnetic (lower panel) form factor. The notation is the same as in Fig.~\ref{fig:GEGMQ2_methods}.}
  \label{fig:GEisConnQ2_methods}
\end{figure}

For the extraction of the connected isoscalar form factors we follow a similar analysis  procedure as in the isovector case.
In Fig.~\ref{fig:GEisConnQ2_methods} we present the connected contribution to isoscalar electric and magnetic form factors 
comparing the plateau, summation and two-state fit methods. Excited states have a smaller effect on the isoscalar form factors being detectable only for the  magnetic at small values of $Q^2$ where the two-state fit yields systematically larger values.  Given that there is agreement between  the plateau values for the largest time separation and the two-state fit we will use the plateau value as the final result for the form factors. The
deviation from the values determined from the two-states fits are then taken as an estimate of the systematic error due to excited states. Since we will be using the plateau values in the case of disconnected since two-state fits are not stable in that case we  do the same for the connected for consistency. 

\subsection{Disconnected contributions}
A major component of this work is the evaluation of the disconnected contributions  shown diagrammatically in 
Fig.~\ref{Fig:ConnDisc} that enter in the evaluation of the isoscalar as well as in the proton and neutron form factors. 

\begin{figure}
  \includegraphics[width=\linewidth]{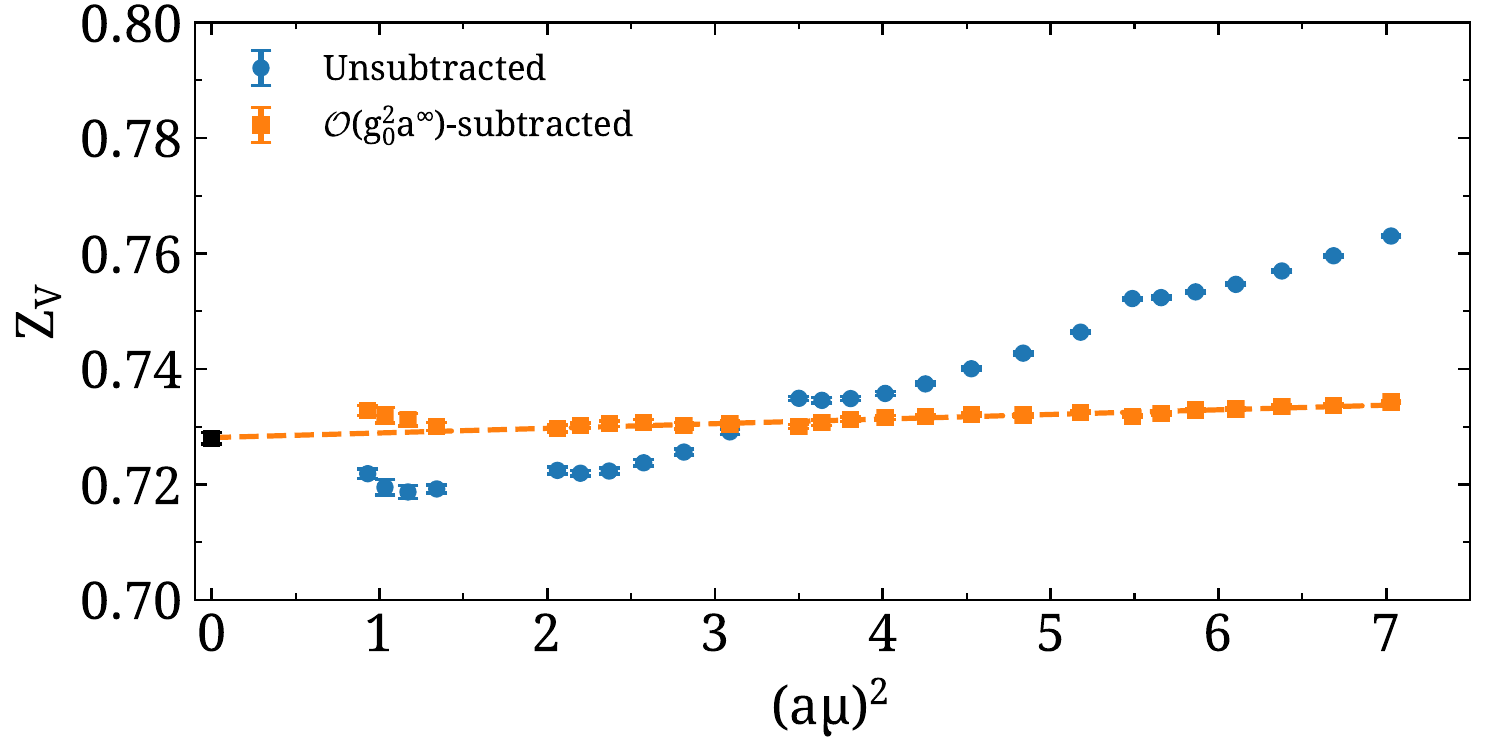}
    \caption{The renormalization constant $Z_V$ as a function of the renormalization scale squared $(a\mu)^2$ before (blue circles) and after (orange squares) performing the subtraction of $\mathcal{O}(g^2a^\infty)$-terms. The dashed line is a linear fit to the latter and the point at $(a\mu)^2=0$ (black square) is the result of the fit.}\label{fig:ZV}
\end{figure}

The  disconnected quark 
loops are computed using  the formalism described in Section~\ref{sec:Stats} with the statistics  summarized in Table~\ref{table:StatsDisc}. 
As already discussed,  the hierarchical probing method, combined with deflation of the low eigenmodes,
 provides  an accurate
 determination of the diagonal of the quark propagator entering in the evaluation of the quark loops. It is thus preferable to use the local vector current for the evaluation of the disconnected contributions    since the conserved current includes non-diagonal terms. We therefore need the renormalization function $Z_V$, which is determined non-perturbatively, in the RI$^{\prime}$-MOM scheme, employing momentum sources. We perform a perturbative subtraction of $\mathcal{O}(g^2a^\infty)$-terms, as described in Refs.~\cite{Alexandrou:2015sea,Alexandrou:2010me}, which subtracts the leading cut-off effects leaving only
a weak dependence on the
renormalization scale $(a\mu)^2$, as shown in Fig.~\ref{fig:ZV}.  We find a value of $Z_V{=}0.728(1)$ where the error is statistical.
Alternatively, $Z_V$ can be determined at $Q^2=0$, by taking the ratio of $G^{u-d}_E(0)$  computed with the  local current to  $G^{u-d}_E(0)$  computed  using the lattice conserved current. This ratio yields a value of 0.715(3). Although this is  2\% smaller than $Z_V$ as determined from the vertex function, the difference between them is still  an order of magnitude smaller as compared to the statistical errors for the disconnected contributions. In what follows we use  $Z_V{=}0.728(1)$ to renormalize the matrix elements computed using the local current, since this determination has taken into account higher order cut-off effects as compared to  the one determined from the ratio. We note that $Z_V$ only enters  in the  disconnected three-point function.
 A more detailed description of the renormalization procedure including other renormalization functions will be provided in a future publication.

Disconnected quark loops are evaluated for every time-slice allowing us to compute the three-point function for every combination of $t_s$ and $t_{\rm ins}$. As in the case of the connected,  we are seeking for a reasonable window in $t_s$ to extract the nucleon matrix elements,
where excited states are sufficiently suppressed and noise is not prohibitively large.
In contrast to the connected diagram, where we have results only for the case $\vec{p}\,'{=}\vec{0}$, for the disconnected diagrams
we have all sink momenta at no additional cost. We analyze, besides $\vec{p}\,'{=}\vec{0}$, the matrix element for the six final momenta with $\vec{p}\,'{=}\pm\frac{2\pi}{L} \hat{n}$, with $\hat{n}=\hat{x}$, $\hat{y}$, or $\hat{z}$, i.e. the unit vector in one of the three spatial directions.
Given that the statistical errors in the case of the disconnected diagrams are larger as compared to the connected diagrams,  we restrict ourselves in  using the plateau method for different values of $t_s$ in order to check for ground state dominance. This is because the two-state fits are problematic given the larger errors of the disconnected diagrams. Whenever they work they yield  large errors and are consistent with the plateau extraction.

\begin{figure}[ht!]
  \includegraphics[width=\linewidth]{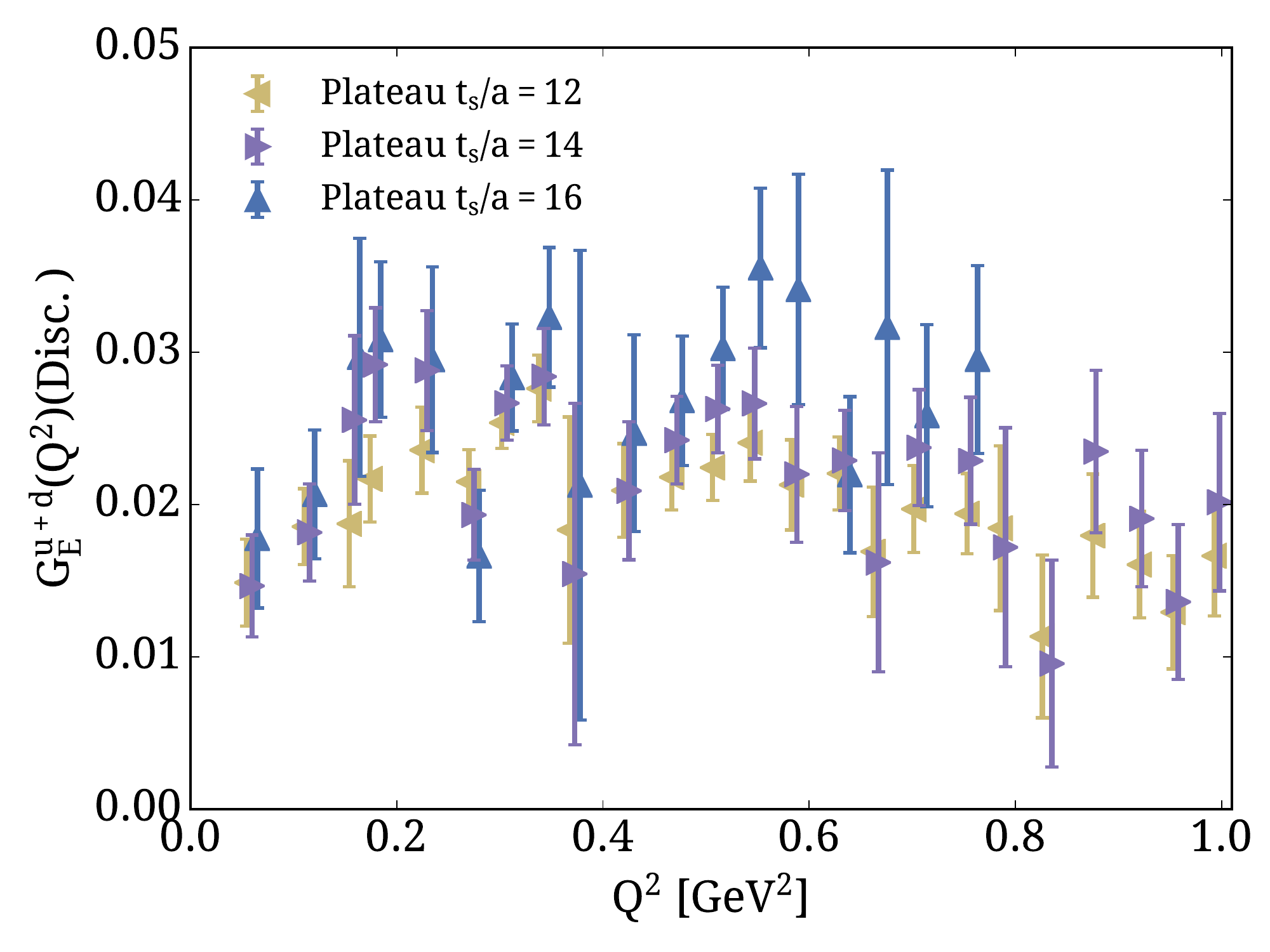}\\[-1.5ex]
    \includegraphics[width=\linewidth]{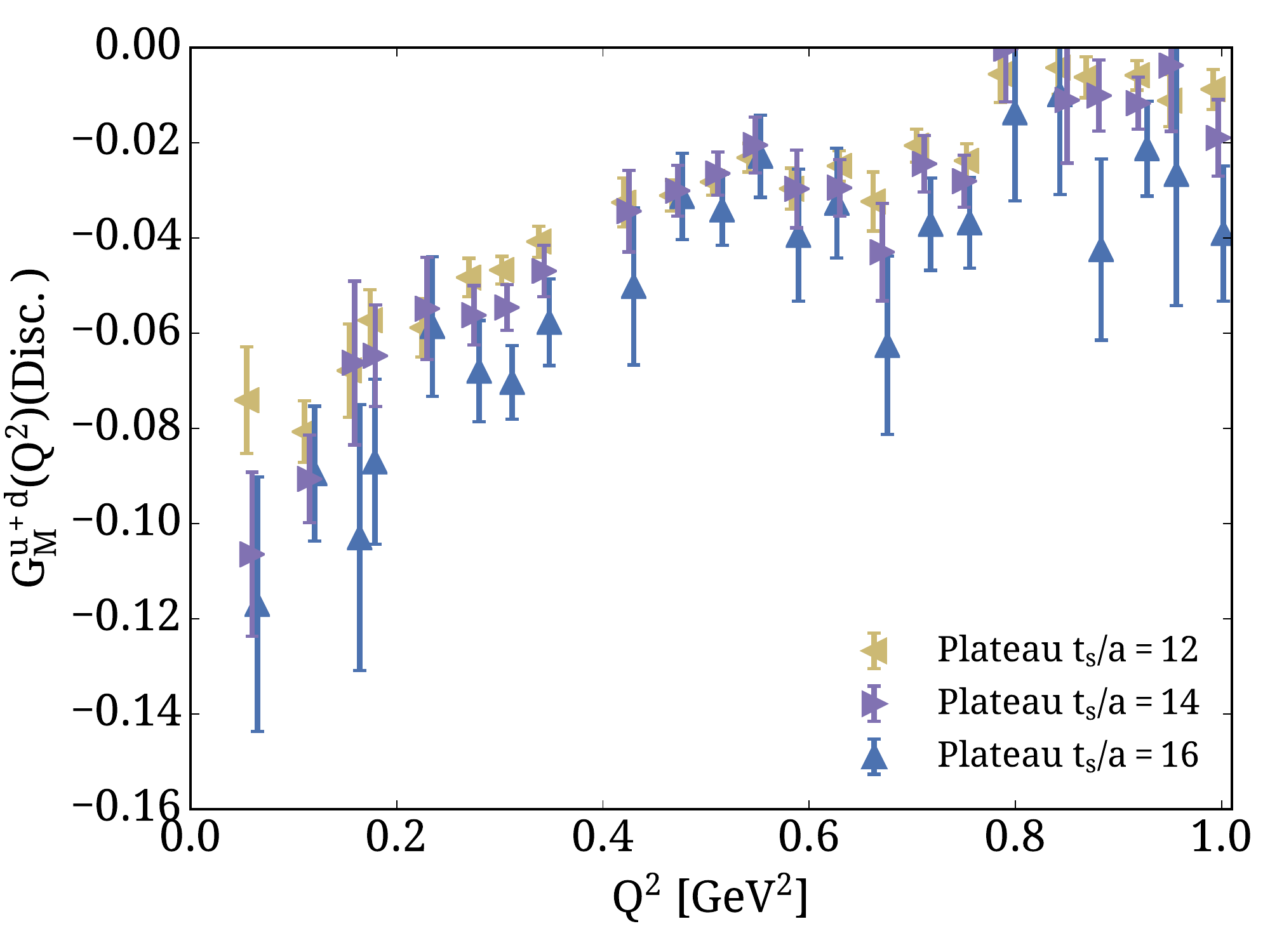}
            \vspace*{-.8cm}
  \caption{The disconnected contributions to $G_E^{u+d}$ (top) and $G_M^{u+d}$ (bottom) using the plateau method for 
   $t_s/a{=}12$ (left triangles), $t_s/a{=}14$ (right triangles) and $t_s/a{=}16$ (upright triangles).
    Points from closely spaced $Q^2$ have been averaged for demonstration, and results from different time separations have
    been slightly shifted to the right for clarity.}
  \label{fig:GEGM_disc_vs_methods}
\end{figure}

In Fig.~\ref{fig:GEGM_disc_vs_methods} we present our results for the disconnected contributions to
$G_E^{u+d}(Q^2)$ and $G_M^{u+d}(Q^2)$ up to $Q^2{=}1$ GeV$^2$ for three time separations, $t_s=0.96, 1.12$ and 1.28~fm.
We can achieve 
a relative statistical error that is less than 20$\%$  for up to $t_s{=}14a {\sim}1.12$~fm, which is unprecedented 
 given that we are using a physical pion mass ensemble.  As we increase the time separation from 
 $t_s/a{=}12$ to $t_s/a{=}14$ we observe, for both  $G_E^{u+d}(Q^2)$ and  $G_M^{u+d}(Q^2)$, that there is a trend for larger values, while
 the results extracted for $t_s/a{=}16$ are in a very good agreement with those extracted for $t_a=14a$ for most $Q^2$ values, albeit with larger errors.
 We, therefore, take as our final result for the disconnected contribution the value extracted
using $t_s/a{=}14$ for both $G_E^{u+d}(Q^2)$ and $G_M^{u+d}(Q^2)$.
We use the  difference between the central value of the results at  $t_s=14a$  and $t_s=16a$ as an estimate of the systematic  error from excited state effects when we quote quantities that include disconnected
contributions. 

\section{Assessment of lattice artifacts}\label{sec:latArtifacts}
We collectively discuss here lattice artifacts that may lead to systematic errors. Since we use simulations with physical values of the light quark masses no chiral extrapolation is needed eliminating one of the biggest uncertainty present in past lattice QCD computations of these quantities. 
\begin{itemize}
  \item{\it Disconnected contributions:}
The main novelty of this work is the accurate computation of the light quarks disconnected contributions using simulations with quark masses tuned to their physical values. This enables us, for the first time, to eliminate this systematic uncertainty in the determination of  the proton and neutron form factors at the physical point. Strange quarks loop contribution is not included in this study, but we know from previous studies \cite{Alexandrou:2018zdf,Sufian:2016pex,Green:2015wqa} that it is much smaller compared to the statistical error of the connected contribution.

\item {\it Quenching effects:} The analysis of the $N_f=2+1+1$ ensemble and two $N_f=2$ ensembles allows us to  check for unquenching effects due to the strange and charm quarks.  In Figs.~\ref{fig:GEGM_comp_uq} and \ref{fig:GEGM_isos_comp_uq} we  compare results using the $N_f{=}2$ cA2.09.48 ensemble to the $N_f{=}2{+}1{+}1$ cB211.072.64 ensemble. The results are extracted from the plateau method with time separation  $t_s=1.3$~fm for  both isovector and isoscalar electric and magnetic form factors.   We observe consistent results  between the two ensembles. Therefore, to the accuracy of our data, no quenching effects due  the strange and charm quarks can be detected. This corroborates our previous study where we found consistent results when comparing an $N_f=2$  and an  $N_f=2+1+1$ ensemble at a pion mass of about 370~MeV~\cite{Alexandrou:2011db}.

\begin{figure}[ht!]
  \includegraphics[width=\linewidth]{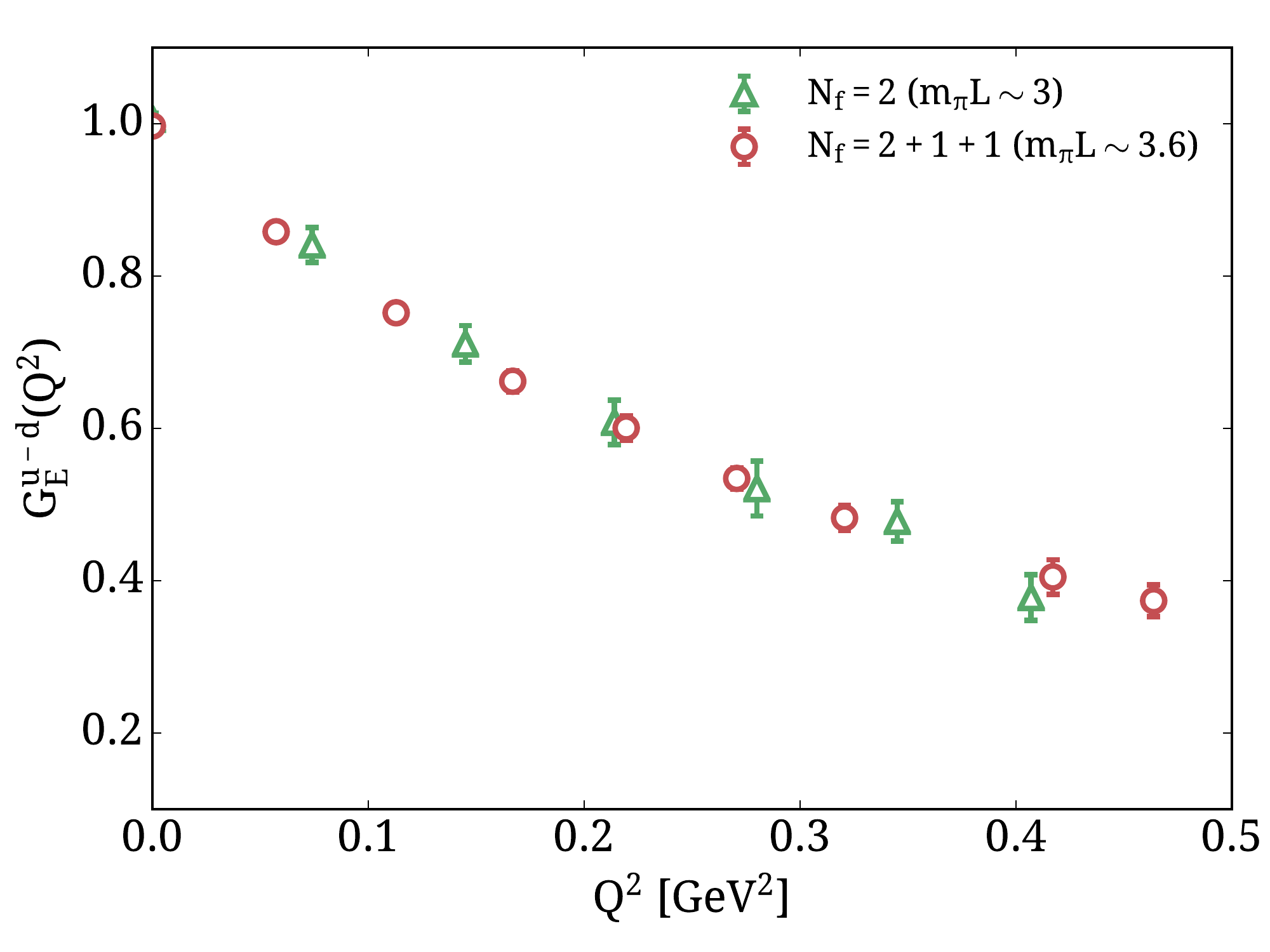}\\[-1.5ex]
    \includegraphics[width=\linewidth]{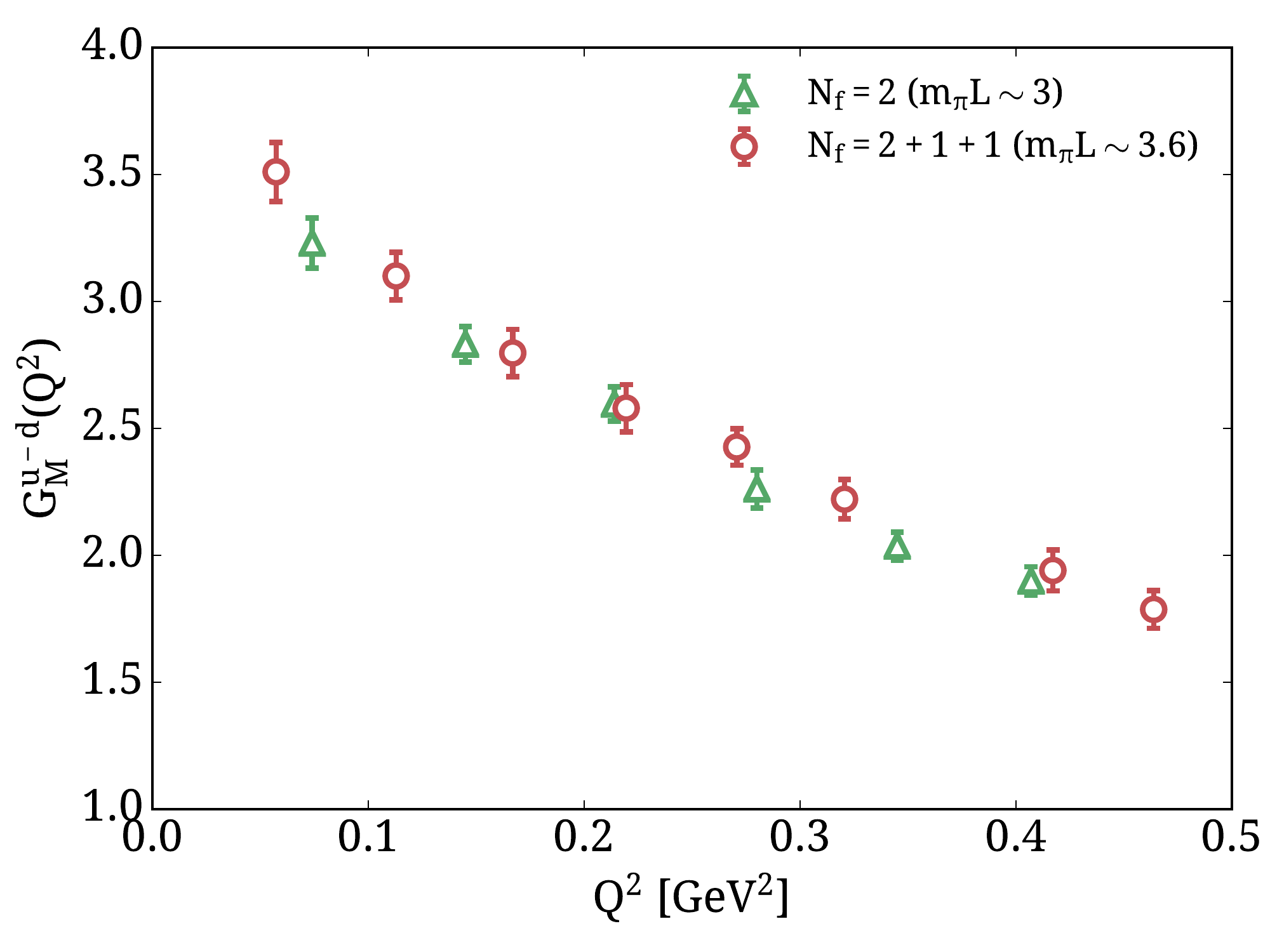}
            \vspace*{-.8cm}
            \caption{Comparison of the isovector electric (top panel) and magnetic (lower panel) form factors between the $N_f{=}2$ cA2.09.48 ensemble~\cite{Alexandrou:2017ypw}
              (green triangles) and the $N_f{=}2{+}1{+}1$ cB211.072.64 ensemble(red circles).
              Results are extracted using the plateau method for sink-source time separation $t_s \simeq 1.3$~fm.}
  \label{fig:GEGM_comp_uq}
\end{figure}

\begin{figure}[ht!]
  \includegraphics[width=\linewidth]{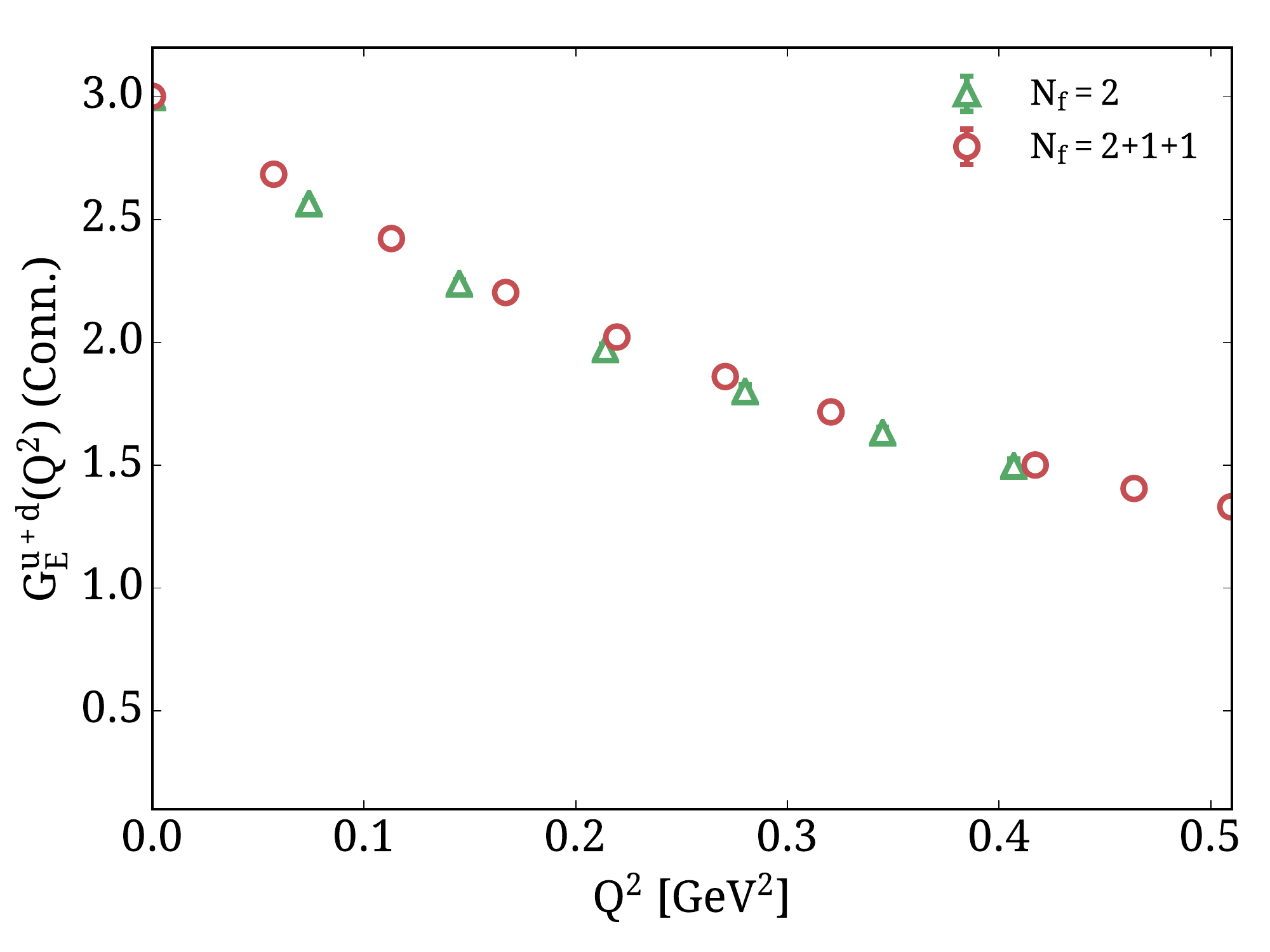}\\[-1.5ex]
    \includegraphics[width=\linewidth]{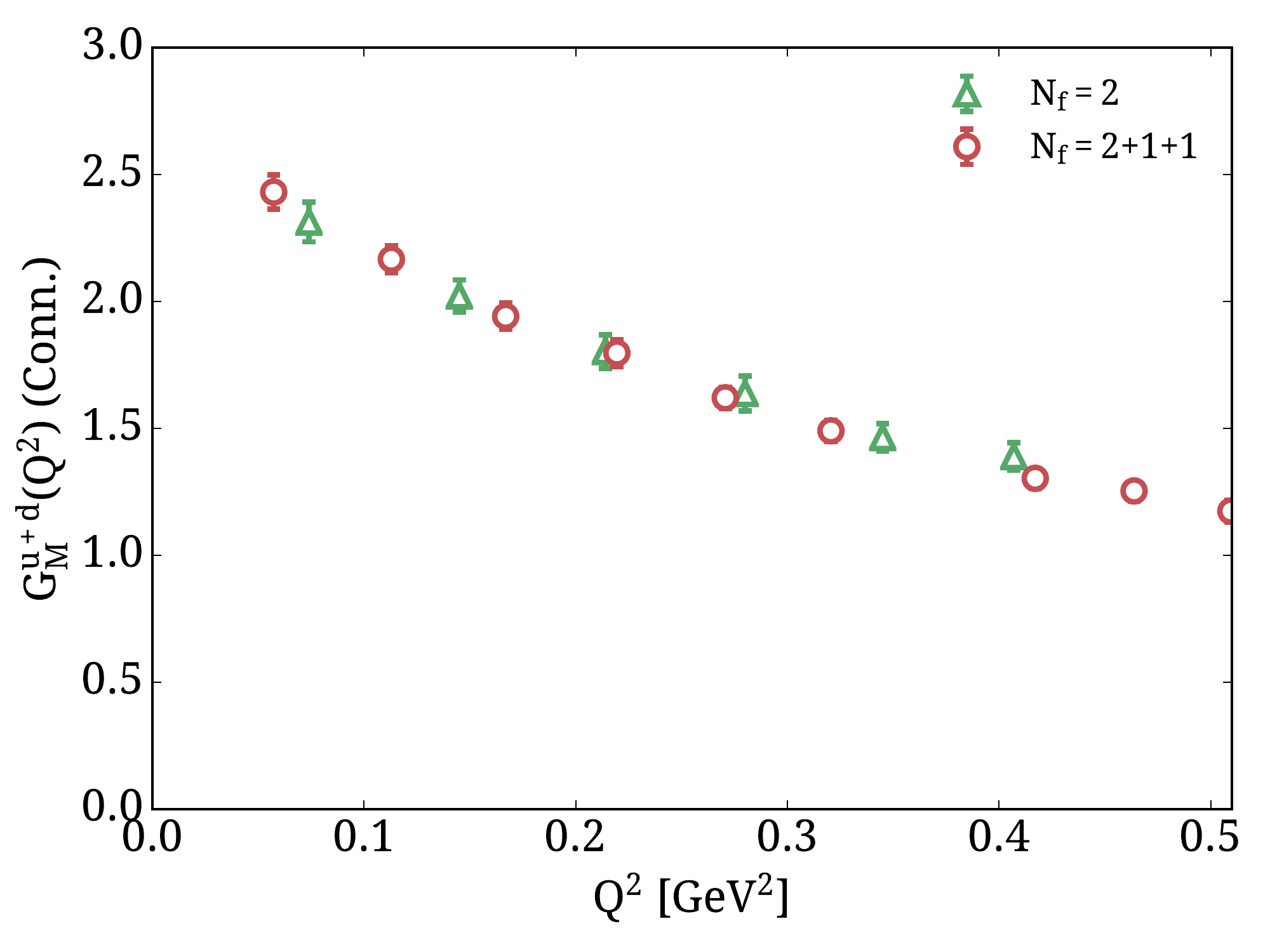}
            \vspace*{-.8cm}
            \caption{Comparison of the connected contribution to the isoscalar electric (top panel) and magnetic (lower panel) form factors between $N_f{=}2$~\cite{Alexandrou:2017ypw}
              and $N_f{=}2{+}1{+}1$ ensembles. The notation is the same as in Fig.~\ref{fig:GEGM_comp_uq}. }
  \label{fig:GEGM_isos_comp_uq}
\end{figure}
\item {\it Isolation of ground state matrix element:} An analysis of excited state contributions is carried out by performing the calculation using several time separations $t_s$. For the target $N_f=2+1+1$ ensemble we use five values of $t_s$ tabulated in  Table~\ref{table:StatsCon211}. We probe convergence to the ground state matrix element by demanding that the matrix element extracted using the plateau and two-state fits are consistent, as explained in detail in Section~\ref{sec:excited states}. The value of  $t_s\sim 1.6$~fm is the largest utilized in this study and to our knowledge in any other study at the physical point. We increase the  statistics as $t_s$  increases  to keep the errors  under control so that a meaningful analysis can be performed to isolate the ground state matrix element. For all of our results we observe agreement between the values extracted using the plateau and two-state fits. Despite this agreement,  residual contamination can still lead to a systematic error within our current statistical.   We give an estimate of such a systematic error by comparing the values obtained with the plateau and two-state fits.  

  \item{Finite volume effects:}
For the assessment of finite volume effects we compare the two $N_f{=}2$ physical point ensembles cA2.09.48 and cA2.09.64 that yield respectively
$m_\pi L \simeq 3$~\cite{Alexandrou:2017ypw} and $m_\pi L \simeq 4$. The lattice spacing and pion mass are the same for these two ensembles. We also use the same time separation $t_s$ for each observable when comparing between the two  ensembles.
The isovector electric and magnetic form factors extracted using the plateau method are shown in Fig.~\ref{fig:GEGM_comp_ve}.
The results fall on the same curve indicating no significant finite volume effects between the two volumes of
$m_\pi L \simeq 3$ and $m_\pi L \simeq 4$. The same behavior is observed for the isoscalar form factors shown in Fig.~\ref{fig:GEGM_isos_comp_ve}.
\begin{figure}[ht!]
  \includegraphics[width=\linewidth]{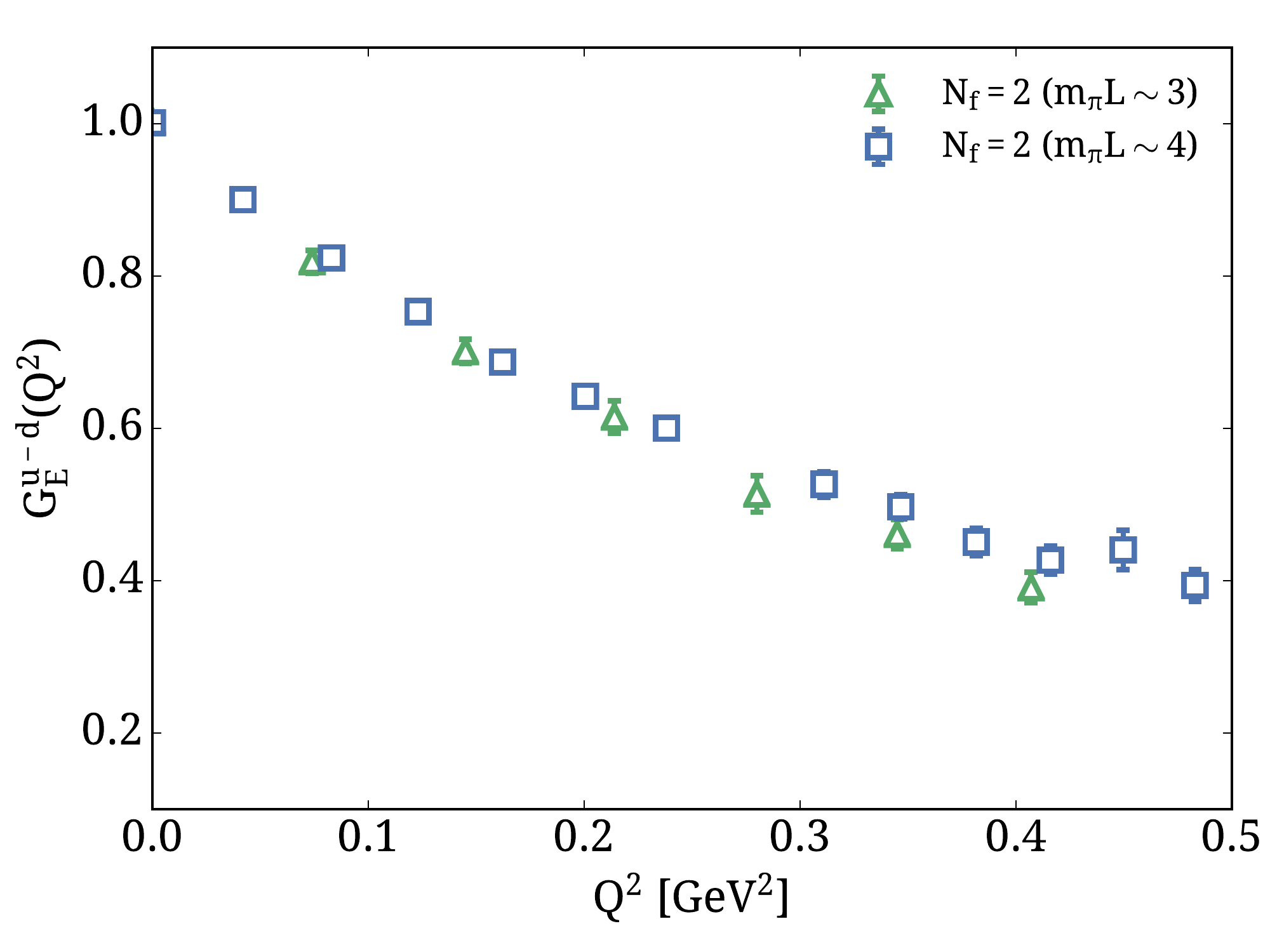}\\[-1.5ex]
    \includegraphics[width=\linewidth]{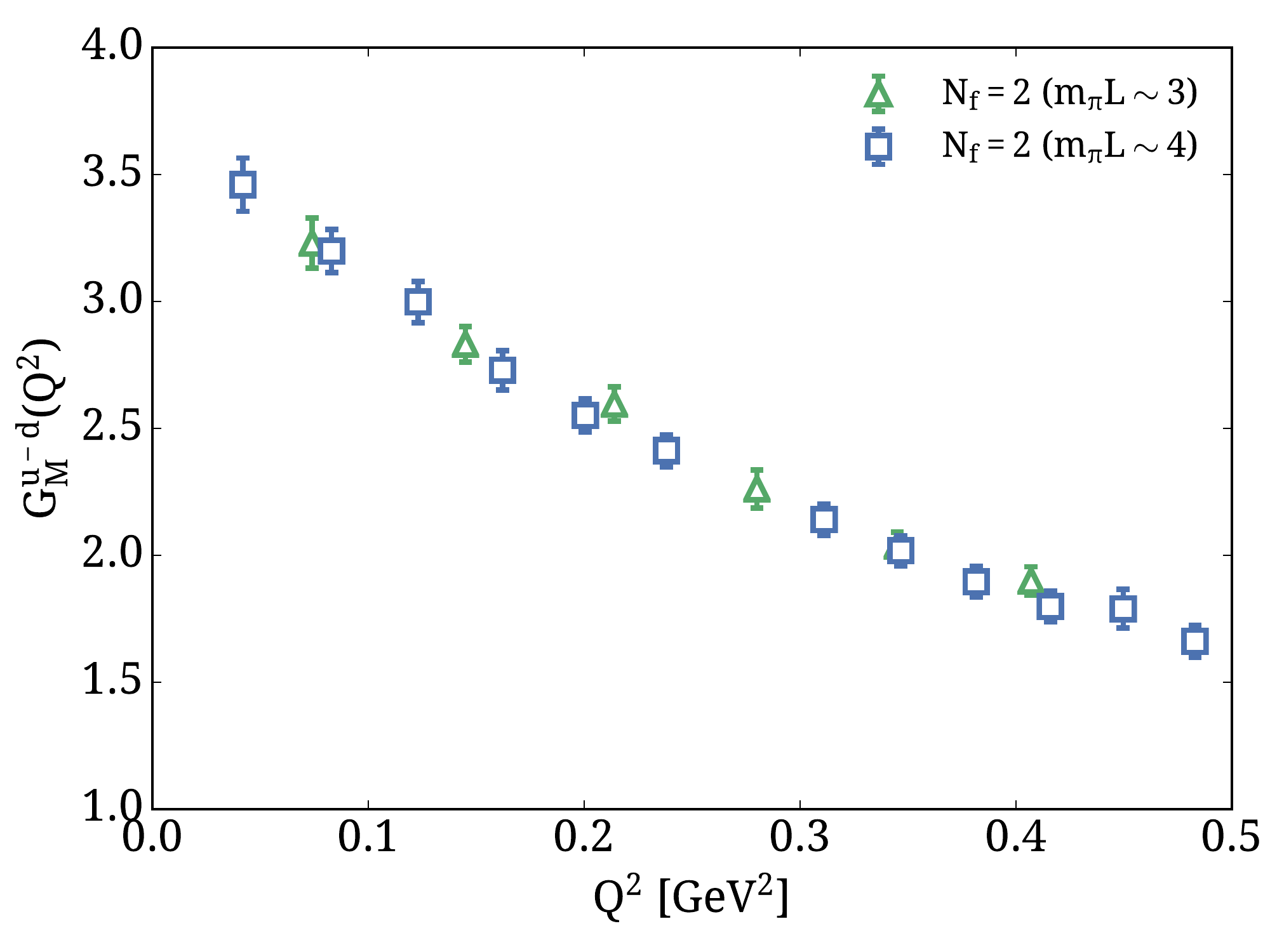}
            \vspace*{-.8cm}
  \caption{The isovector electric (top panel) and magnetic (lower panel) form factor for two different physical spatial volumes. With
    green triangles we show the results for the cA2.09.48 with $m_\pi L \simeq 3$~\cite{Alexandrou:2017ypw}
    and  with blue squares the cA2.09.64 
    with $m_\pi L \simeq 4$. Results are extracted using the plateau method for sink-source separation $t_s \simeq 1.5$~fm for the electric and
    $t_s \simeq 1.3$~fm for the magnetic form factors.}
  \label{fig:GEGM_comp_ve}
\end{figure}
\begin{figure}[ht!]
  \includegraphics[width=\linewidth]{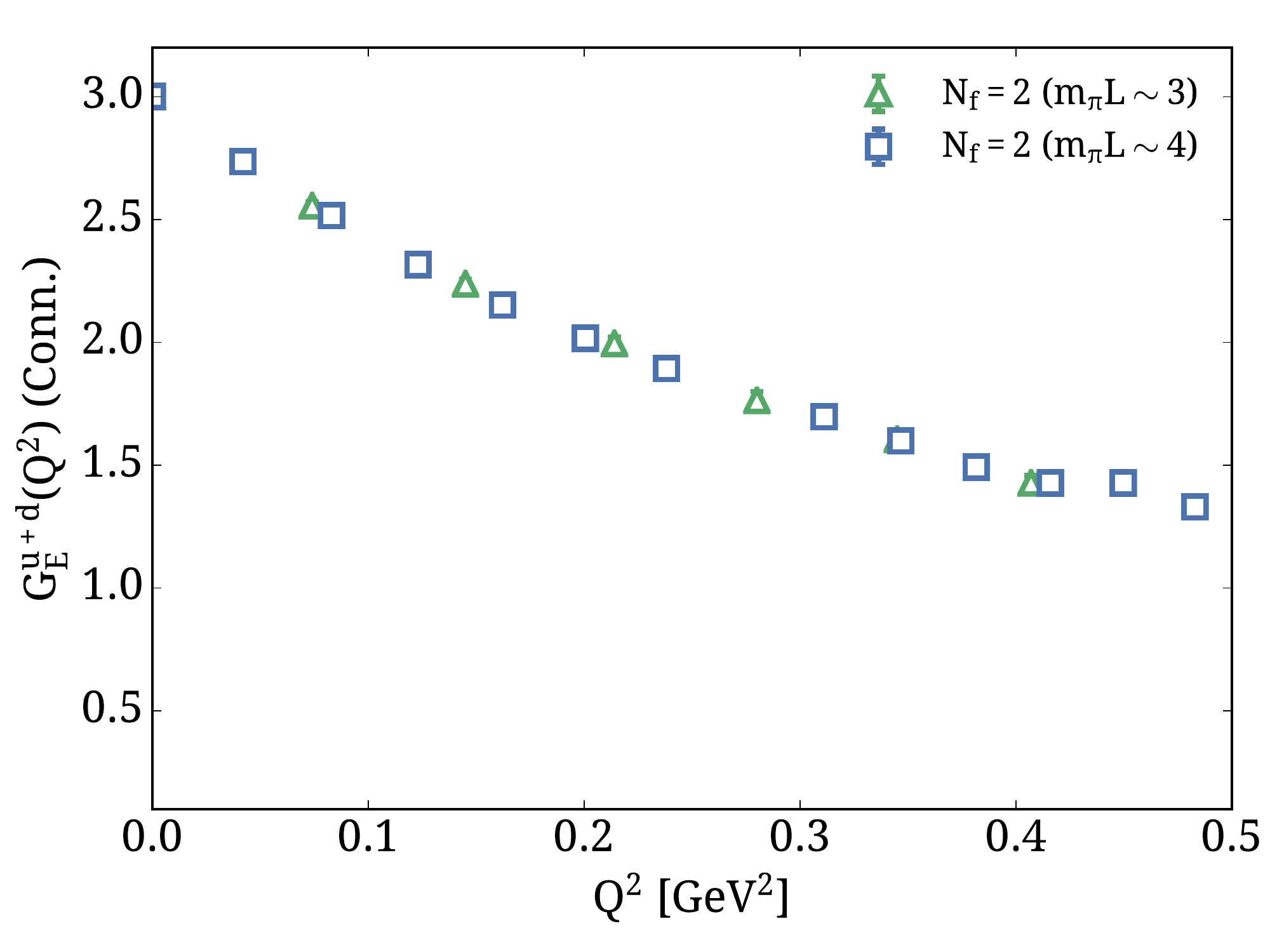}\\[-1.5ex]
    \includegraphics[width=\linewidth]{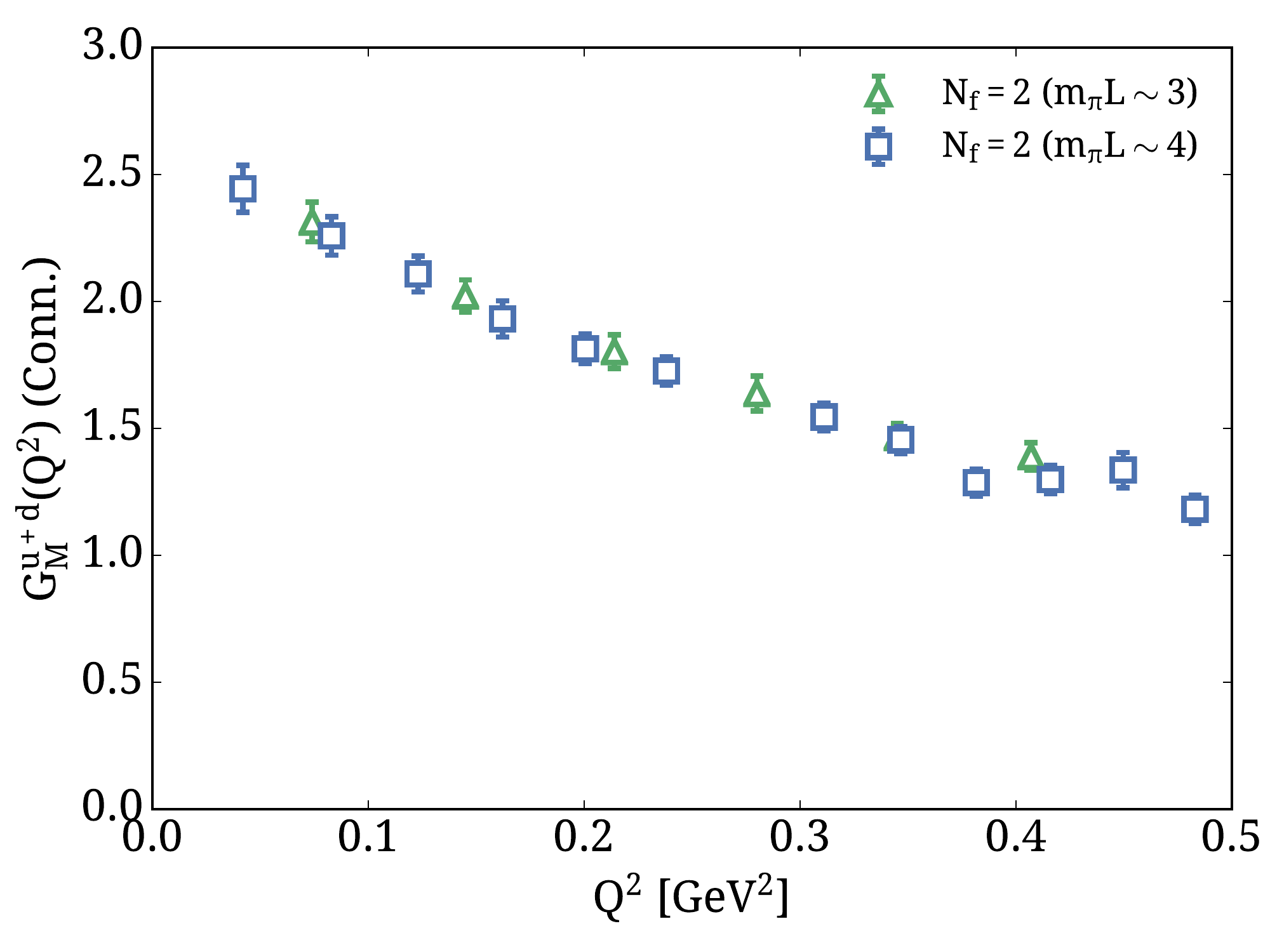}
            \vspace*{-.8cm}
            \caption{The isoscalar electric (top panel) and magnetic (lower panel) form factor for two different physical spatial volumes.
            The notation is as in Fig.~\ref{fig:GEGM_comp_ve}.}
  \label{fig:GEGM_isos_comp_ve}
\end{figure}
We would like to stress once more that our statement of detecting no volume effects  can only be made within the current accuracy and some residual volume effects can still lead to a systematic effect.
One complication as the volume increases is the  contamination due to higher excited states, since the number of  multi-hadron states allowed increases~\cite{Bar:2018xyi}. Such multi-hadron states are not expected to affect results at larger pion masses but are expected to be more severe at the physical pion mass.
Such effects can be modeled within chiral perturbation theory for the axial form factors~\cite{Bar:2018xyi}. For the electromagnetic form factors these effects are not known but an interplay between volume and excited state effects may account for the deviations observed between lattice QCD data and experimental values.
\item{\it Finite lattice spacing, a:} Since in this work we are using the twisted mass formulation at maximal twist our results are automatically ${\cal O}(a)$ improved without any need to improve the current. This is different from clover fermions where the current must be improved in order to eliminate oder $a$ contributions. Therefore, our results only  have corrections of ${\cal O}(a^2)$.
  Continuum extrapolation cannot be performed given that
  we have analyzed only one  $N_f=2+1+1$ ensemble. The two $N_f=2$ ensembles analyzed have the same lattice spacing and so again finite lattice spacings affects cannot be assessed. Previous studies done using ensembles with pion mass spanning about 460~MeV to 260~MeV and three values of the lattice spacings have indeed demonstrated that the   ${\cal O}(a^2)$ correction is negligible~\cite{Alexandrou:2011db}. We thus do not expect large systematic cut-off effects on our results. However, an analysis of cut-off effects will need to be carried out in the future when additional ensembles are available.
\end{itemize}
In summary, there maybe a slow convergence as a function of the volume in conjunction with residual excited state effects.  This may explain the few $\sigma$ discrepancy observed between lattice QCD results and the experimental values. In particular, we note that the electric form factors has increasing excited state effects for larger values of $Q^2$, whereas for $G_M$ these effects are bigger at small $Q^2$. As we will see these are the ranges of momenta where we see discrepancies with the experimental values.

\section{$Q^2$-dependence of the isovector and isoscalar form factors}\label{sec:Isov Isos}
In this section we discuss the $Q^2$-dependence of the form factors using  standard parameterizations as described in the next section.
\subsection{Parameterizations of the $Q^2$-dependence}\label{sec:Q2-forms}

Assuming vector meson pole dominance for $Q^2 < 0$, one expects
that for small $Q^2 > 0$ the behavior will be dominated by the poles in the time-like
region. One would then expect a  dipole form  given by~\cite{Perdrisat:2006hj}
\begin{equation}
  G(Q^2) = \frac{G(0)}{(1 + \frac{Q^2}{M^2})^2},
  \label{Eq:Dipole}
\end{equation}
where $M$ is the mass of the vector meson that parameterizes the $Q^2$ dependence. The value of the form factor
at zero momentum transfer gives the electric charge in the case of
the electric form factor and the magnetic moment in the case of
the magnetic form factor.
Combining Eq.~(\ref{Eq:Dipole}) and Eq.~(\ref{Eq:radius}) one
can relate $M$ to the mean square radius as
\begin{equation}
 \langle r^2 \rangle = \frac{12}{M^2}.  
\end{equation}

The neutron electric form factor and disconnected contributions to
the electric form factors are zero for $Q^2{=}0$ and we treat
them as special cases, fitting them using the Galster-like
parameterization~\cite{Galster:1971kv,Alberico:2008sz}, given by
\begin{equation}
  G(Q^2) = \frac{Q^2 A}{4 m_N^2 + Q^2 B} \frac{1}{(1+\frac{Q^2}{0.71 {\rm GeV}^2})^2},
  \label{Eq:Galster-like}
\end{equation}
with $A$ and $B$ fit parameters. In this case the radius is given by
\begin{equation}
  \langle r^2 \rangle = - \frac{3A}{2 m_N^2}.
\end{equation}

Another fit form, which has been applied recently to experimental data of both electromagnetic 
and axial form factors, is the model independent z-expansion~\cite{Hill:2010yb}. In this case,
the form factor is expanded in a series given by,
\begin{equation}
  G(Q^2) = \sum_{k = 0}^{k_{\rm max}} a_k z^k,
  \label{Eq:zExp}
\end{equation}
where
\begin{equation}
  z = \frac{\sqrt{t_{\rm cut} + Q^2} - \sqrt{t_{\rm cut}}}{\sqrt{t_{\rm cut} + Q^2} + \sqrt{t_{\rm cut}}}
\end{equation}
and $t_{\rm cut}$ is the time-like cut of the form factor. We take $t_{\rm cut} {=} 4 m_\pi^2$ for
the isovector combination $G^u-G^d$ and $t_{\rm cut} {=} 9 m_\pi^2$ for the isoscalar combination $G^u+G^d$~\cite{Hill:2010yb}. For convergence of the truncated series of Eq.~(\ref{Eq:zExp}), the coefficients $a_k$ should be 
bounded in size and convergence should be demonstrated by increasing $k_{\rm max}$. The interested reader is referred to Ref.~\cite{Alexandrou:2018zdf}
for details about our procedure. The mean square radius is given by
\begin{equation}
  \langle r^2 \rangle = - \frac{3 a_1}{2 a_0 t_{cut}},
\end{equation}
while the value of the form factor at zero momentum transfer is $G(0){=}a_0$.

\subsection{Fits to lattice QCD results}
\begin{figure}[ht!]
  \includegraphics[width=\linewidth]{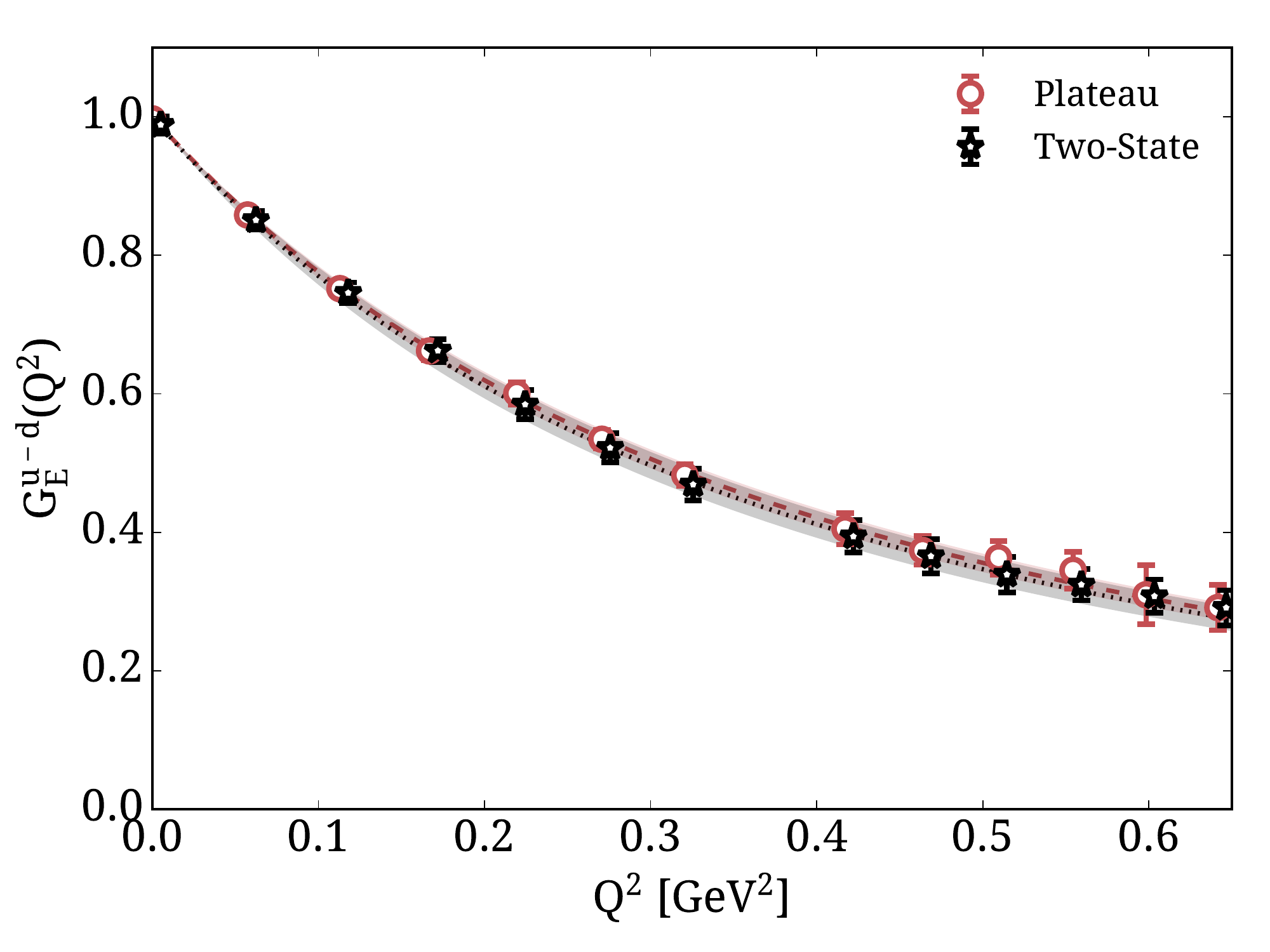}
  \vspace*{-.8cm}
  \caption{$G_E^{u-d}(Q^2)$ from the plateau method for $t_s/a=20$ (circles) and two-state fits (stars). The dashed (dotted) curve and corresponding band is a dipole fit to the plateau (two-state) fit results, which overlap.}
  \label{fig:GE_plt_tsf}
\end{figure}

\begin{figure}[ht!]
  \includegraphics[width=\linewidth]{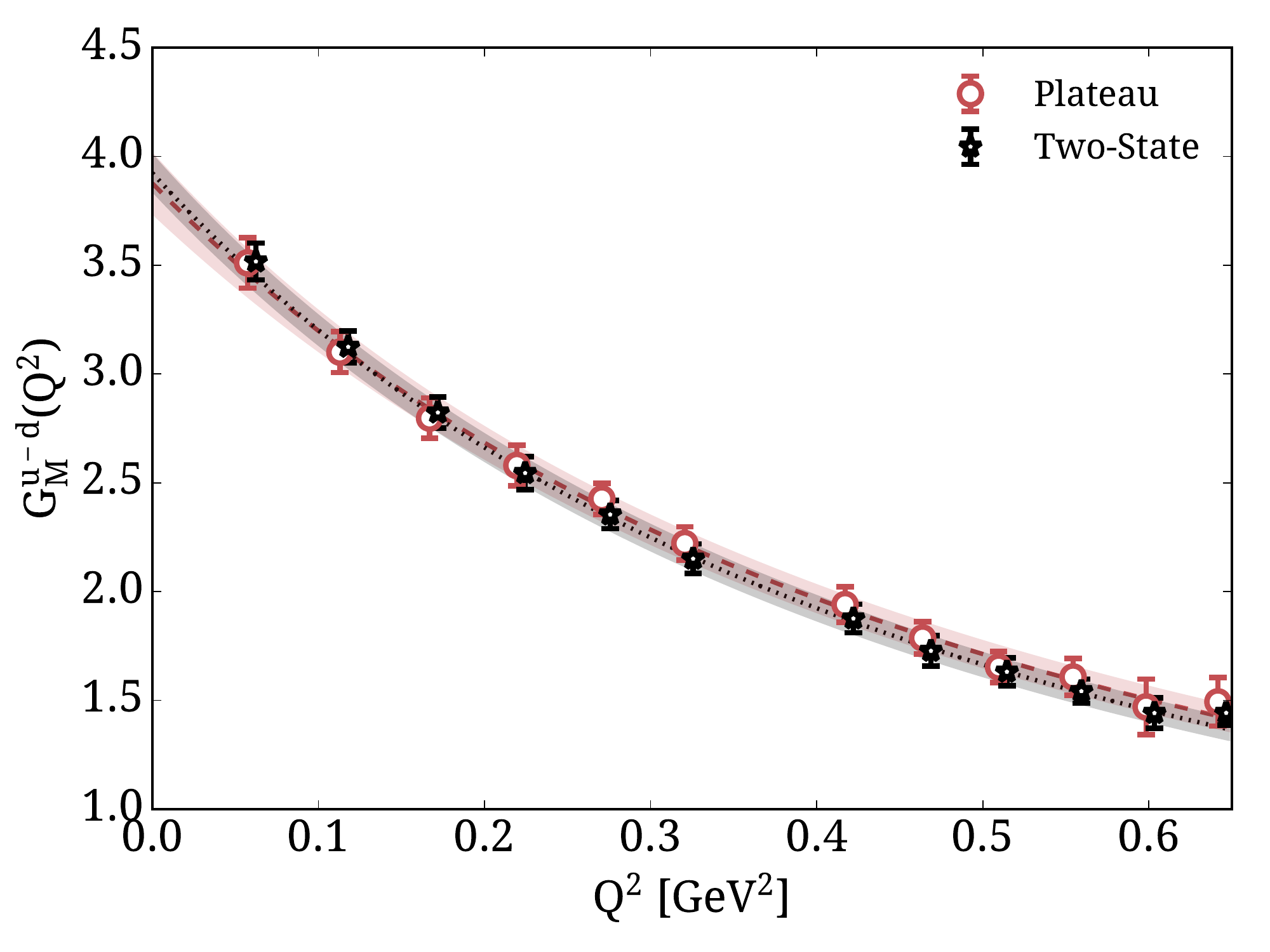}
  \vspace*{-.8cm}
  \caption{$G_M^{u-d}(Q^2)$ from the plateau method for $t_s/a=20$ (circles) and two-state fits (stars). The rest of the notation is as in Fig.~\ref{fig:GE_plt_tsf}.}
  \label{fig:GM_plt_tsf}
\end{figure}

We consider first the isovector form factors where only the connected diagram contributes.
In Figs.~\ref{fig:GE_plt_tsf} and~\ref{fig:GM_plt_tsf} we show fits using the dipole form, comparing between results from the plateau method at $t_s/a=20$ and from two-state fits for $G_E^{u-d}(Q^2)$ and $G_M^{u-d}(Q^2)$, respectively. As can be seen, fits using the plateau and two-state methods are fully consistent and do not show any significant systematic effect on the determination of the $Q^2$-dependence of the form factors,  indicating that excited states are sufficiently suppressed.  Since results are in agreement,
from now on we will use the plateau method at $t_s/a=20$
to extract final results on the form factors, r.m.s radii, and magnetic moment.
We will use the results extracted from the two-state fits to estimate the systematic error due to excited states.

\begin{figure}[ht!]
  \includegraphics[width=\linewidth]{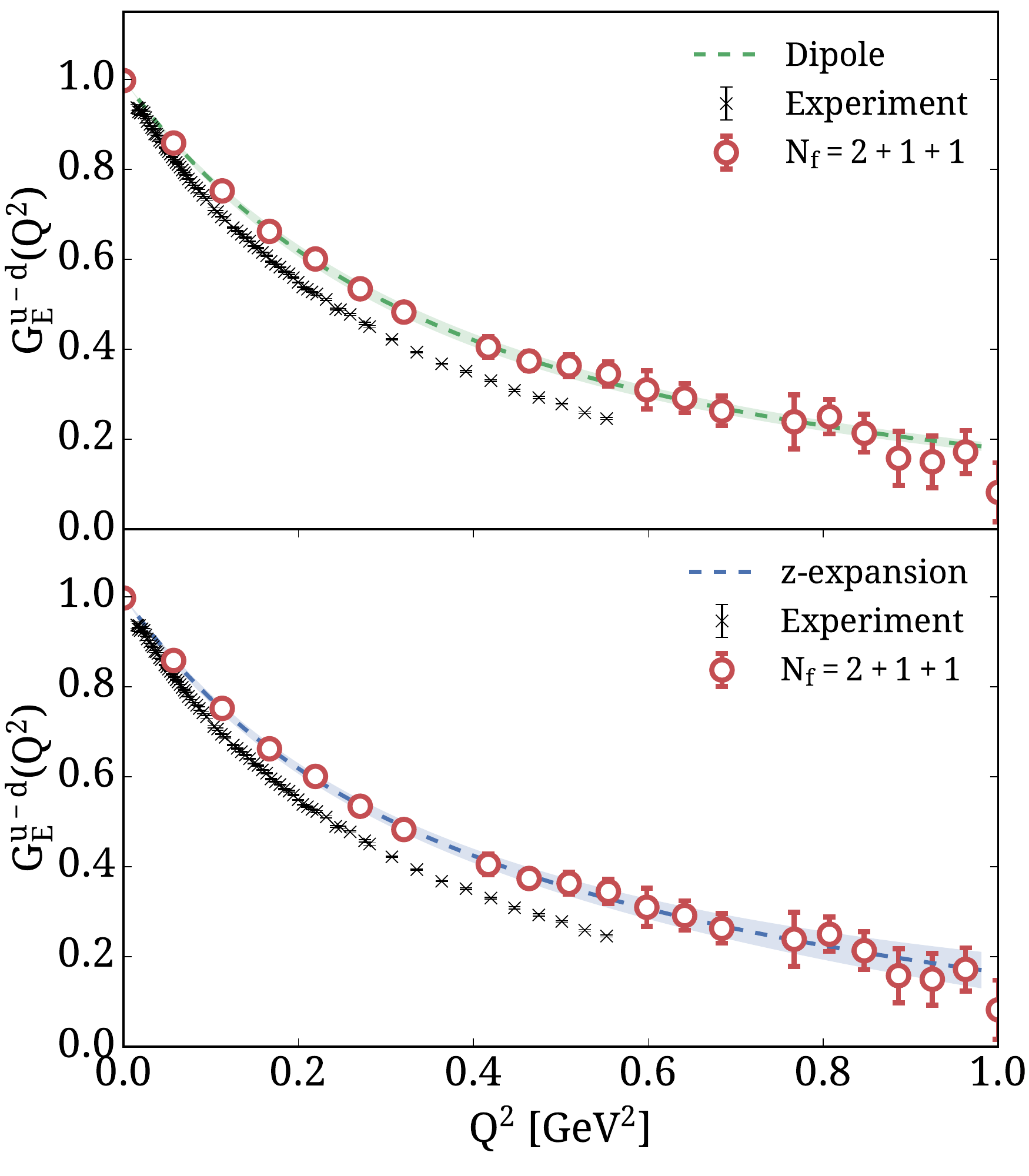}
          \vspace*{-.8cm}
  \caption{The isovector electric form factor as a function of $Q^2$ (circles). We show fits to our results using a dipole form (top) and
    using the z-expansion (bottom) for $k_{\rm max}{=}4$. Black crosses are experimental results taken from the A1 collaboration~\cite{Bernauer:2013tpr} for the proton and from Refs.~\cite{Golak:2000nt,Becker:1999tw,Eden:1994ji,Meyerhoff:1994ev,Passchier:1999cj,Warren:2003ma,Zhu:2001md, Plaster:2005cx, Madey:2003av, Rohe:1999sh, Bermuth:2003qh,Glazier:2004ny,Herberg:1999ud,Schiavilla:2001qe,Ostrick:1999xa} for the neutron.}
  \label{fig:GE_dip_zExp}
\end{figure}

\begin{figure}[ht!]
  \includegraphics[width=\linewidth]{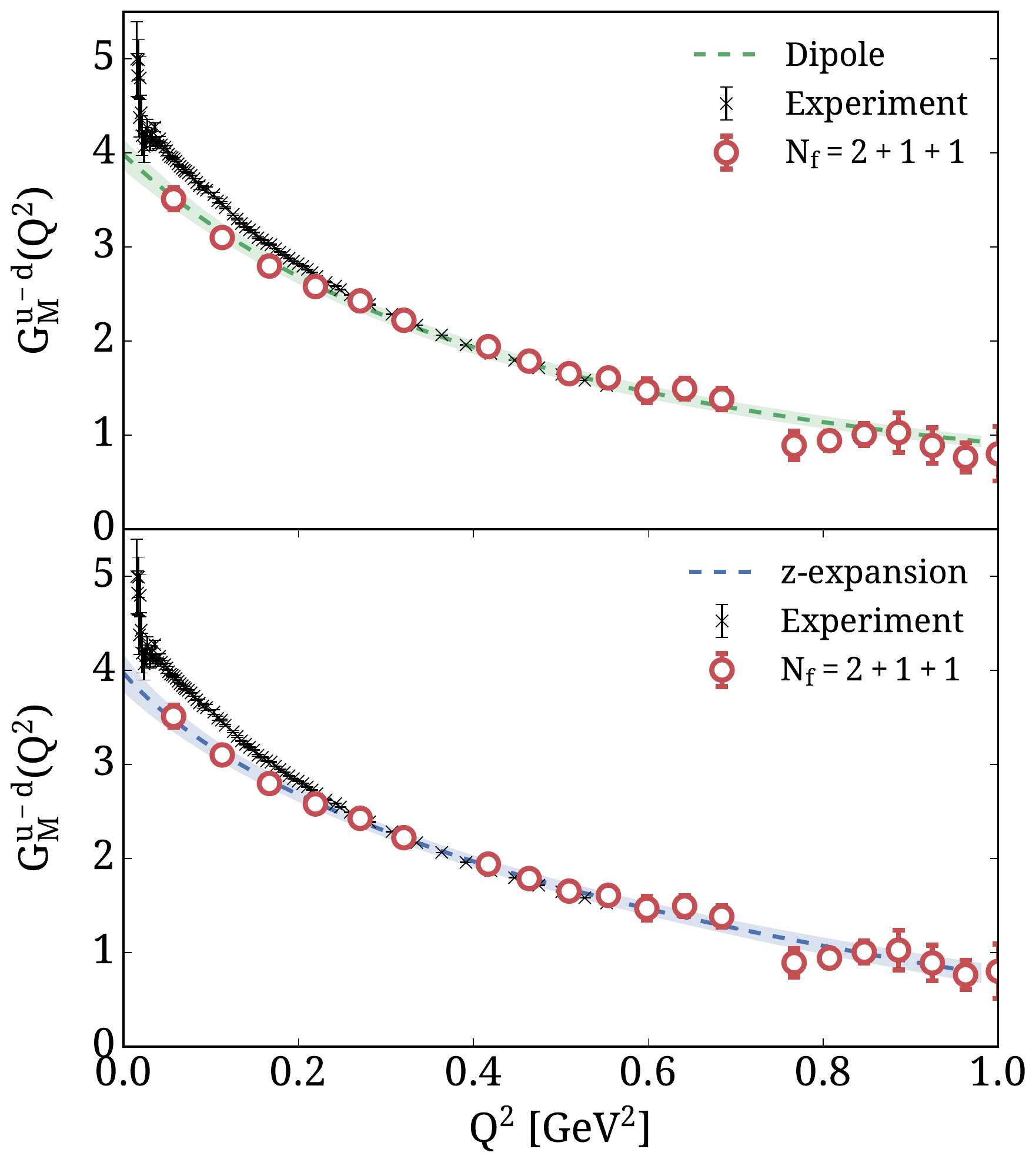}
          \vspace*{-.8cm}
          \caption{The isovector magnetic form factor fitted using a dipole form (top) and
    using the z-expansion (bottom) with the notation of Fig.~\ref{fig:GE_dip_zExp}.}
  \label{fig:GM_dip_zExp}
\end{figure}

\begin{figure}[ht!]
  \includegraphics[width=\linewidth]{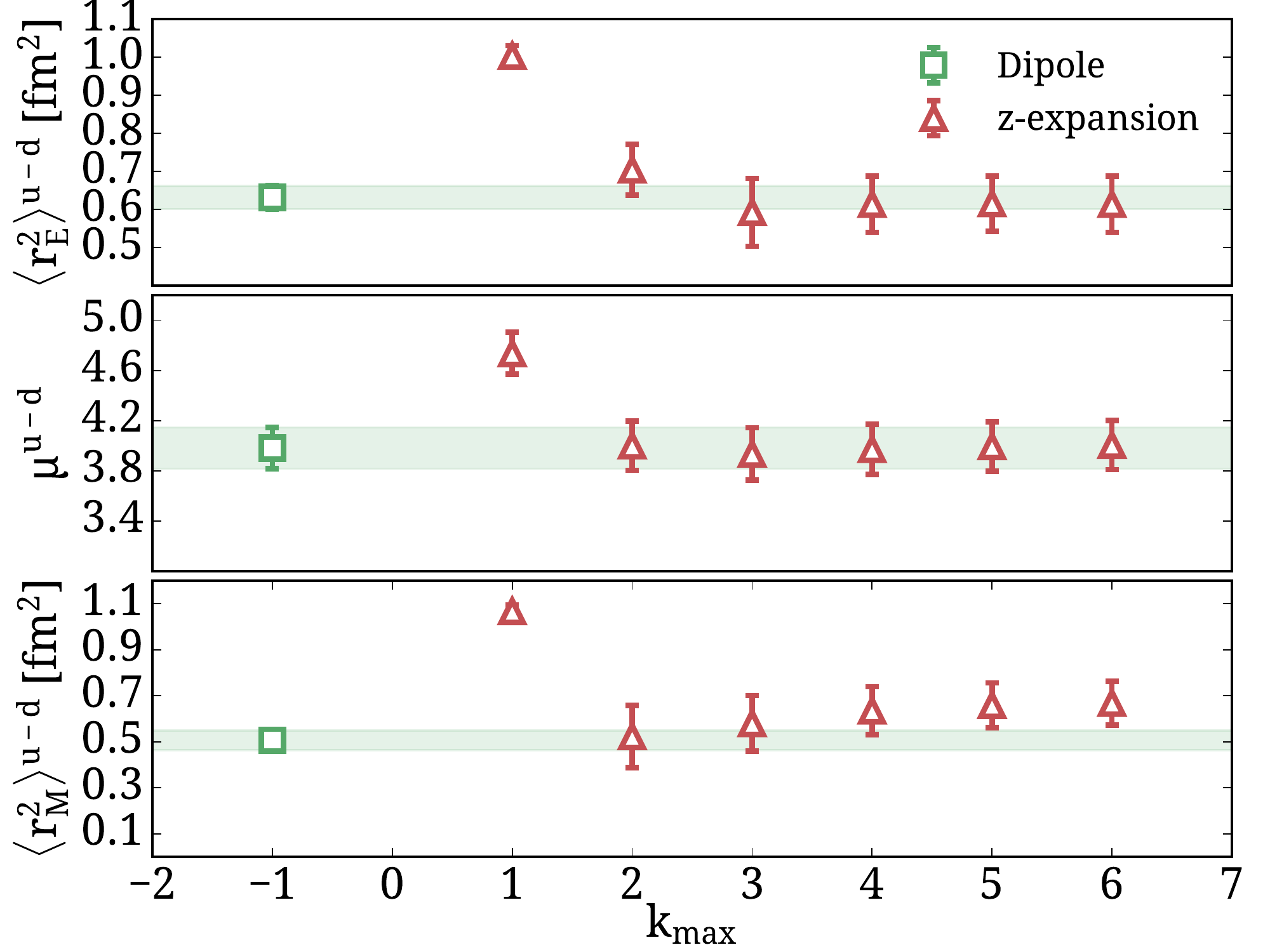}
          \vspace*{-.8cm}
  \caption{Results for the isovector charge radius $\langle r_E^2 \rangle^{u-d}$, magnetic moment $\mu^{u-d}$ and magnetic radius $\langle r_M^2 \rangle^{u-d}$ from the plateau method using $t_s/a{=}20$ as extracted from a dipole fit (green square) and z-expansion (red triangles). The latter are shown as a function of $k_{\rm max}$. The green band is the statistical error on the value extracted from the dipole fit.}
  \label{fig:rE2_muM_rM2_isov}
\end{figure}

In Figs.~\ref{fig:GE_dip_zExp} and~\ref{fig:GM_dip_zExp}, we show fits to $G_E^{u-d}(Q^2)$ and $G_M^{u-d}(Q^2)$, respectively using the dipole form and the z-expansion of Eq.~(\ref{Eq:zExp}) and compare to experiment. For the z-expansion, we check convergence by
increasing $k_{\rm max}$. The resulting magnetic moment and r.m.s radii are shown in Fig.~\ref{fig:rE2_muM_rM2_isov}, where we observe convergence for $k_{\rm max}{=}4$. For $G_E^{u-d}(Q^2)$, we see from Fig.~\ref{fig:GE_dip_zExp} that the  slope
of the lattice QCD data is less as compared to the experimental values.
Therefore, although both dipole form and z-expansion describe very well our data, shown in separate panels for clarity, they lie consistently  above the experimental values. A study using a larger volume with a careful examination of excited state effects is planned to understand  the origin of this remaining discrepancy. In extracting the r.m.s radius, we see from Fig.~\ref{fig:rE2_muM_rM2_isov} that results obtained from using the dipole fit and z-expansion are compatible, and yield
\begin{equation}
  \sqrt{\langle r_E^2 \rangle^{u-d}} {=} 0.796(19)(12)(12) \; {\rm fm},
  \label{rms electric}
\end{equation}
 where the central value and the statistical error are taken from the dipole fit, the second error is a systematic
computed as the difference in the mean values between dipole and z-expansion for $k_{\rm max}{=}4$ and the third error
is the systematic error due to excited states obtained from the difference when fitting the form factor extracted from the plateau and from the two-state fit method. 
Subsequent quantities given in the paper  will have statistical and systematic errors quoted using the same convention as in Eq.~\ref{rms electric}.

For $G_M^{u-d}$, shown in Fig.~\ref{fig:GM_dip_zExp}, we observe that our results are in agreement with the experimental values for $Q^2 > 0.2$~GeV$^2$, whereas for small $Q^2$ they tend to be lower. A possible explanation for this discrepancy is that effects from the pion cloud, expected to be prominent for small momenta~\cite{Sato:2003rq}, are suppressed in our calculation due to our finite volume. The fact that
we have seen no volume effects when  we increase the volume from $L m_\pi {\simeq} 3$ to $L m_\pi {\simeq} 4$ for our two $N_f=2$ ensembles may indicate that pion cloud suppression may not be detectable for these volume sizes  requiring larger volumes to unfold.
Indeed, preliminary results  by PACS using a physical point ensemble  with $L m_\pi {\simeq} 7.4$~\cite{Shintani:2018ozy} finds a higher value that may point to a finite volume effect. This would need further investigation to confirm.

The isovector magnetic moment and mean square  magnetic radii are shown in Fig.~\ref{fig:rE2_muM_rM2_isov}. As  can be seen, the mean values extracted for $\mu^{u-d}$ 
using the dipole and z-expansion are compatible, while for $\langle r_M^2 \rangle^{u-d}$ the z-expansion produces a slightly higher mean value, which, however, is consistent within errors. Quoting the values from the dipole fit, we find
\begin{eqnarray}
  \mu^{u-d} &=& 3.97(15)(2)(5)(^1_0) \\
  \sqrt{\langle r_M^2 \rangle^{u-d}} &=& 0.712(27)(87)(5)(^1_0) \; {\rm fm}\,. 
\end{eqnarray}
Here we have included a fourth systematic error computed as  the difference in the values of $\mu^{u-d}$ and $\langle r_M^2 \rangle^{u-d}$ when fitting $G_M^{u-d}(Q^2)$ including and excluding the lowest $Q^2$ value from the fit. The error is asymmetric, since the expectation is that pion cloud effects will increase the value of the magnetic form factor. It is also small compared to the systematic error due to excited states. In what follows we will not include this fourth systematic error. 

\begin{figure}[ht!]
  \includegraphics[width=\linewidth]{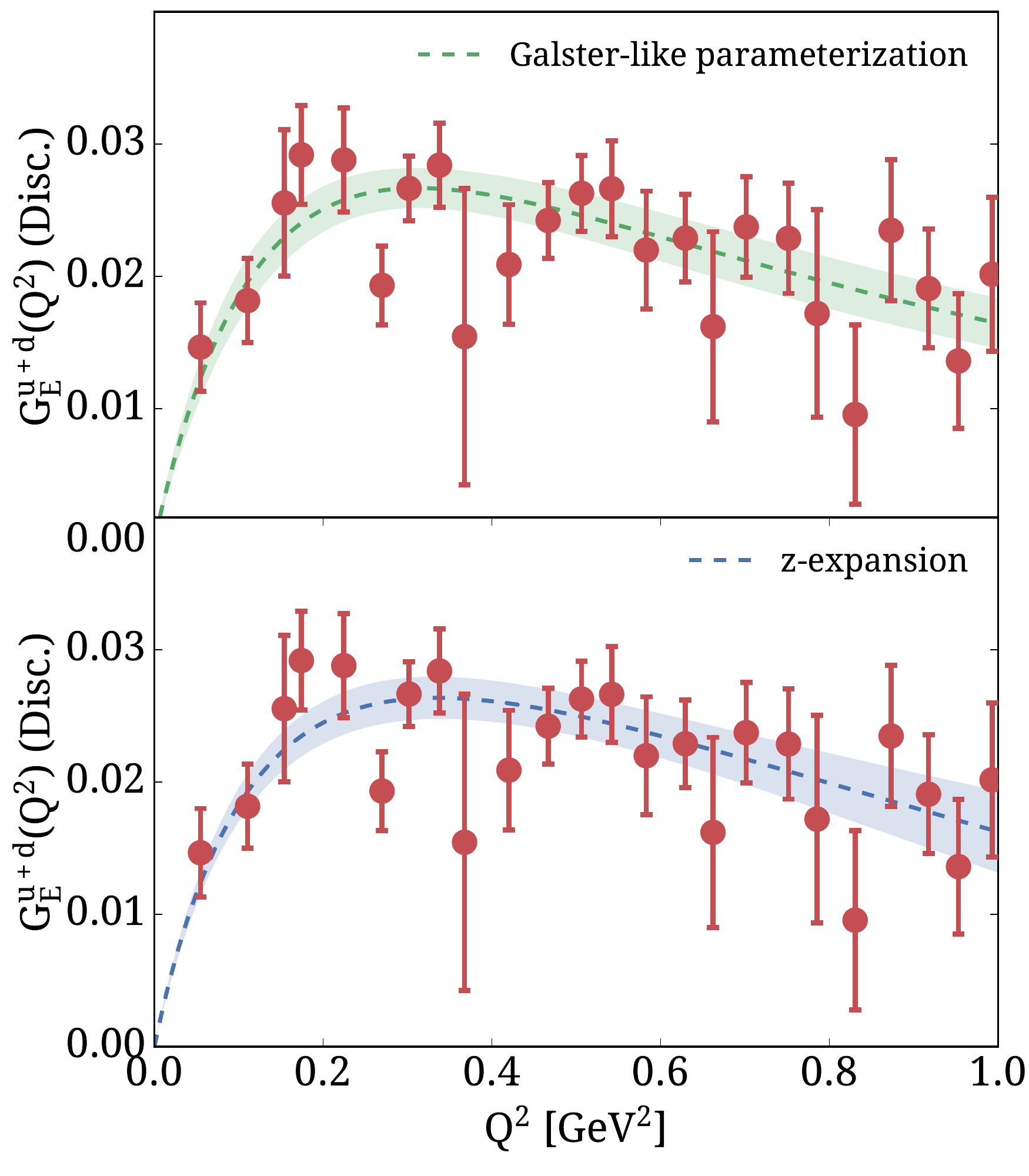}
          \vspace*{-.8cm}
  \caption{Disconnected contributions to the isoscalar electric form factor  (circles) as a function of $Q^2$. The fits using the  Galster-like parameterization of Eq.~(\ref{Eq:Galster-like}) and the z-expansion 
    for $k_{\rm max}{=}3$ are shown in the upper and lower panels, respectively.}
  \label{fig:GE_light_disc_std_zExp}
\end{figure}

Before presenting fits to the total isoscalar form factors we discuss separately  the $Q^2$-dependence of the disconnected contributions.
  In Fig.~\ref{fig:GE_light_disc_std_zExp} we show the  disconnected contribution to the isoscalar electric form factor
$G_E^{u+d}$, accompanied by fits to the  Galster-like parameterization and z-expansion.  We note that in the case of the z-expansion we take $a_0{=}0$, since 
$G_E^{u+d}(0){=}0$ for the disconnected contribution. Both parameterizations 
describe well our results with  the z-expansion yielding a larger error for
 the larger $Q^2$ values.

\begin{figure}[ht!]
  \includegraphics[width=\linewidth]{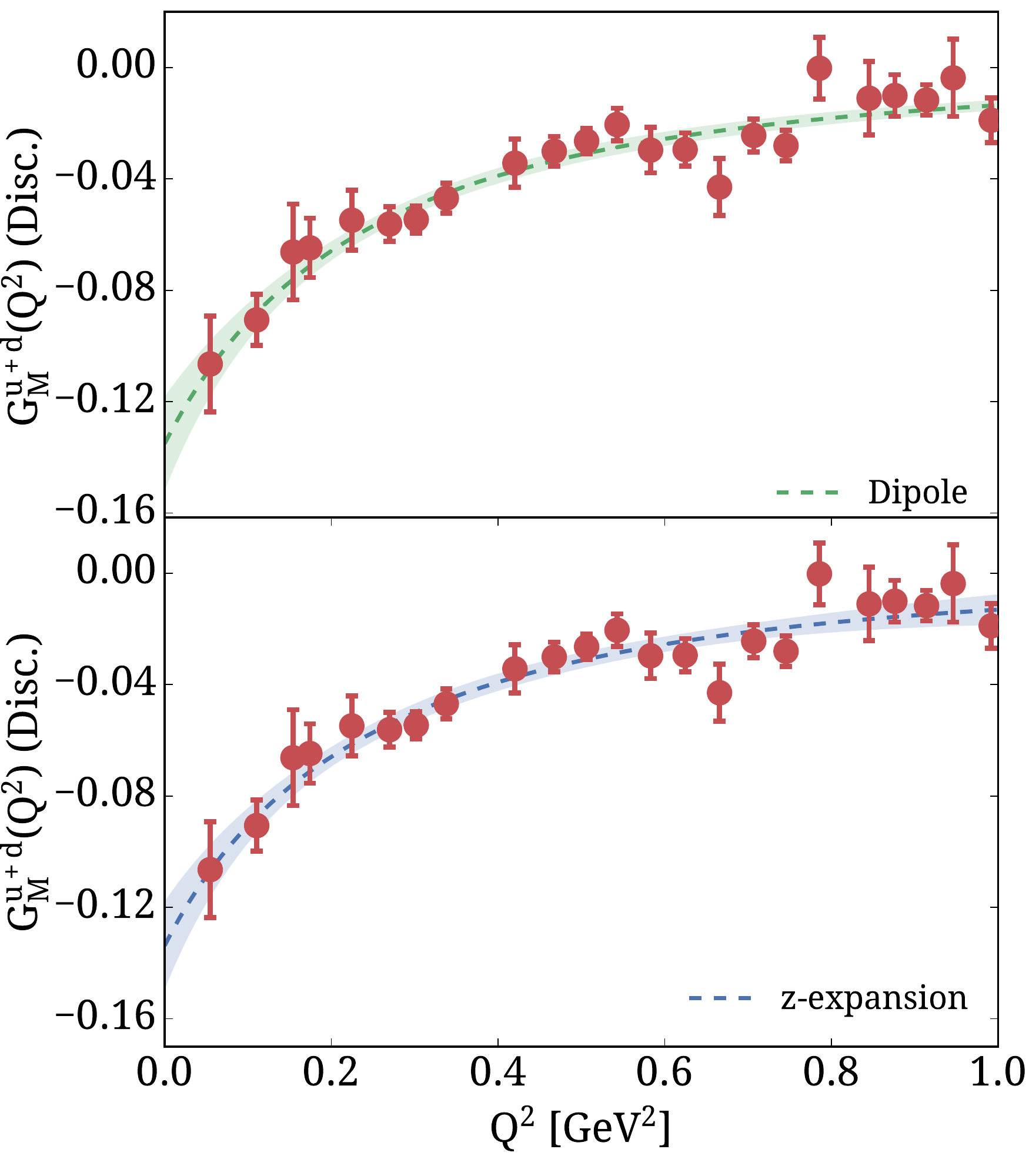}
          \vspace*{-.8cm}
  \caption{Disconnected contributions to the isoscalar magnetic form factor. The notation is the same as in 
  Fig.~\ref{fig:GE_light_disc_std_zExp}.}
  \label{fig:GM_light_disc_std_zExp}
\end{figure}

The disconnected contribution to  $G_M^{u+d}(Q^2)$ is shown in Fig.~\ref{fig:GM_light_disc_std_zExp}. We find that both
dipole and z-expansion are in good agreement. In particular, they yield compatible values at zero momentum transfer.
Like in the case of  the disconnected contribution to $G_E^{u+d}(Q^2)$, for large $Q^2$ the dipole fit has a smaller error band as compared to the z-expansion.
The values extracted from fitting  the disconnected contributions alone are 
\begin{eqnarray}
\langle r^2_E\rangle^{u+d} ({\rm Disc.}) &=& -0.071(6)(4)(6) \; {\rm fm}^2, \\
\mu^{u+d}({\rm Disc.}) &=& -0.134(17)(1)(10), \\
\langle r^2_M\rangle^{u+d} ({\rm Disc.}) &=& -0.136(30)(2)(12) \; {\rm fm}^2, 
\end{eqnarray}
where we have not normalized with the value of the form factor at zero momentum transfer, i.e. the radii are extracted from $\langle r^2 \rangle = -6 \frac{\partial G(Q^2)}{\partial Q^2} \Big \vert_{Q^2=0}\,$ rather than from Eq.~\ref{Eq:radius}.

\begin{figure}[ht!]
  \includegraphics[width=\linewidth]{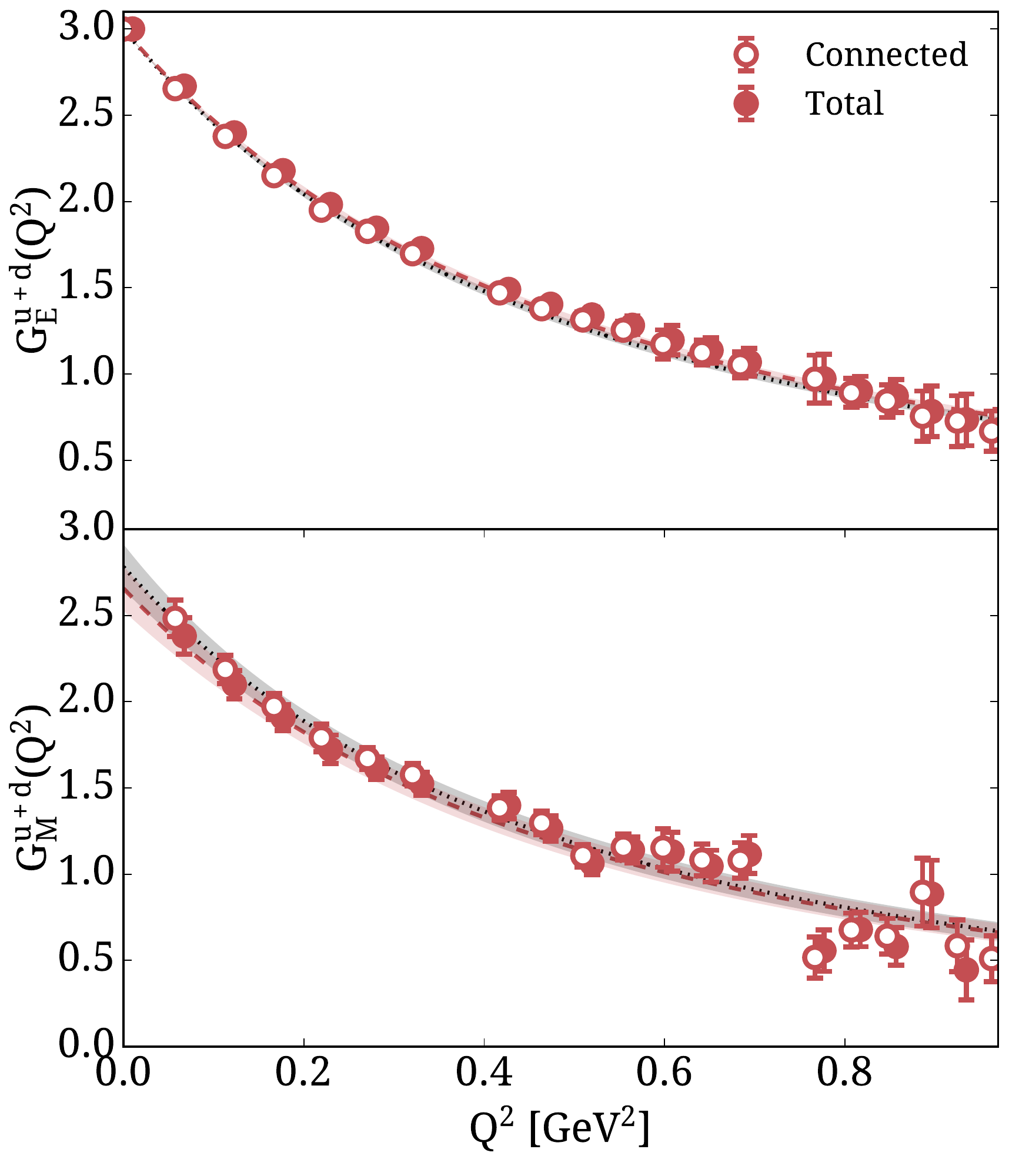}
          \vspace*{-.8cm}
  \caption{Comparison of the connected (open circles) and total  (filled circles) contributions to the isoscalar electric (top) and magnetic (bottom) form factors. Dipole fits to the connected and total contributions are shown with the dotted and dashed curves respectively.}
  \label{fig:GEM_isos_CI_tot_bands}
\end{figure}

In Fig.~\ref{fig:GEM_isos_CI_tot_bands} we show the isoscalar form factors when including and excluding disconnected
contributions. Although the effect is small for both $G_E^{u+d}(Q^2)$ and $G_M^{u+d}(Q^2)$ there is  a systematic  shift affecting the parameters of the fits.
 This comparison shows that disconnected contributions although small are important to include and that their omission would result in an uncontrolled systematic error comparable to the statistical uncertainty. Such systematics need to be under control for precision results required for distinguishing e.g. the two experimental  determinations of the charge radius of the proton.

\begin{figure}[ht!]
  \includegraphics[width=\linewidth]{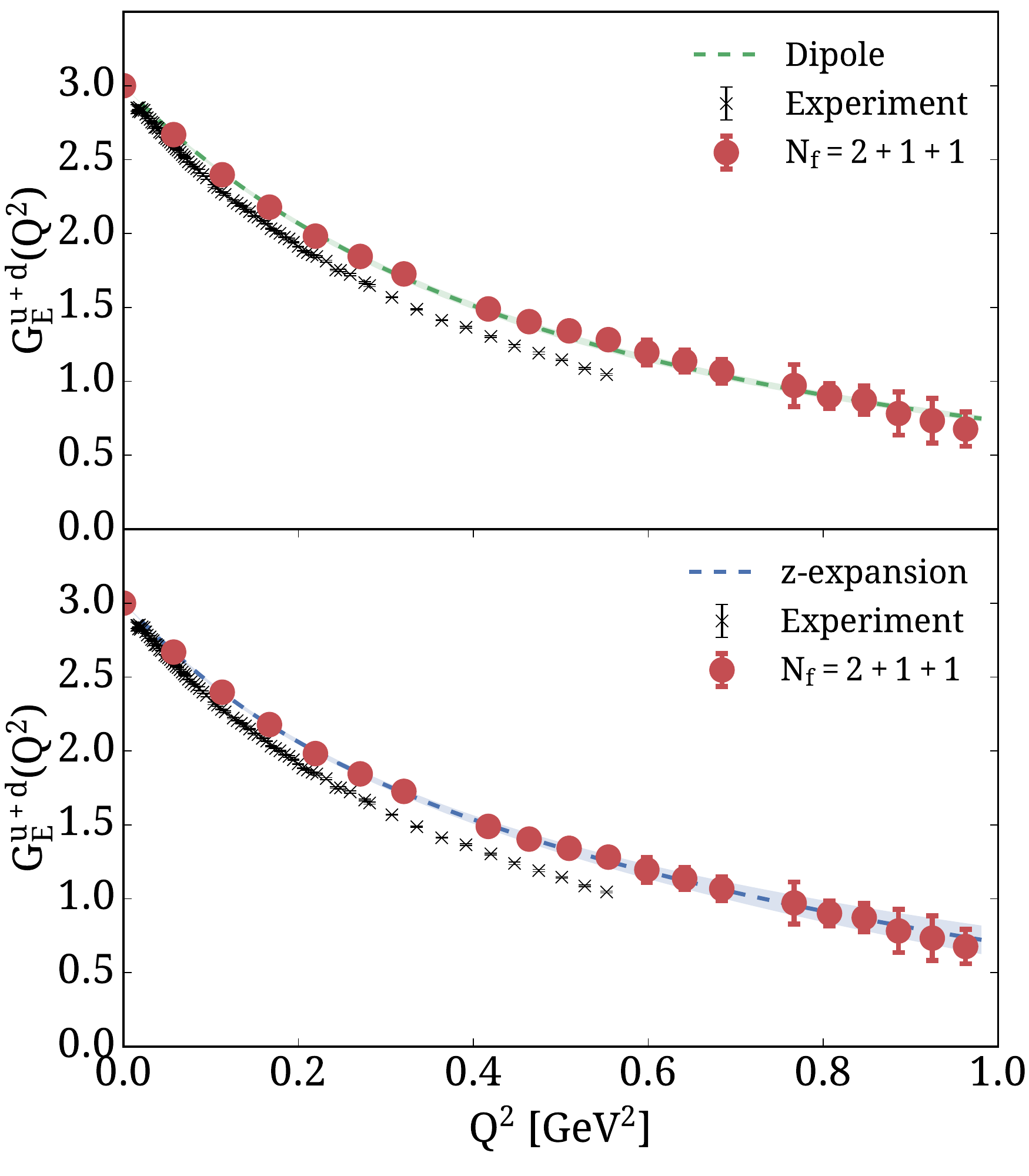}
          \vspace*{-.8cm}
  \caption{Isoscalar electric form factor (circles) as a function of $Q^2$. We combine the connected contribution from the plateau for $t_s/a{=}20$ with the disconnected contribution for $t_s/a{=}14$.
    The remaining notation is as in Fig.~\ref{fig:GE_dip_zExp}.}
  \label{fig:GE_isos_dip_zExp}
\end{figure}

\begin{figure}[ht!]
  \includegraphics[width=\linewidth]{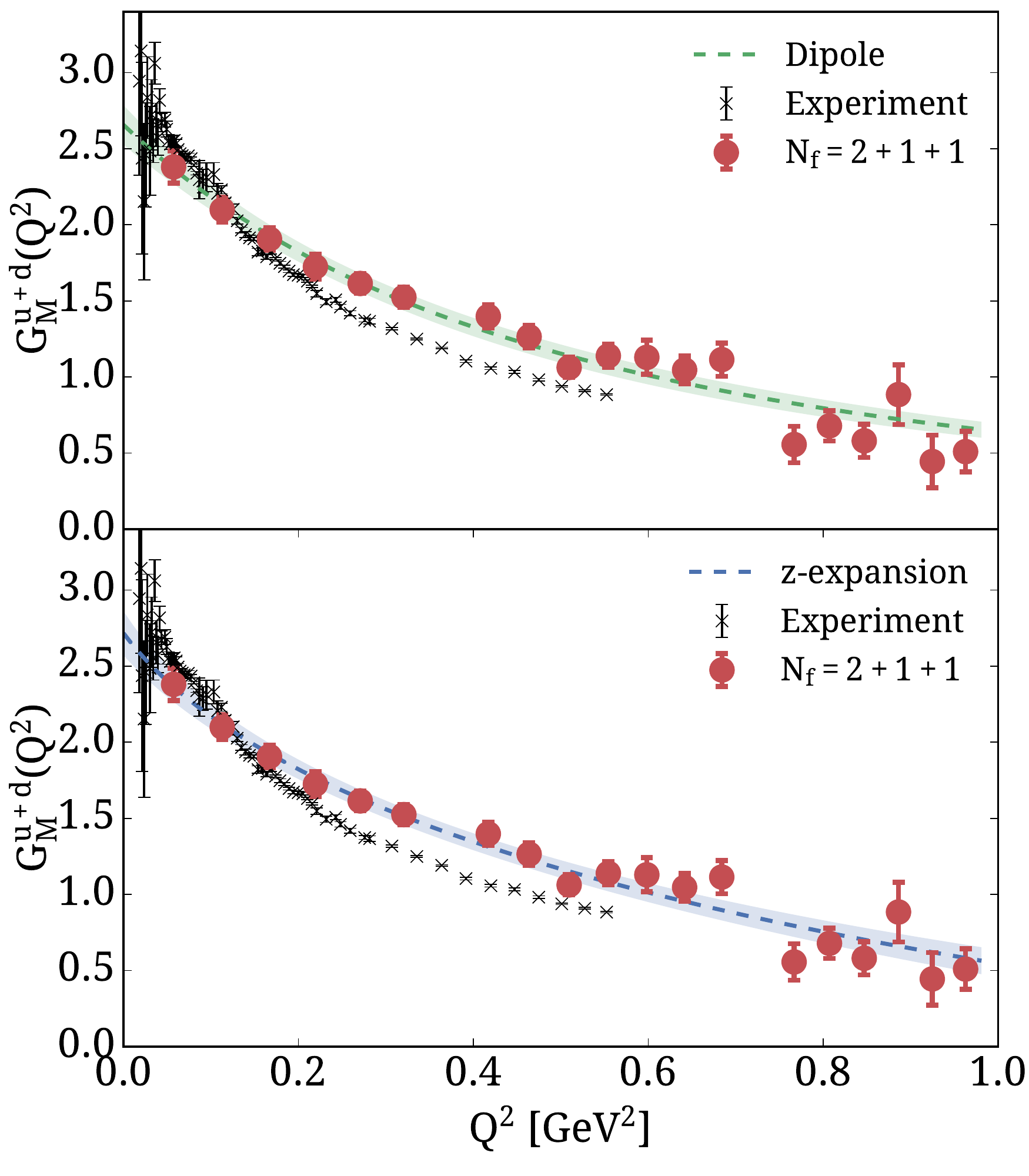}
          \vspace*{-.8cm}
  \caption{Isoscalar magnetic form factor. The notation is as in Fig.~\ref{fig:GE_isos_dip_zExp}.}
  \label{fig:GM_isos_dip_zExp}
\end{figure}

In Figs.~\ref{fig:GE_isos_dip_zExp} and~\ref{fig:GM_isos_dip_zExp} we show the fits of the total isoscalar
electric and magnetic form factors using the dipole form and z-expansion. Both fits describe well the data with the dipole fit being more precise at larger $Q^2$, a behavior also observed for the isovector form factors.
For intermediate $Q^2$ values our results are systematically higher compared to experiment, which is then
reflected in the fit bands. Since for low $Q^2$ there is agreement, the extracted value for the isoscalar magnetic moment agrees with the experimental value.
On the other hand, the slope of our lattice data is not as steep as in the experimental results, which  leads to a smaller value for the corresponding radii.


\begin{figure}[ht!]
  \includegraphics[width=\linewidth]{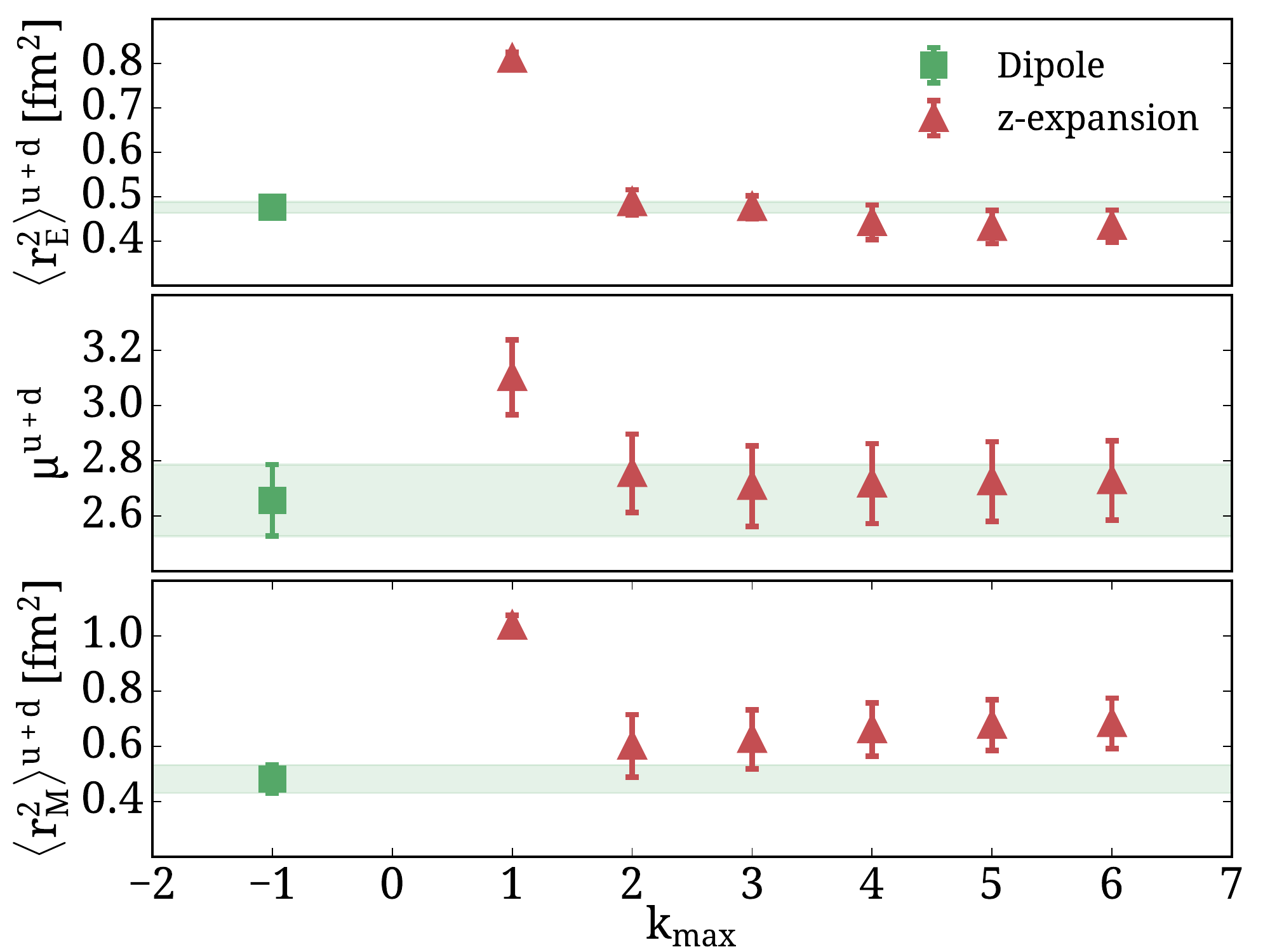}
          \vspace*{-.8cm}
  \caption{Results for the isoscalar charge square radius, magnetic moment, and magnetic square radius. The notation is as in Fig.~\ref{fig:rE2_muM_rM2_isov}.}
  \label{fig:rE2_muM_rM2_isos}
\end{figure}
In the top panel of Fig.~\ref{fig:rE2_muM_rM2_isos} we show the isoscalar electric  square radius. As can be seen, the z-expansion fit yields values that are within errors for $k_{\rm max}>1$ but with twice larger errors than the
dipole. In Fig.~\ref{fig:rE2_muM_rM2_isos} we also show results for the magnetic moment and the magnetic radius where convergence of the z-expansion is
observed already for $k_{\rm max}{=}2$. In general, there is agreement
between the results extracted from the dipole and z-expansion. 
In what follows we will  quote the  values determined from the dipole fits and quote as a systematic error the difference between the mean values of the dipole  and the z-expansion fits. We find
\begin{eqnarray}
  \sqrt{\langle r^2_E \rangle^{u+d}} &=& 0.691(9)(7)(14) \; {\rm fm}\,\\
  \mu^{u+d} &=& 2.66(13)(9)(9) \;\; \text{and} \\
  \sqrt{\langle r^2_M \rangle^{u+d}} &=& 0.695(33)(80)(13)\; {\rm fm}\,.
\end{eqnarray}
Note that from our definition of the isoscalar combination, the proton plus neutron magnetic moment is obtained by: $\mu^p+\mu^n=\mu^{u+d}/3$.


\section{Comparison with other studies} \label{sec:comp}

Before we discuss our final results for the proton and neutron form factors we compare 
with results by other groups  using different lattice QCD ensembles and discretization schemes. These mainly exist for the isovector electromagnetic form factors allowing us  to qualitatively assess lattice artifacts.
This is useful since most groups  use a single 
 ensemble and thus  infinite volume and continuum extrapolations are lacking.
We summarize  the lattice QCD discretized actions used by different groups for the computation of the electromagnetic form factors, restricting ourselves only to published works and results that were obtained using  simulations with pion mass less than 170~MeV:
\begin{itemize}
\item  LHPC analyzed one ensemble of $N_f{=}2{+}1$ with two levels of  HEX-smeared clover fermions with $m_\pi{=}149$ MeV,
  lattice spacing $a{=}0.116$ fm and $L m_\pi {=}4.21$ at   three sink-source time separations from 0.93~fm to 1.39~fm~\cite{Green:2014xba}. 
  They give as their final results the ones extracted using the summation method, which leads to larger statistical errors.
  Additionally, they analyzed an $N_f{=}2{+}1$ ensemble with two levels of  HEX-smeared clover fermions  with $m_\pi = 135$~MeV, lattice spacing
  $a{=}0.093$~fm and $L m_\pi {=} 4$~\cite{Hasan:2017wwt}. They analyzed three lattice separations from 0.93~fm to 1.5~fm and they have extracted results using the summation method. A momentum  derivative method has been used to extract the magnetic moment and the electric radius directly from the correlation functions avoiding a fitting procedure.
\item The PACS collaboration analyzed an ensemble of  $N_f{=}2{+}1$ stout-smeared  clover
  fermions with $m_\pi{=}146$ MeV, $a \simeq 0.085$ fm and a spatial extend of 8.1 fm or $L m_\pi \simeq 6$ allowing access  to relatively small momenta~\cite{Ishikawa:2018rew,Shintani:2018ozy}.  PACS has computed three-point functions for one sink-source time separation of 1.27~fm and they used the plateau method to identify the ground state matrix element. 
\item The $\chi$QCD collaboration~\cite{Sufian:2017osl} computed only the disconnected contributions to the nucleon electromagnetic form factors using a hybrid action of overlap valence quarks and  $N_f{=}2{+}1$ domain wall sea quarks produced by RBC/UKQCD. Their analysis includes an ensemble with pion mass $m_\pi=139$~MeV, $a{=}0.1141(2)$~fm and $L m_\pi {=} 3.86$. They computed  nucleon two-point functions  stochastically using $Z_3$-noise grid sources and disconnected quark loops with $Z_4$-noise grids applying even-odd and time dilution as well as low-mode average.
\item Our results obtained using  the three ensembles of Table~\ref{table:sim}  simulated by  the Extended Twisted Mass Collaboration (ETMC). These include the two analyses of this work, namely the $N_f$=2+1+1 ensemble with $m_\pi$=139~MeV, $a$=0.0801(4)(3)~fm and $Lm_\pi{\simeq}3.6$ and the $N_f{=}2$ cA2.09.64 ensemble with $m_\pi{=}130(1)$~MeV, $a{=}0.0938(3)(1)$~fm and $L m_\pi{\simeq}4$ as well as our results from Refs.~\cite{Alexandrou:2018zdf,Alexandrou:2017ypw}, which were obtained using the $N_f{=}2$ cA2.09.48 ensemble with $L m_\pi{\simeq}3$ and same pion mass and lattice spacing.
\end{itemize}

\begin{figure}[ht!]
  \includegraphics[width=\linewidth]{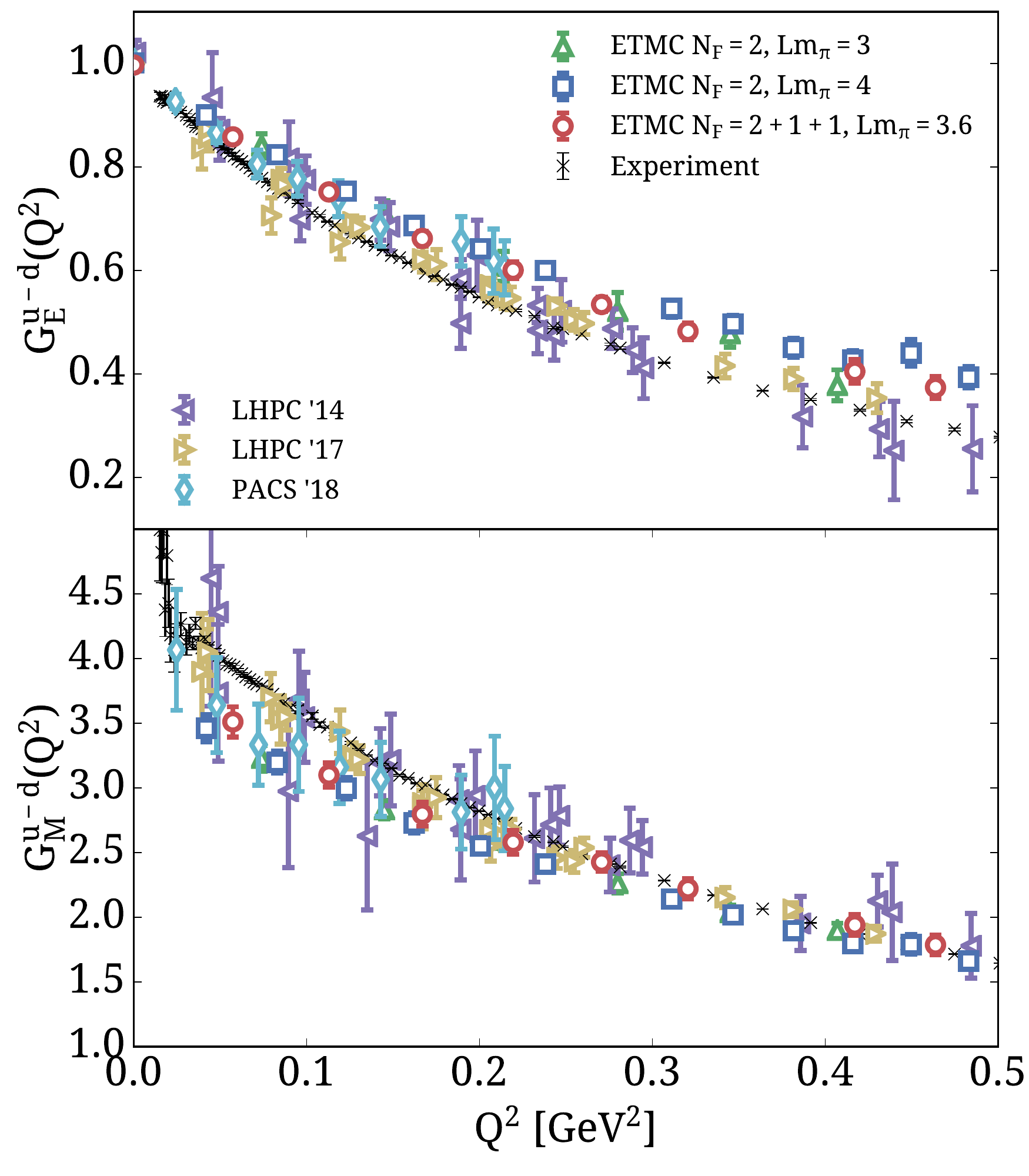}
  \vspace*{-.8cm}\caption{Comparison of results for $G_E^{u-d}(Q^2)$ (upper panel) and $G_M^{u-d}(Q^2)$ (lower panel) from the $N_f{=}2{+}1{+}1$ twisted mass ensemble of this work (red circles), the $N_f{=}2$ twisted mass ensemble with $m_\pi L\simeq 4$ of this work (blue squares), the $N_f{=}2$ twisted mass ensemble with $m_\pi L\simeq 3$ from Ref.~\cite{Alexandrou:2017ypw} (green triangles), LHPC using $N_f{=}2{+}1$ stout-smeared clover fermions from Ref.~\cite{Green:2014xba} (left purple triangles) and Ref.~\cite{Hasan:2017wwt} (right yellow triangles) and from PACS using $N_f{=}2{+}1$ stout-smeared clover fermions from Ref.~\cite{Ishikawa:2018rew} (cyan rhombuses).}
  \label{fig:GEM_comp}
\end{figure}

\begin{figure}[ht!]
  \includegraphics[width=\linewidth]{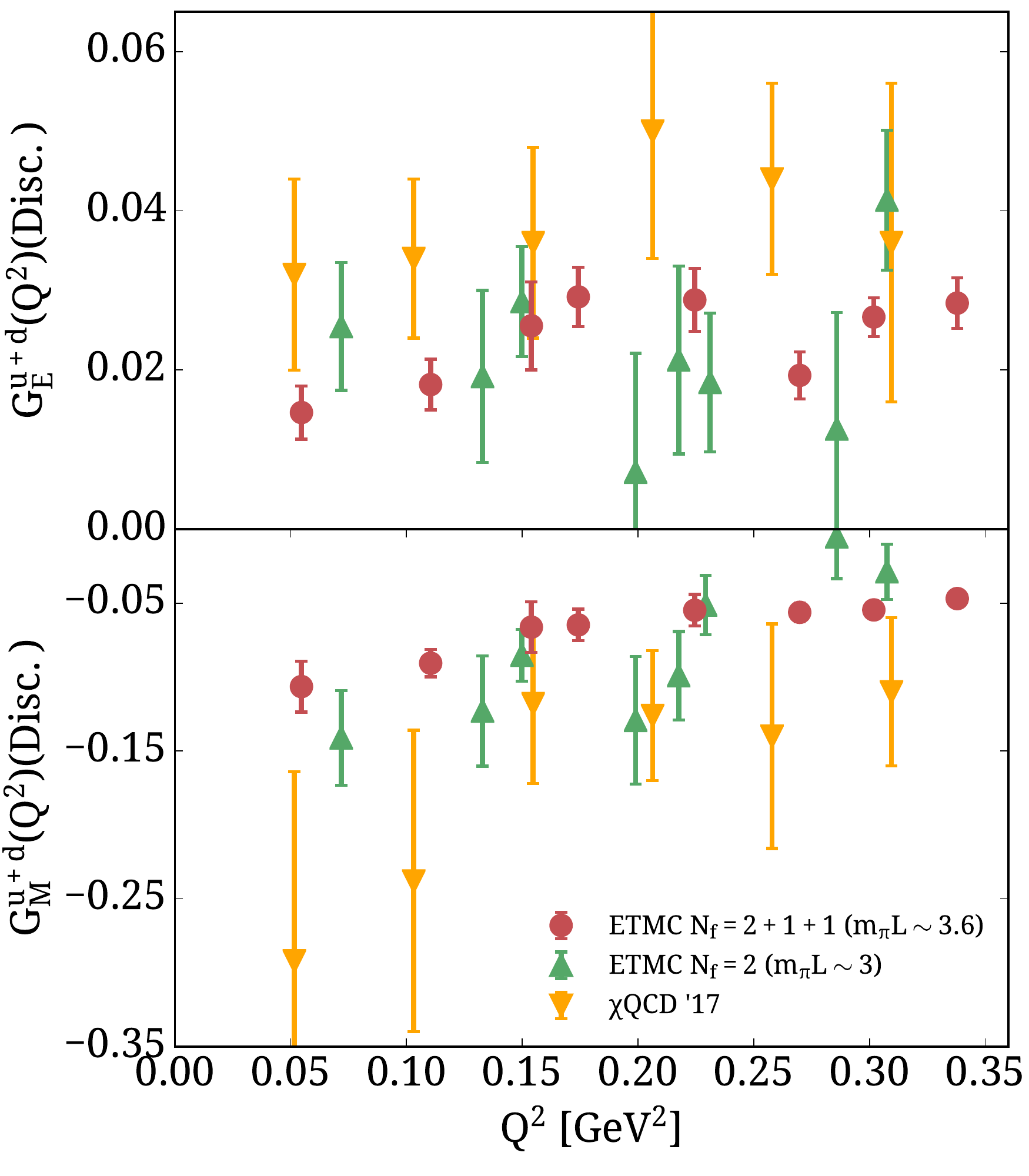}
  \caption{Comparison of the disconnected contributions to $G_E^{u+d}$ (top) and $G_M^{u+d}$ (bottom) from the twisted mass ensemble using $N_f$=2+1+1 from this work (circles) compared to the twisted mass ensemble of $N_f$=2 results of Ref.~\cite{Alexandrou:2018zdf} (up triangles) and the results from the $\chi$QCD collaboration (down triangles) from Ref.~\cite{Sufian:2017osl}.}
  \label{fig:GEGM_disc_comparison}
\end{figure}

\begin{figure}[ht!]
  \includegraphics[width=\linewidth]{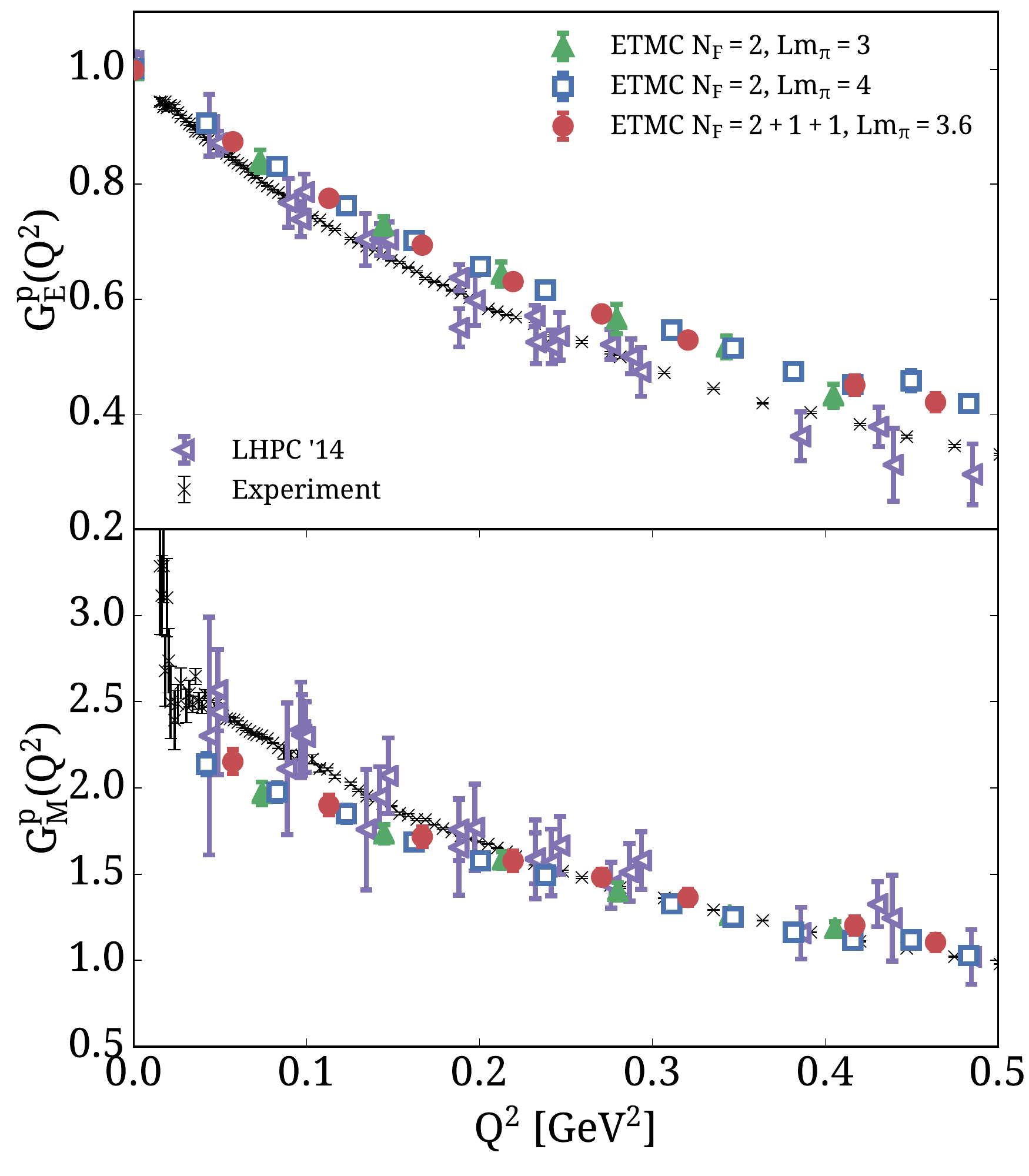}
  \vspace*{-.8cm}\caption{Comparison of results for $G_E^{p}(Q^2)$ (upper panel) and $G_M^{p}(Q^2)$ (lower panel) from ETMC and LHPC following the notation of Fig.~\ref{fig:GEM_comp}. Filled symbols are used for results that include disconnected contributions and open symbols for results without disconnected contributions. Black crosses are experimental results from the A1 collaboration~\cite{Bernauer:2013tpr}.
    }
  \label{fig:GEM_p_comp}
\end{figure}
\begin{figure}[ht!]
  \includegraphics[width=\linewidth]{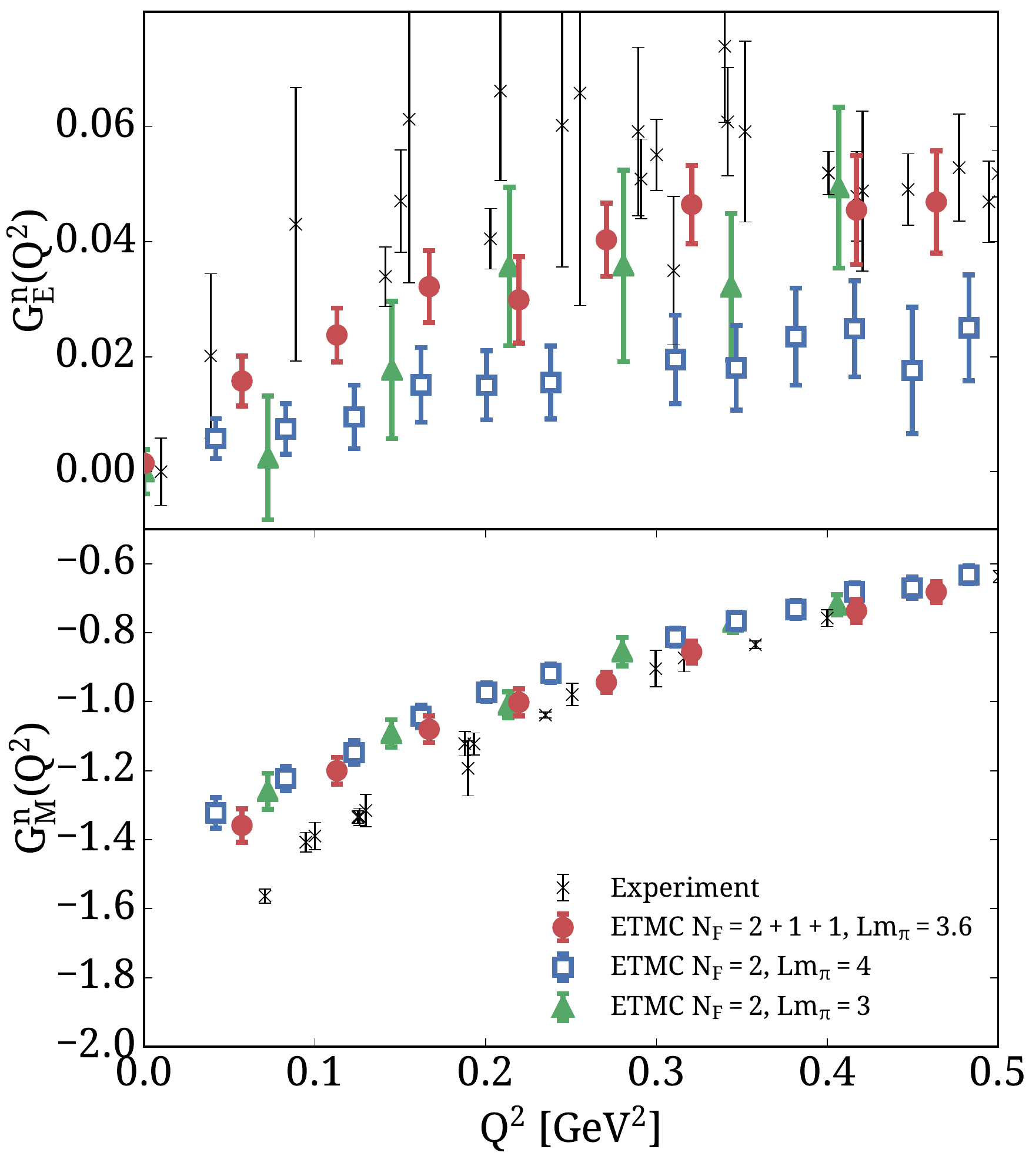}
  \vspace*{-.8cm}\caption{Comparison of results for $G_E^{n}(Q^2)$ and $G_M^{n}(Q^2)$ using the $N_f$=2+1+1 results of this work (red circles), using the $N_f$=2 results with $m_\pi L\simeq4$ of this work (blue squares), and using the $N_f$=2 ensemble with $m_\pi L\simeq3$ from Ref.~\cite{Alexandrou:2018zdf} (green triangles). Filled symbols are used for results that include disconnected contributions and open symbols for results without disconnected contributions. Crosses are experimental results  taken from Refs.~\cite{Golak:2000nt,Becker:1999tw,Eden:1994ji,Meyerhoff:1994ev,Passchier:1999cj,
      Warren:2003ma,Zhu:2001md, Plaster:2005cx, Madey:2003av, Rohe:1999sh, Bermuth:2003qh,Glazier:2004ny,Herberg:1999ud,Schiavilla:2001qe,Ostrick:1999xa} 
    for the electric form factor and from
    Refs.~\cite{Anderson:2006jp, Gao:1994ud,Anklin:1994ae, Anklin:1998ae, Kubon:2001rj, Alarcon:2007zza} 
    for the magnetic form factor.}
  \label{fig:GEM_n_comp}
\end{figure}
\begin{figure}[ht!]
  \includegraphics[width=\linewidth]{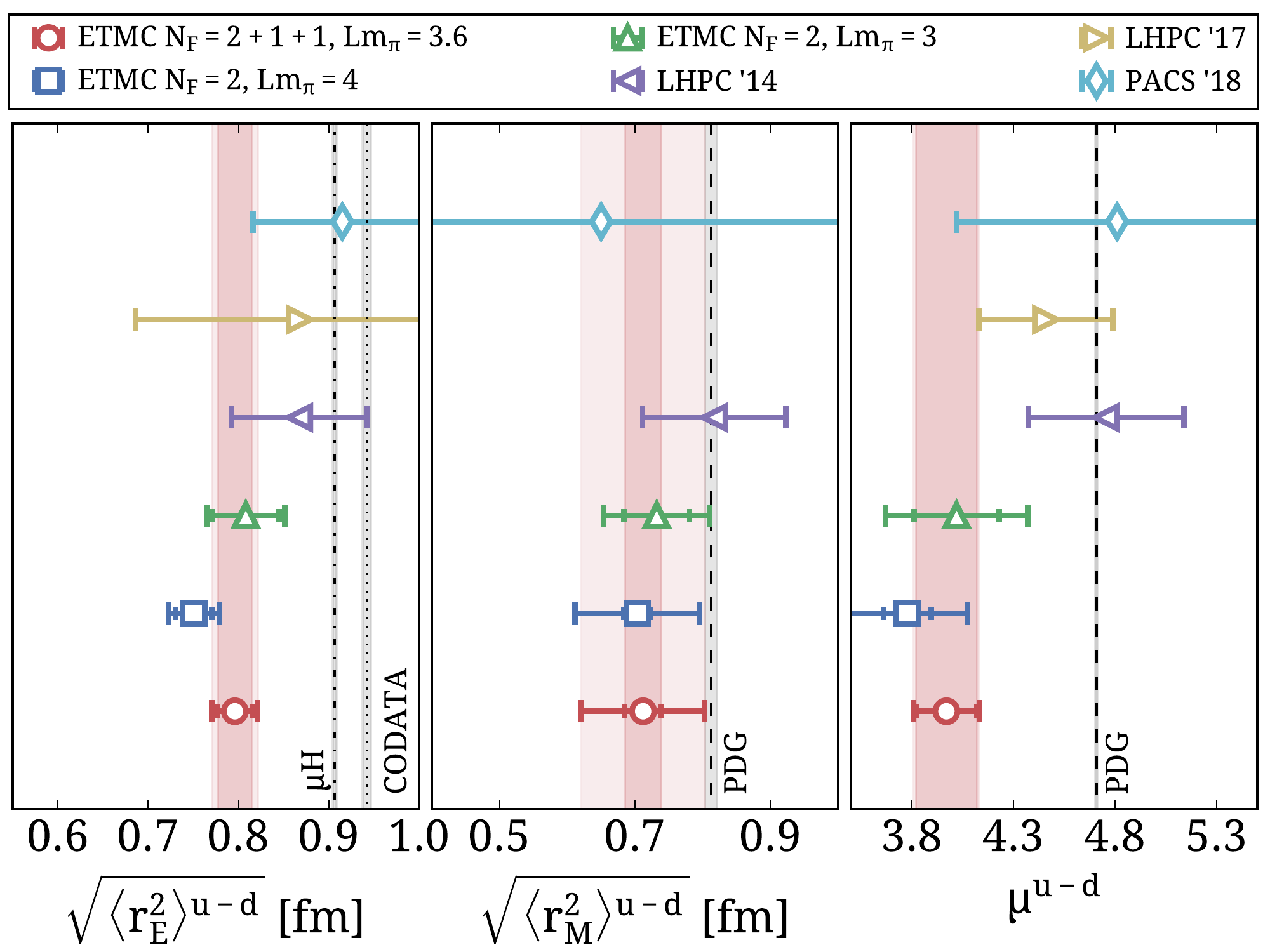}
  \vspace*{-.8cm}\caption{Isovector $\sqrt{\langle r_E^2\rangle^{u-d}}$, $\sqrt{\langle r_M^2\rangle^{u-d}}$ and $\mu^{u-d}$ with lattice QCD results following the notation of Fig.~\ref{fig:GEM_comp}.
    The  experimental result extracted from muonic hydrogen~\cite{Pohl:2010zza}  is shown by the vertical dashed-dotted line and from CODATA~\cite{Mohr:2015ccw} by the dotted vertical line. The PDG value~\cite{Patrignani:2016xqp} is shown with the dashed vertical line. The red vertical inner band denotes the statistical error extracted using the $N_f{=}2{+}1{+}1$ twisted mass ensemble of this work and the outer lighter band is the total error adding statistical and systematic errors in quadrature.}
  \label{fig:rE_rM_muM_isov_comp}
\end{figure}
\begin{figure}[ht!]
  \includegraphics[width=\linewidth]{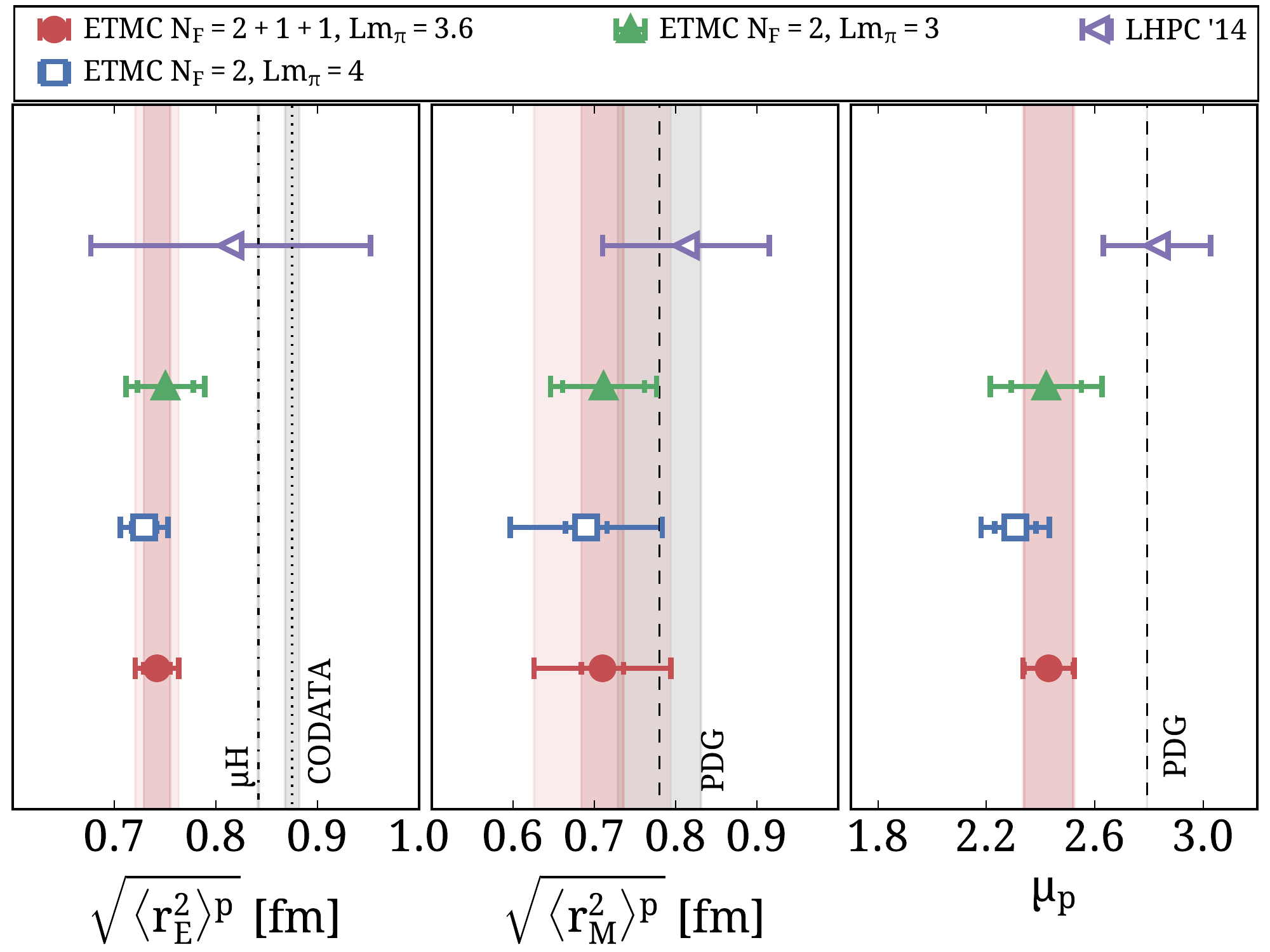}
  \vspace*{-.8cm}\caption{Results for $\sqrt{\langle r_E^2\rangle^{p}}$ and  $\sqrt{\langle r_M^2\rangle^{p}}$ using the same notation as in Fig.~\ref{fig:rE_rM_muM_isov_comp}. Filled symbols denote results that include all contributions whereas open symbols are those where disconnected contributions are neglected. The rest of the notation follows that of Fig.~\ref{fig:rE_rM_muM_isov_comp}.}
  \label{fig:rE_rM_muM_p_comp}
\end{figure}
\begin{figure}[ht!]
  \includegraphics[width=\linewidth]{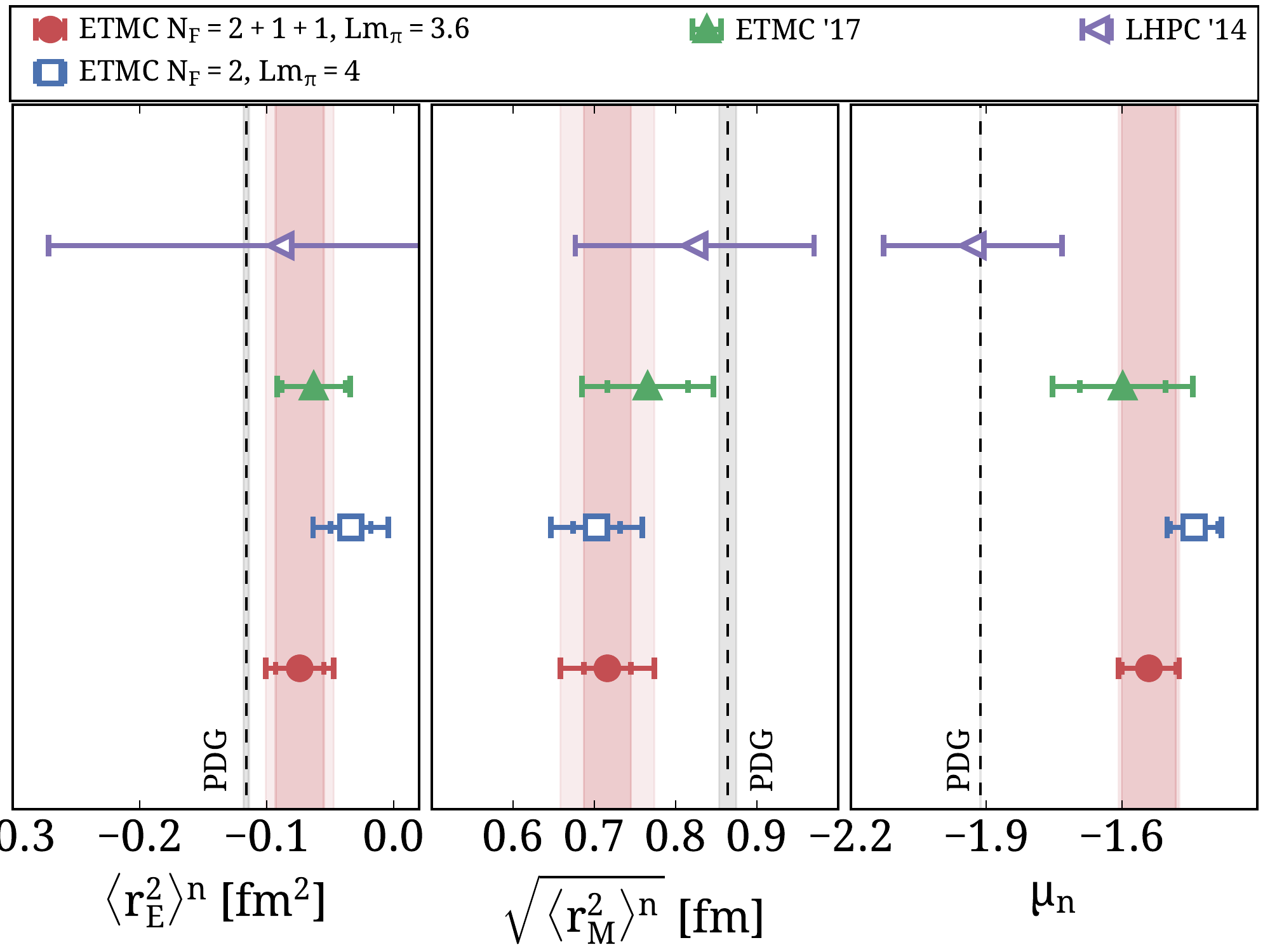}
  \vspace*{-.8cm}\caption{Lattice QCD results for $\langle r_E^2\rangle^{n}$, $\sqrt{\langle r_M^2\rangle^{n}}$ and $\mu_{n}$. The notation is
  as in Fig.~\ref{fig:rE_rM_muM_p_comp}.}
  \label{fig:rE2_rM_muM_n_comp}
\end{figure}

In  Fig.~\ref{fig:GEM_comp} we show a comparison of lattice QCD results  for $G_E^{u-d}(Q^2)$  up to $Q^2{=}0.5$ GeV$^2$ from the analyses  mentioned above.
As can be seen, ETMC and PACS results are in good agreement but systematically higher than the experimental values.
LHPC results  were obtained using the summation method and in general have larger statistical errors making them compatible with both our results and the experimental values. 

In  Fig.~\ref{fig:GEM_comp}, we  also show the corresponding results  for $G_M^{u-d}(Q^2)$.
The  ETMC results of this work are the most precise and in good agreement with those obtained from other studies. We note the very good agreement of lattice QCD results and experiment for $Q^2{>}0.2$~GeV$^2$. As pointed out, the underestimation of lattice QCD results compared to experimental values at smaller $Q^2$ may indicate that a larger spatial volume is required to develop fully the pion contributions. Although our study using two ensembles of $N_f=2$ showed no detectable volume effects when we increase the spatial extent from 4.5~fm to 6~fm (or equivalently from $Lm_\pi\sim 3$ to $Lm_\pi\sim 4$) the volume dependence could be weak and require a larger volume to manifest itself. The new PACS results may indicate such a trend~\cite{Shintani:2018ozy}. 
A conclusion that we can, however, draw from these lattice QCD studies is  that there is agreement among them for both the electric and magnetic form factors. Given  the different discretization schemes
employed, this agreement  indicates that cut-off effects are smaller than the statistical errors.

In Fig.~\ref{fig:GEGM_disc_comparison} we show a comparison of the disconnected contributions to $G_E^{u+d}$ and $G_M^{u+d}$ using results obtained from our $N_f=2+1+1$ and $N_f=2$  twisted mass ensembles and from the hybrid action as analyzed by the $\chi$QCD collaboration~\cite{Sufian:2017osl}. We would like to stress the accuracy of the results of the current work  using the $N_f$=2+1+1 twisted mass ensemble.  In our previous evaluation of the disconnected contributions for the $N_f=2$ twisted mass ensemble we used 2120 configurations with 100 source positions for the computation of the two-point functions and 2250 stochastic vectors for the disconnected loops~\cite{Alexandrou:2018zdf}. This is approximately the same number of inversions (and thus cost) as for the $N_f$=2+1+1 ensemble (see Table~\ref{table:StatsDisc}), which demonstrates  the effectiveness of the hierarchical probing method  employed in the current analysis of the $N_f$=2+1+1 ensemble.

The proton and neutron form factors can be extracted from the isovector and isoscalar form factors discussed in 
Section~\ref{sec:Isov Isos}, using the linear combinations
\begin{eqnarray}
  G^p(Q^2) &=& \frac{1}{2} \left[ \frac{G^{u{+}d}(Q^2)}{3} + G^{u{-}d}(Q^2) \right], \\
  G^n(Q^2) &=& \frac{1}{2} \left[ \frac{G^{u{+}d}(Q^2)}{3} - G^{u{-}d}(Q^2) \right].
\end{eqnarray}
In Fig.~\ref{fig:GEM_p_comp}, we show lattice QCD results for the proton electromagnetic form factors. To extract these, one needs both the isovector and isoscalar combinations. The latter includes disconnected contributions, which have only been computed by ETMC for ensembles with  physical pion masses. We still provide a comparison with the lattice results by LHPC
which however do not include these disconnected contributions. We use filled symbols to indicate lattice QCD results that include disconnected  contributions. For both the proton electric and magnetic form factors LHPC results  are in agreement with ours, with the LHPC results exhibiting larger errors  due to the usage  of the summation method.  The accurate ETMC results are higher than the experimental values for  $G_E^p(Q^2)$, while  for $G_M^p(Q^2)$ they are in agreement except for the two lowest $Q^2$ values.  Unfortunately, LHPC results  carry large errors and in general are compatible both with our values and the experimental ones prohibiting any definite conclusions as to the nature of the discrepancy with the experimental values. As discussed volume and residual excited state effects may lead to a slow convergence of the lattice data that can account for the discrepancies with the experimental values.

Results for the neutron electromagnetic form factors are only provided by the ETMC for pion masses below 170~MeV. They are compared to the experimental values in Fig.~\ref{fig:GEM_n_comp}.
We observe that results for the electric form factor extracted from the cB211.072.64 ensemble that includes disconnected contributions  are in agreement with the experimental values. This is also true for the cA2.09.48 ensemble that includes  disconnected contributions although they carry larger errors. For the cA2.09.64 ensemble, where disconnected contributions have not been included, underestimate the electric neutron form factor. This clearly indicates the significance of including disconnected contributions, especially for this quantity, an observation consistent with the conclusion reached also in Ref.~\cite{Sufian:2017osl}. For the magnetic form factor, results using the cB211.072.64 twisted mass ensemble with disconnected contributions are closer to experiment as to compared to the $N_f=2$ ensembles, but there is still a discrepancy with the experiment for small $Q^2$ values that needs to be further investigated.

In  Fig.~\ref{fig:rE_rM_muM_isov_comp},  we compare the lattice QCD values of  the isovector r.m.s radii  $\sqrt{\langle r_E^2\rangle^{u-d}}$, 
and $\sqrt{\langle r_M^2\rangle^{u-d}}$ finding  agreement among them.
As expected by the less steep fall-off  of the electric isovector form factor, lattice QCD results are systematically lower than the experimental values. We note that the ETMC results have errors that are already the same as the difference between the two experimental  determinations showing that the statistical accuracy required can be achieved. A high-statistics dedicated study to better assess the remaining systematics can thus yield valuable insights on the r.m.s. charge radius from a first principles calculation. 
In the case of   $\langle r_M^2\rangle^{u-d}$ the errors are larger and lattice QCD results are both in good agreement among them and compatible with the PDG value~\cite{Patrignani:2016xqp}.

In Fig.~\ref{fig:rE_rM_muM_p_comp} we show the corresponding  quantities for the proton. Only the ETMC  results include disconnected contributions, which, although small, have a systematic effect. We observe a similar behavior as for the isovector case, namely smaller values for the electric and magnetic r.m.s radii.  LHPC results extracted using the summation method have larger errors and are thus compatible with both the muonic and electron scattering determinations of
the r.m.s. radii.
For the  neutron radii we have only results from ETMC and LHPC. They are displayed in Fig.~\ref{fig:rE2_rM_muM_n_comp}.  ETMC results on the electric r.m.s. radius are determined at high accuracy and include all contributions. Although they are still smaller in magnitude than the experimental values, the discrepancy is  within one standard deviation.
 We note that including disconnected contributions brings better agreement in particular in the case of $\langle r_E^2\rangle^{n}$.


\section{Proton and neutron electromagnetic form factors} \label{sec:PNFF}

\begin{figure}[ht!]
  \includegraphics[width=\linewidth]{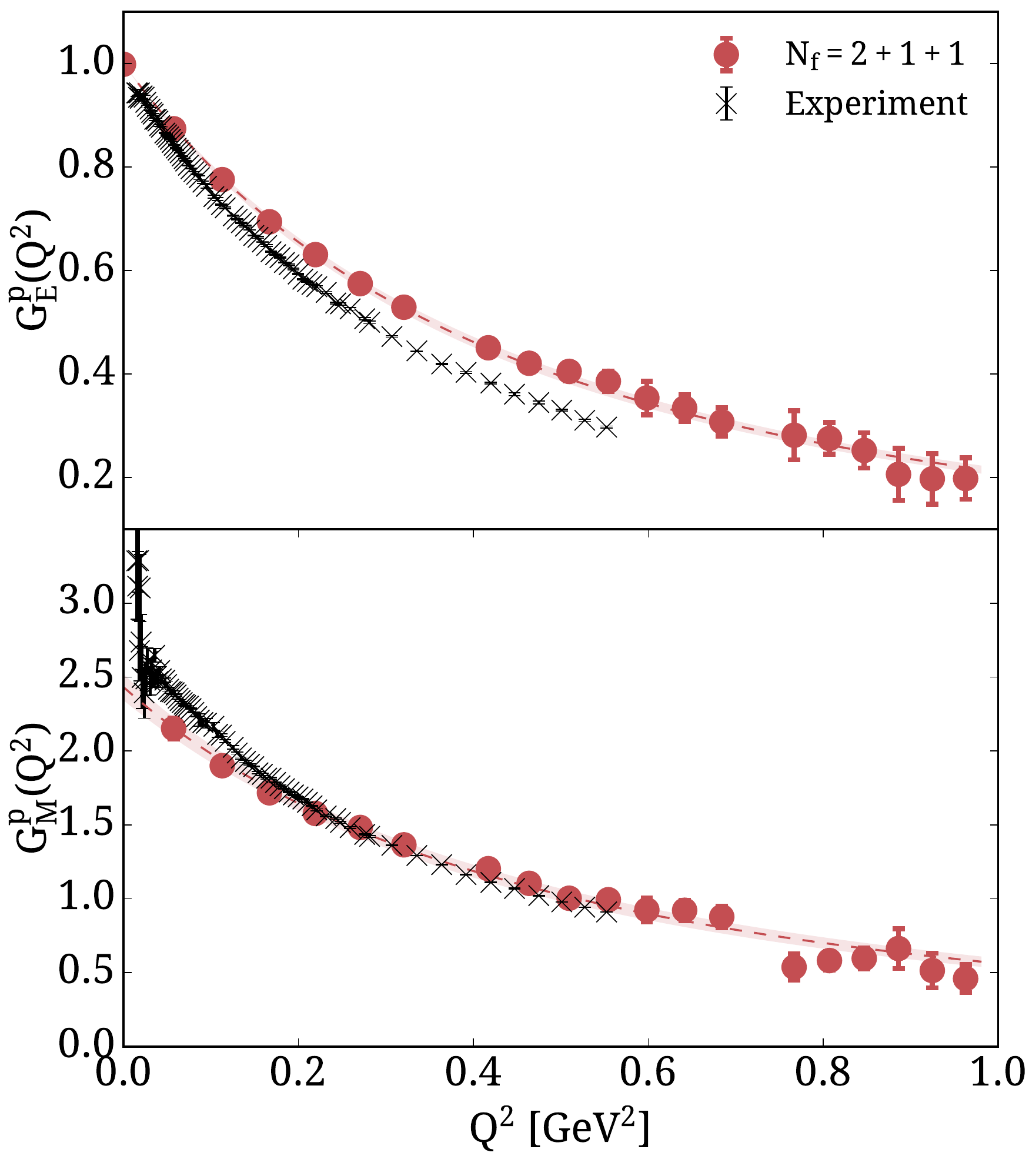}
  \vspace*{-.8cm}\caption{Proton electric (upper panel) and magnetic (lower panel) form factors as a function of $Q^2$.
    Filled circles show the lattice QCD results of this work and black crosses are experimental results from the A1 collaboration~\cite{Bernauer:2013tpr}.
    The band is the fit to our results using Eq.~(\ref{Eq:Dipole}).}
  \label{fig:GEM_p}
\end{figure}

Having compared with other groups and with the $N_f$=2 results from  ETMC, we collect here our final results on the proton and neutron form factors using the $N_f$=2+1+1 ensemble, which has the most accurate results at the physical point.
In Fig.~\ref{fig:GEM_p} we show our results for the proton electric and magnetic form factors compared
to experimental data. As expected from the behavior observed for the isovector and isoscalar electric form factors,
 the proton electric form factor is consistently higher than the  experimental results. The  proton magnetic form
factor agrees with the experiment for all $Q^2$ except the lowest two. This may be due to finite volume or residual excited state effects as discussed in Section~\ref{sec:latArtifacts}. 

\begin{figure}[ht!]
  \includegraphics[width=\linewidth]{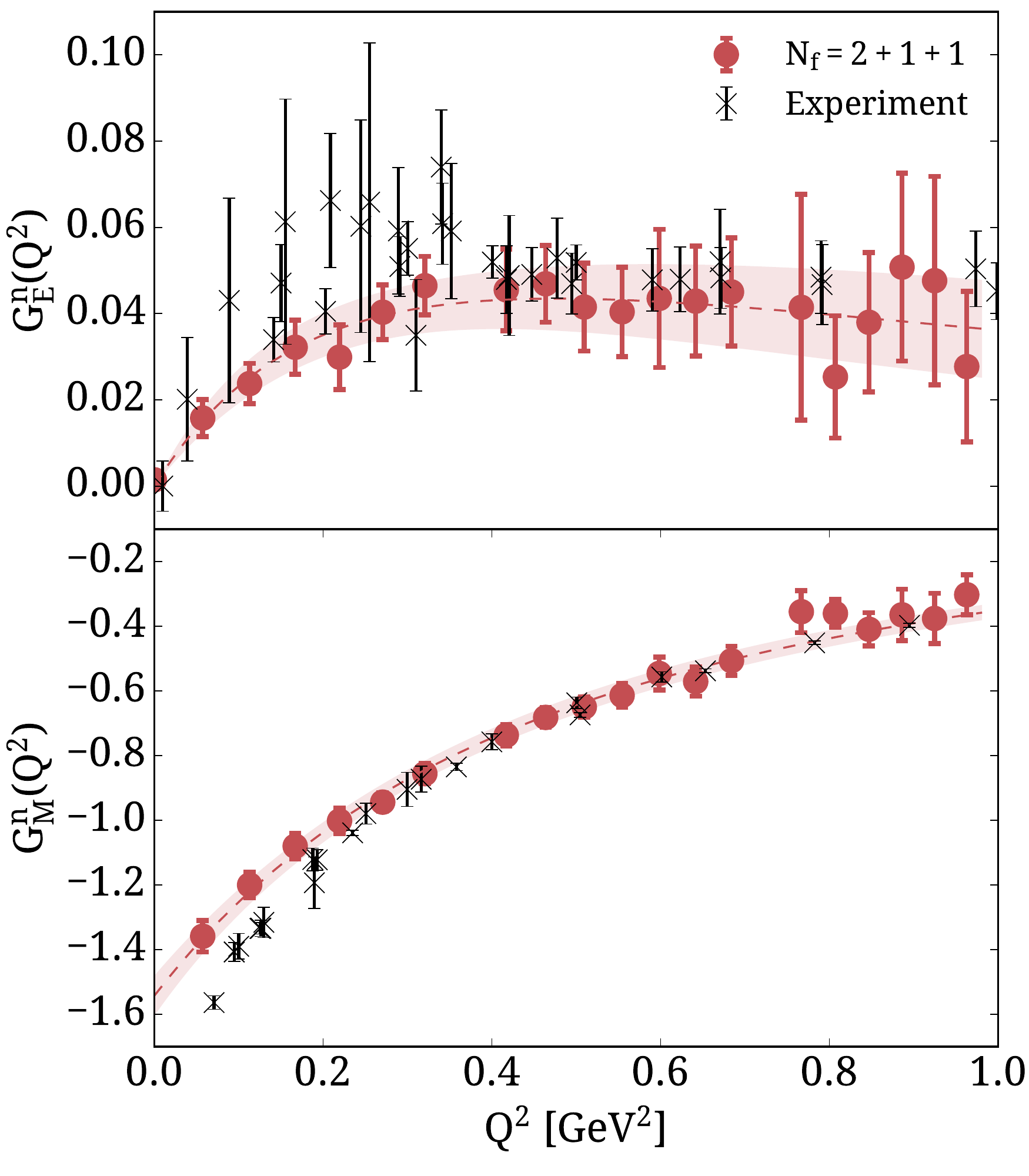}
  \vspace*{-.8cm}\caption{Neutron electric (upper panel)  and magnetic (lower panel) form factors as a function of $Q^2$.
  Filled circles show the lattice QCD results of this work and black crosses are experimental results taken from Refs.~\cite{Golak:2000nt,Becker:1999tw,Eden:1994ji,Meyerhoff:1994ev,Passchier:1999cj,
      Warren:2003ma,Zhu:2001md, Plaster:2005cx, Madey:2003av, Rohe:1999sh, Bermuth:2003qh,Glazier:2004ny,Herberg:1999ud,Schiavilla:2001qe,Ostrick:1999xa}
    for the case of the electric form factor and from
    Refs.~\cite{Anderson:2006jp, Gao:1994ud,Anklin:1994ae, Anklin:1998ae, Kubon:2001rj, Alarcon:2007zza} for the case of the magnetic form factor. The fits to
    our results use Eq.~(\ref{Eq:Galster-like}) for the electric form factor and Eq.~(\ref{Eq:Dipole}) for the magnetic form factors.}
  \label{fig:GEM_n}
\end{figure}

In Fig.~\ref{fig:GEM_n} we show our results for the neutron form factors. The determination of  $G_E^n(Q^2)$ directly from lattice QCD is very promising:
 we find  good agreement with the experimental values 
  but more importantly, at low $Q^2$, the errors from lattice QCD are smaller by up to a factor of four in some cases, allowing for a more precise description of its $Q^2$ dependence. The lattice QCD determination yields also accurate results for 
$G_M^n(Q^2)$ that are in agreement with experiment for $Q^2{>}0.2$ GeV$^2$. At small $Q^2$ we observe the same discrepancy as that observed for the isovector case. Such an underestimation have been seen also for the induced pseudo-scalar form factor where  leading order chiral perturbation theory can show that is due to multi-hadron state contributions with pions. Whether this is the explanation also for the neutron magnetic form factor  remains an open question.

Our results for the proton radii and magnetic moment, as extracted from the dipole fit, are 
\begin{eqnarray}
  \sqrt{\langle r^2_E \rangle^p} &=& 0.742(13)(9)(14)\; {\rm fm}, \nonumber \\
  \mu_p &=& 2.43(9)(1)(3)(^1_0), \nonumber \\
  \sqrt{\langle    r^2_M \rangle^p}    &=& 0.710(26)(80)(6)(^2_0)\; {\rm fm}.
\end{eqnarray}
The corresponding quantities for the neutron using the Galster-like parameterization for the electric and the dipole form for the magnetic are
\begin{eqnarray}
  \langle r^2_E \rangle^n &=& -0.074(16)(16)(8)\; {\rm fm^2}, \nonumber \\
  \mu_n &=& -1.54(6)(2)(3), \nonumber \\
  \sqrt{\langle    r^2_M \rangle^n}    &=& 0.716(29)(44)(24) \; {\rm fm}.
\end{eqnarray}
As already explained, the first error is statistical, the second is an estimate of the systematic due to  the Ans\"atz chosen for the fit  and
the third an estimate of  excited state effects. 
We note here that  disconnected contributions to 
$\langle r^2_E \rangle^n$   are  non-negligible.
If we were to neglect them we would obtain $\langle r^2_E \rangle^{n, conn.}{=}{-}0.063(15)$ fm$^2$, namely more than a 15\% shift in the mean value, i.e. comparable to the other quoted systematic errors.

\section{Summary and conclusions} \label{sec:conclusions}
The nucleon electromagnetic Sachs form
factors are computed using an $N_f{=}2{+}1{+}1$ ensemble of maximally twisted mass fermions with quark masses
tuned to their physical values as well as an ensemble of $N_f{=}2$ twisted mass fermions simulated at a pion mass of 130~MeV. Comparing results calculated using $N_f$=2 and $N_f$=2+1+1 twisted mass ensembles leads to the conclusion that no quenching effects are detected within the accuracy of the results that is within a couple of a percentage. 

A main novelty of this work is the  computation to an unprecedented accuracy of  the disconnected light quark contributions,
allowing us to extract the individual proton and neutron electromagnetic form factors. 
This is accomplished by using state-of-the-art techniques that combine
hierarchical probing and deflation of the lowest eigen-modes and a large number of randomly distributed
smeared point sources in order to suppress gauge noise.
In particular, we find that disconnected contributions to the neutron electric form factor are non-negligible and need to be taken into account to bring agreement with the experimental values. 

Excited states are thoroughly investigated using 
five sink-source time separations in the range of [0.96-1.60]~fm allowing the identification of the ground state to good precision and the determination of a  systematic error due to the excited states by comparing results from the plateau method with the two-state fit method. The summation method is used as a confirmation of the results extracted from the plateau and two-states fits. 

 Our values for the electric and magnetic r.m.s. radii as well as the magnetic moments for the isovector, isoscalar, proton and neutron are collected in Table~\ref{tab:res}.
\begin{table}[ht!]
  \footnotesize
  \caption{Our results for the electromagnetic radii and the magnetic moment using the $N_f{=}2{+}1{+}1$ ensemble for
    the isovector combination  $(p{-}n)$, isoscalar $(p{+}n)$, the proton and neutron. The first error is statistical, the second is a systematic due to the fit Ans\"atz, and the third a systematic due to excited states, derived as explained in the text.}
  \label{tab:res}
  \addtolength{\tabcolsep}{-1.pt}
  \begin{tabular}{c| c | c | c}
    \hline\hline
    & $\sqrt{\langle r_E^2\rangle}$ [fm] & $\sqrt{\langle r_M^2\rangle}$ [fm] & $\mu$ \\
    \hline
    $p-n$ & 0.796(19)(12)(12) & 0.712(27)(87)(5) & 3.97(15)(2)(5)\\
    \hline
    $p+n$ & 0.691(9)(7)(14) & 0.695(36)(80)(13) & 0.89(4)(3)(3) \\
    \hline
    $p$   & 0.742(13)(9)(14)  & 0.710(26)(80)(6) & 2.43(9)(1)(3) \\
    \hline
    \multirow{2}{*}{$n$} & $\langle r_E^2\rangle$ [fm$^2$] & \multirow{2}{*}{0.716(29)(44)(24)} & \multirow{2}{*}{-1.54(6)(2)(3)} \\
    & -0.074(16)(16)(8) & &  \\
    \hline\hline
  \end{tabular}
\end{table}
The results are extracted using the dipole ans\"atz or the Galster-like parameterization and a systematic error on the parameterization used is extracted by comparing with the model independent z-expansion.  
Our result for the proton electric r.m.s  radius is underestimated due to the slower decay of $G_E^p(Q^2)$. Similarly there is an underestimation of the magnetic moments for the proton and neutron.
A most plausible explanation for these remaining discrepancies may come from a combination  of residual volume and multi-hadron contributions. 
Finite volume effects are investigated in this work by comparing two $N_f{=}2$ twisted mass ensembles with pion mass of 130~MeV with the same lattice spacing but $Lm_\pi {\simeq} 3$ and $Lm_\pi {\simeq} 4$. Although we  observe consistent results between these two volumes,  we cannot exclude finite volume effects that may affect the magnetic form factor for small $Q^2$ values as well as the electric form factor. A slow convergence of the results as a function of the volume in combination with effects of multi-hadron states maybe difficult to detect. A theoretical investigation within chiral perturbation theory can shed light on multi-hadronic contributions. Furthermore,  a study on a   larger volume will also  help to probe  adequately volume effects. Thus, further  studies are required to be able to take the infinite volume limit and make definite conclusions on the small $Q^2$ behavior of the magnetic form factor and on the slope of the electric form factor. Finite lattice spacing effects, although are expected to be small, need to also be investigated. Before this program is completed one cannot make final  statements 
 on the two experimental results for the proton charge radius. The ETM collaboration is generating further ensembles in order to enable the investigation of these issues that will require large computational resources.

\begin{acknowledgements}
We would like to thank all members of ETMC for a very constructive and enjoyable collaboration. 
M.C. acknowledges financial support by the U.S. National Science Foundation under Grant No.\ PHY-1714407.
This project has received funding from the Horizon 2020 research and innovation program
of the European Commission under the Marie Sk\l{}odowska-Curie grant agreement No 642069.
S.B. is supported by this program as well as from the project  COMPLEMENTARY/0916/0015 funded by the Cyprus Research Promotion Foundation.
The authors gratefully acknowledge the Gauss Centre for Supercomputing e.V. (www.gauss-centre.eu)
for funding the project pr74yo by providing computing time on the GCS Supercomputer SuperMUC
at Leibniz Supercomputing Centre (www.lrz.de).
Results were obtained using Piz Daint at Centro Svizzero di Calcolo Scientifico (CSCS),
via the project with id s702.
We thank the staff of CSCS for access to the computational resources and for their constant support.
This work also used computational resources from Extreme Science and Engineering Discovery Environment (XSEDE), 
which is supported by National Science Foundation grant number TG-PHY170022. This work used computational resources from the John von Neumann-Institute for Computing on the Jureca system at the research center in J\"ulich, under the project with id ECY00.
\end{acknowledgements}

\bibliography{refs.bib}

\appendix
\widetext

\section{Expressions relating nucleon vector matrix elements to electromagnetic form factors}
\label{sec:appendix equations}

In this Appendix we give a summary of the expressions relating  the Sachs form factors $G_E{\equiv}G_E(Q^2)$ and $G_M{\equiv}G_M(Q^2)$ to the ratio of three-point and two-point functions.
The expressions are given for
 a general frame with initial (final) momentum ${\vec p}$ (${\vec p}^\prime$)
and initial (final) energy $E$ ($E^\prime$). All expressions are given in Euclidean space.

\begin{eqnarray}
\Pi_\mu(\Gamma_0,\vec{p}\,',\vec{p}) &=& \frac{-i C\, G_E}{2 m (4 m^2 + Q^2)} \left( (p^\prime_\mu + p_\mu) \left[m\left(E^\prime  + E  +m\right) -p^\prime_\rho p_\rho\right] \right) \nonumber \\
&&   + \frac{ C\, G_M}{4 m^2(4m^2+Q^2)} \Big( \delta_{\mu 0} \big( 4m^4 + m^2Q^2 + 4m^2 p^\prime_\rho p_\rho + Q^2 p^\prime_\rho p_\rho \big)  \nonumber \\
&&   + 2 i m^2p^\prime_\mu \big(E^\prime  - E \big) - 2 i m^3(p^\prime_\mu+p_\mu) - i E Q^2 p^\prime_\mu - i E^\prime Q^2p_\mu \nonumber \\
&&  - i m Q^2 (p^\prime_\mu+p_\mu) - 2 i m^2 p_\mu \big(E^\prime  - E \big) - 2 i m p^\prime_\rho p_\rho (p^\prime_\mu+p_\mu) \Big)\,,
\label{Eq:GEGMEQ1}
\end{eqnarray}

\begin{eqnarray}
\Pi_\mu(\Gamma_k,\vec{p}\,',\vec{p}) &=&  \frac{- C\, G_E}{2 m (4 m^2 + Q^2)} \Big( m^2 \,\varepsilon_{\mu k 0 \rho} (p_\rho^\prime - p_\rho) - i \,\varepsilon_{\mu k \rho \sigma} p_\rho^\prime p_\sigma \big(E^\prime  + E \big) \nonumber \\
&& + \,\varepsilon_{\mu 0 \rho \sigma}p_\rho^\prime p_\sigma (p_k^\prime + p_k) - \,\varepsilon_{\mu k 0 \rho} p_\sigma^\prime p_\sigma (p_\rho^\prime - p_\rho) \Big) \nonumber \\
&&  - \frac{ C\, G_M}{4 m^2(4m^2+Q^2)} \Big( m \,\varepsilon_{\mu k 0 \rho} (p_\rho^\prime - p_\rho) (2m^2 + Q^2)  \nonumber \\
&&  + 2 i m \,\varepsilon_{\mu k \rho \sigma} p^\prime_\rho p_\sigma \big(2m+E^\prime  + E +\frac{Q^2}{2m} \big) \nonumber \\
&&  - 2 m\,\varepsilon_{\mu 0 \rho \sigma}p^\prime_\rho p_\sigma (p_k^\prime + p_k) + 2 m\,\varepsilon_{\mu k 0 \rho} p^\prime_\sigma p_\sigma (p_\rho^\prime - p_\rho) \Big)
\label{Eq:GEGMEQ2},
\end{eqnarray}

where $C$ is a kinematic factor given by
\begin{equation}
  C= \frac{2m}{E (E(\vec{p}\,')+m)} \sqrt{\frac{E (E(\vec{p}\,')+m)}{E(\vec{p}\,')(E +m)}}\,.
\end{equation}

In the case where $\vec{p}\,'=\vec{0}$ the expressions  simplify as follows
\begin{equation}
  \Pi_0(\Gamma_0,\vec{p}) = C \frac{E + m}{2m} G_E(Q^2),
  \label{Eq:GEQ2_1}
\end{equation}
\begin{equation}
  \Pi_i(\Gamma_0,\vec{p}) = C \frac{p_i}{2m} G_E(Q^2),
    \label{Eq:GEQ2_2}
\end{equation}
\begin{equation}
  \Pi_i(\Gamma_k,\vec{p}) = C \frac{\epsilon_{ijk} p_j}{2m} G_M(Q^2)\,
    \label{Eq:GMQ2}
\end{equation}
and
\begin{equation}
  C= \sqrt{\frac{2m^2}{E(E+m)}}\,.
\end{equation}

\section{Numerical results for the electromagnetic form factors} \label{sec:appendix results}

\begin{table}[!ht]
  \caption{Results for the electromagnetic form factors using the cB211.072.64 ensemble for the isovector combination $G_{E,M}^p-G_{E,M}^n$,  for the proton $G_{E,M}^p$ and neutron $G_{E,M}^n$, including the disconnected contributions for the latter two form factors.}
  \begin{center}
    \addtolength{\tabcolsep}{10pt}
    \begin{tabular}{l|c|c|c|c|c|c}
      $Q^2[GeV^2]$ & $G_E^{p-n} (Q^2)$ & $G_E^p (Q^2)$ & $G_E^n (Q^2)$ & $G_M^{u-d} (Q^2)$ & $G_M^p (Q^2)$ & $G_M^n (Q^2)$ \\ 
\hline\hline
0.000 & 0.997(3) & 0.998(2) & 0.001(1) & NA & NA & NA \\ 
0.057 & 0.858(10) & 0.874(6) & 0.016(5) & 3.516(101) & 2.156(62) & -1.361(43) \\ 
0.113 & 0.752(11) & 0.775(8) & 0.023(5) & 3.105(78) & 1.903(48) & -1.202(33) \\ 
0.167 & 0.662(14) & 0.694(9) & 0.032(7) & 2.801(80) & 1.719(47) & -1.082(35) \\ 
0.219 & 0.601(17) & 0.631(10) & 0.030(8) & 2.583(82) & 1.580(50) & -1.003(35) \\ 
0.270 & 0.534(14) & 0.575(9) & 0.040(7) & 2.430(62) & 1.485(38) & -0.945(26) \\ 
0.320 & 0.482(16) & 0.529(11) & 0.046(7) & 2.224(70) & 1.367(43) & -0.857(29) \\ 
0.417 & 0.405(23) & 0.450(16) & 0.045(10) & 1.943(78) & 1.200(48) & -0.743(32) \\ 
0.464 & 0.374(21) & 0.420(15) & 0.047(9) & 1.789(71) & 1.104(44) & -0.684(29) \\ 
0.510 & 0.363(25) & 0.404(18) & 0.041(11) & 1.655(69) & 1.012(42) & -0.644(31) \\ 
0.554 & 0.345(27) & 0.385(21) & 0.040(11) & 1.610(84) & 0.994(50) & -0.616(37) \\ 
0.598 & 0.310(43) & 0.351(33) & 0.041(16) & 1.472(127) & 0.923(79) & -0.549(51) \\ 
0.642 & 0.291(33) & 0.336(27) & 0.045(13) & 1.495(109) & 0.926(67) & -0.570(44) \\ 
0.684 & 0.263(34) & 0.308(27) & 0.046(13) & 1.386(112) & 0.866(71) & -0.521(44) \\ 
0.767 & 0.239(60) & 0.283(48) & 0.043(27) & 0.893(149) & 0.532(90) & -0.361(65) \\ 
0.807 & 0.250(39) & 0.278(31) & 0.028(15) & 0.942(101) & 0.583(62) & -0.361(43) \\ 
0.847 & 0.213(43) & 0.249(35) & 0.035(16) & 1.006(117) & 0.605(70) & -0.402(51) \\ 
0.886 & 0.158(61) & 0.203(51) & 0.048(22) & 1.028(211) & 0.662(135) & -0.367(80) \\ 
0.925 & 0.150(58) & 0.205(50) & 0.056(24) & 0.891(190) & 0.539(115) & -0.353(79) \\ 
0.963 & 0.172(48) & 0.200(41) & 0.029(18) & 0.765(153) & 0.461(94) & -0.302(61) \\ 

      \hline
    \end{tabular}
  \end{center}
\end{table}

\end{document}